\documentclass[12pt,titlepage]{article}

\usepackage[default]{jasa_harvard}    % for formatting citations in text
\usepackage{JASA_manu}

\newcommand {\ctn}{\citeasnoun} % change to \citet if using natbib
\usepackage{graphicx,subfigure,amsmath,latexsym,amssymb}
\usepackage{float,epsfig,multirow,rotating,times}
\usepackage{upgreek,wrapfig}
\usepackage{comment}
\usepackage{slashbox}

\newcommand{\blambda}{\boldsymbol{\lambda}}

\newcommand{\bLambda}{\boldsymbol{\Lambda}}

\newcommand{\bbeta}{\boldsymbol{\beta}}

\newcommand{\bxi}{\boldsymbol{\xi}}

\newcommand{\bSigma}{\boldsymbol{\Sigma}}

\newcommand{\bmu}{\boldsymbol{\mu}}

\newcommand{\bC}{\boldsymbol{C}}

\newcommand{\bD}{\boldsymbol{D}}

\newcommand{\bG}{\boldsymbol{G}}

\newcommand{\bA}{\boldsymbol{A}}

\newcommand{\bE}{\boldsymbol{E}}
\newcommand{\bp}{\boldsymbol{p}}
\newcommand{\bP}{\boldsymbol{P}}

\newcommand{\bv}{\boldsymbol{v}}

\newcommand{\bx}{\bm{x}}
\newcommand{\bX}{\boldsymbol{X}}

\newcommand{\bzero}{\boldsymbol{0}}

\newcommand{\bm}{\mathbf}

\numberwithin{equation}{section}
\numberwithin{algo}{section}
\numberwithin{table}{section}
\numberwithin{figure}{section}

\begin{document}

\normalsize

\title{\vspace{-0.8in}
Effects of Gene-Environment and Gene-Gene Interactions
in Case-Control Studies: A Novel Bayesian Semiparametric Approach}
\author{Durba Bhattacharya and Sourabh Bhattacharya\thanks{
Durba Bhattacharya is an Assistant Professor in St. Xavier's College, Kolkata, pursuing PhD in 
Interdisciplinary Statistical Research Unit, Indian Statistical
Institute, 203, B. T. Road, Kolkata 700108. Sourabh Bhattacharya is an Associate Professor in
Interdisciplinary Statistical Research Unit, Indian Statistical Institute, 203, B. T. Road, Kolkata 700108.
Corresponding e-mail: sourabh@isical.ac.in.
}}
\date{\vspace{-0.5in}}
\maketitle%

\begin{abstract}
Present day bio-medical research is pointing towards the fact that virtually almost all 
diseases are manifestations of complex interactions of genetic susceptibility factors 
and modifiable environmental conditions. Cognizance of gene-environment interactions may help 
prevent or detain the onset of complex diseases like cardiovascular disease, cancer, type2 diabetes, 
autism or asthma by adjustments to lifestyle. 

In this regard, we extend the Bayesian semiparametric gene-gene interaction model of \ctn{Bhattacharya16}
to detect not only the roles of genes and their interactions, but also the possible influence of environmental
variables on the genes in case-control studies. Our model also accounts for the unknown number
of genetic sub-populations via finite mixtures composed of Dirichlet processes, which are
related to each other through a hierarchical matrix-normal structure, incorporating gene-gene and gene-environment
interactions.
An effective parallel computing methodology, developed by us harnesses the power of
parallel processing technology to increase the efficiencies of our conditionally independent Gibbs sampling
and Transformation based MCMC (TMCMC) methods.

Applications of our model and methods to simulation studies with biologically realistic
case-control genotype datasets obtained under five distinct set-ups of gene-environment interactions action yield encouraging results in each case.
We followed these up by application of our ideas to a real, case-control based genotype dataset on early onset of myocardial infarction. Beside being in broad agreement with the reported literature on
this dataset, the results obtained give some interesting insights to the differential effect of gender on MI.
\\[2mm]
{{\bf Keywords}: \it
Case-control study; Dirichlet process; Gene-gene and gene-environment interaction; Matrix normal; 
Parallel processing; Transformation based MCMC. 
}
\end{abstract}

%\begin{keywords}
%Case-control study; Dirichlet process; Gene-gene and gene-environment interaction; Matrix normal; 
%Parallel processing; Transformation based MCMC. 
%\end{keywords}

\maketitle

\tableofcontents

\newpage

\section{{\bf Introduction}}
\label{sec:intro}

%\subsection{{\bf Importance of studying gene-environment interaction along with gene-gene interaction}}
%\label{subsec:importance of ge}

%With the completion of the Human Genome Project and rapid development in the field of molecular biology in the recent years, %there has been a dramatic increase in our understanding of the genetic influences on many diseases. Unfortunately, inspite of %many genetic association studies, including genome-wide association studies (GWAS), only a small portion of the genetic %heritability of many complex diseases like cardiovascular diseases, autism or cancer has so far been described.

Although many people tend to classify the cause of a disease as either genetic or environmental, only a few diseases like Huntington's Disease(HD) or GM2 gangliosidosis have so far been identified as purely genetic disorders.  As indicated by many epidemiological studies, a different effect of a genotype is often observed on disease risk in persons with different environmental exposures (See \ctn{Mapp03}, \ctn{Khouri05}). Also there may be multiple genes which interact with each other to cause a disease only when an environmental factor passes a given threshold, implying thereby that presence of a risk allele may not be exposing all individuals to the same risk.
%For example, an individual having a predisposing polymorphism for bronchial asthma may or may not develop asthma, depending %on the environmental risk factors like place of upbringing, diet, work history and psycological events he is exposed to.

\ctn{Hunter05} and \ctn{Mather76}, point out that estimation of only the separate contributions of genes and environment to a disease, ignoring their interactions, will lead to incorrect estimation of the proportion of the disease (the ``population attributable fraction'') that is explained by the genes, the environment, and their joint effect.

Study of gene-environment interaction is important to the field of pharmacogenetics also, since the efficacy and side-effects of some medications can vary depending on an individual's genotype (see \ctn{Scott11}). Hence, extensive study of gene-environment interactions through sophisticated statistical modelling is necessary to devise new methods of disease prevention, detection and intervention.

%\subsection{{\bf Statistical definition and mechanism of gene-environment interaction}}
%\label{subsec:statistical_definition}

Gene-environment interaction is often conceptualized as genetic control of sensitivity to different environments (\ctn{Purcell02}).   
According to \ctn{Mather76} (see also \ctn{Ottman10}) gene-environment interaction is defined as ``a different effect of an environmental exposure on disease risk in persons with different genotypes''.
As genes are the fundamental units of change in an environmental response system, in order to model the gene-environment interaction effectively, it is important to understand the mechanism through which genes and environment interact together to bring about a physiological change in an individual.
An environmental exposure could trigger a physiological change in a number of ways. Exposure to certain environmental stimuli may directly or indirectly alter the epigenome of an individual. Exposure to mutagens like high doses of x-ray or nuclear radiation, smoking etc. can enter into the body through tissues and directly interfere with the DNA sequence or replication mechanism. %For example, association of cigarette smoking with lung cancer has been indicated in many studies. %On the other hand chronic stress, may stimulate the body to produce its own inherent epigenetic factors. 
Some environmental stimuli may affect DNA indirectly by altering transcription factors and hence changing the expressions of certain genes. 
Many gene-gene interactions have been shown to be started by some environmental exposure. For example, excessive alcohol intake has been shown to suppress TACE gene, which then activates less MTHFR, resulting in reduced folate metabolism, causing depression.

Although the study of gene-environment interaction has become essential to the understanding of the aetiology of 
almost every disease, very little success has so far been achieved in this field. This want of success may be 
attributed to many causes like inadequacy of models incorporating the complex mechanism through 
which genes and environment may affect a disease risk (\ctn{Wang10}).
Indeed, given the complexity involved in the
gene-environment interactions, no simple linear or additive relationship alone can model the relationship effectively. 
According to \ctn{Wright02} and \ctn{Wang10}, although statistical definition of gene environment interaction may 
lack clear biological interpretations, quantification of biological interaction should be based on statistical 
concepts of interaction. 
Furthermore, inadequacy of data regarding 
environmental exposure of individuals and stratified population structure are also important factors
impeding success of the existing methods in this field. 
Association tests based on a pooled set of genetically diverse subpopulations 
(i.e., having differences in allele frequencies across subpopulations) may result in extremely inflated 
rates of false positives (see \ctn{Bhattacharjee10}). 

The above discussion points towards the fact that the widely-used log-linear models (see, for example, 
\ctn{Mukherjee08a}, \ctn{Mukherjee08b}, \ctn{Mukherjee10}, \ctn{Mukherjee12}, \ctn{Sanchez12}, 
\ctn{Ahn13}, \ctn{Ko13}) 
are perhaps not quite adequate for modeling complex gene-gene and gene-environment interactions. Moreover, such models
consider quite restrictive and ad-hoc association structures for simplifying computation
and only attempt to test whether or not the interaction 
is present without being able to quantify the strength of the interaction. Uncertainty regarding unknown 
number of subpopulations are also not generally accounted for in the existing interaction models.

Our Bayesian hierarchical mixture model framework is aimed at incorporating all the aforementioned 
desirable mechanisms through which gene-environment interaction, along with the isolated effects of genes 
and their interactions may affect an 
individual's risk of being affected by a disease, taking into account the fact that the underlying 
population may be stratified in nature. Since the number of sub-populations is not usually known, one must 
coherently and carefully account for the uncertainty associated with the unknown number of sub-populations. 
An additional feature of our model is learning about the number of underlying genetic sub-populations.
%Our model and the associated methodologies generalize the work of \ctn{Bhattacharya16} (henceforth, BB), 
%who address gene-gene interaction, but not gene-environment interaction. 
%Our model generalizes that of \ctn{Bhattacharya16} (henceforth, BB) whose goal was to study gene-gene interactions 
%when environmental variables are known to be irrelevant.
%To detect the roles of genes, environment, gene-gene and gene-environment interactions, we also extend the 
%Bayesian hypotheses testing methods of BB, and for the purpose of computation
%we develop a powerful parallel Markov chain Monte Carlo (MCMC) algorithm which exploits the conditional 
%independence structures inherent in our Bayesian model, and combines the efficiencies of our Gibbs sampling method associated
%with the mixtures and Transformation based MCMC (TMCMC) of \ctn{Dutta14}.

%The main differences of our work with theirs are as follows. 
Because of dependence on environmental variables, our Bayesian semiparametric 
model comprises Dirichlet process based finite mixture models even at the individual subject level, in addition
to genetic and case-control status. The mixtures share a complex dependence structure between themselves
through suitable hierarchical matrix-normal distributions, suitably taking account of the dependence
induced by the environmental variable.
To detect the roles of genes, environment, gene-gene and gene-environment interactions, we extend the gene-gene
interaction model and the associated
Bayesian hypotheses testing methods of \ctn{Bhattacharya16} (henceforth, BB), and for the purpose of computation
we develop a powerful parallel Markov chain Monte Carlo (MCMC) algorithm which exploits the conditional 
independence structures inherent in our Bayesian model, and combines the efficiencies of our Gibbs sampling method associated
with the mixtures and Transformation based MCMC (TMCMC) of \ctn{Dutta14}.

The rest of our paper is structured as follows. 
We introduce our proposed Bayesian semiparametric gene-environment interaction model
in Section \ref{sec:proposal}. 
%and in Section \ref{sec:computation} we propose a fast and efficient
%parallel MCMC methodology for fitting our model.
%The methodology combines efficient Gibbs sampling strategies with an efficient TMCMC scheme, while
%exploiting the conditional independence structure of our model to build a highly efficient parallel MCMC
%algorithm.
In Section \ref{sec:detection} we extend the Bayesian hypothesis testing procedures proposed in BB
to learn about the roles of genes, environmental variables and their interactions in case-control studies.
In Section \ref{sec:simulation_study} we demonstrate the validity of our model and methods with successful applications
to five biologically realistic simulated data sets associated with five different set-ups. 
We also analysed a case-control type myocardial infarction data set obtained from dbGap with our 
model and methods, the results of which we  
report and discuss in detail in Section \ref{sec:realdata}. As we point out, our results broadly agree
with and in some cases contrast the existing results on this data set.
Finally, we summarize our work with concluding remarks in Section \ref{sec:conclusion}.
Further details are provided in the supplement, whose sections and figures have the prefix
``S-" when referred to in this paper.

\section{{\bf A new Bayesian semiparametric model for gene-gene and gene-environment interactions}}
\label{sec:proposal}

\subsection{{\bf Case-control genotype data}}
\label{subsec:data}

For $s=1,2$ denoting the two chromosomes, let $x^s_{ijkr}=1/0$  indicate respectively
the presence and absence of the minor allele at $r$-th locus of the $j$-th gene for the $i$-th individual belonging to 
the $k$-th group of case/control, where $k=0,1$, with $k=1$ denoting case; 
$i=1,\ldots,N_k$; $r=1,\ldots,L_j$ and $j=1,\ldots,J$; let $N=N_1+N_2$.
Let $\bE_i$ denote a set of environmental variables associated with the $i$-th individual.
In what follows, we model this case-control genotype data, along with the information on the environmental variables using our
Bayesian semiparametric model, described in the next few sections.

%In this paper, we shall concern ourselves with data sets of the aforementioned type. However,
%for our model, which we introduce below, it is obvious that data sets consisting of only minor allele counts at each locus
%contains exactly the same information as the above described data type.

\subsection{{\bf Mixture models based on Dirichlet processes}}
\label{subsec:mixtures}
Let $\bx_{ijkr}=(x^1_{ijkr},x^2_{ijkr})$ represent the genotype at the $r$-th locus of the $j$-th gene for the 
$i$-th individual belonging to the $k$-th group of case/control, and let
$\bX_{ijk}=(\bx_{ijk1},\bx_{ijk2},\ldots,\bx_{ijkL_j})$ denote the genotype information of the $i$-th individual 
of the $k$-th group at all the $L_j$ loci corresponding to the $j$-th gene.
We assume that for every triplet $(i,j,k)$,  $\bX_{ijk}$ 
%are independently distributed with
%mixture probability mass function with a {\it maximum} of $M$ components, given by 
have the mixture distribution
\begin{equation}
[\bX_{ijk}]=\sum_{m=1}^M\pi_{m ijk}\prod_{r=1}^{L_j}f\left(\bx_{ijkr}\vert p_{m ijkr}\right),
\label{eq:mixture1}
\end{equation}
where $f\left(\cdot\vert p_{m ijkr}\right)$ 
%is the probability mass function of independent Bernoulli
%distributions, given by
is the Bernoulli mass function given by
\begin{equation}
f\left(\bx_{ijkr}\vert p_{m ijkr}\right)=
\left\{p_{m ijkr}\right\}^{x^1_{ijkr}+x^2_{ijkr}}
\left\{1-p_{m ijkr}\right\}^{2-(x^1_{ijkr}+x^2_{ijkr})},
\label{eq:pmf1}
\end{equation}
and $M$ denotes the {\it maximum} number of mixture components possible.

Allocation variables $z_{ijk}$, with probability distribution
\begin{equation}
[z_{ijk}=m]=\pi_{m ijk},
\label{eq:alloc_z}
\end{equation}
for $i=1,\ldots,N_k$ and $m=1,\ldots,M$, allow representation of (\ref{eq:mixture1}) as
\begin{equation}
[\bX_{ijk}|z_{ijk}]=\prod_{r=1}^{L_j}f\left(\bx_{ijkr}\vert p_{z_{ijk} ijkr}\right).
\end{equation}
Following \ctn{Majumdar13}, BB, we set $\pi_{m ijk}=1/M$, for $m=1,\ldots,M$,
and for all $(j,k)$.
%We may assume appropriate Dirichlet distribution priors on $\left(\pi_{1ijk},\ldots,\pi_{Mijk}\right)$
%for $j=1,\ldots,J$; $i=1,\ldots,N_k$; $k=0,1$. 
%However, as investigated in \ctn{Majumdar13}, the Dirichlet distribution often yields very small values
%of the probabilities $\bpi_{m ijk}$, thereby tending to underestimate the true number of mixture components.
%On the other hand, setting $\bpi_{m ijk}=1/M$ exhibited much better performance.
%Therefore, in this work, we set $\pi_{m ijk}=1/M$, for $m=1,\ldots,M$,
%and for all $(j,k)$. 

Letting $\bp_{m ijk}=\left(p_{m ijk1},p_{m ijk2},\ldots,p_{m ijkL_j}\right)$ denote the vector of minor allele frequencies 
at the $L_j$ loci of the $j$-th gene for the $i$-th individual of the $k$-th group of case/control corresponding to 
the $m$-th subpopulation (note that the vector depends upon the chromosomes through the respective genes), we next assume that 
\begin{align}
\bp_{1ijk},\bp_{2ijk},\ldots,\bp_{Mijk}&\stackrel{iid}{\sim} \bG_{ijk};\label{eq:dp1}\\
\bG_{ijk}&\sim \mbox{DP}\left(\alpha_{ijk}\bG_{0,ijk}\right),\label{eq:dp2}
\end{align}
where $\mbox{DP}\left(\alpha_{ijk}\bG_{0,ijk}\right)$ stands for Dirichlet process
with expected probability measure $\bG_{0,ijk}$ having precision parameter $\alpha_{ijk}$.
We specify the base probability measure $\bG_{0,ijk}$ as follows: for $m=1,\ldots,M$ and $r=1,\ldots,L_j$, 
\begin{equation}
%\bG_{0,jk}=\prod_{r=1}^{L_j}\mbox{Beta}\left(\nu_{1jkr},\nu_{2jkr}\right).
p_{mijkr}\stackrel{iid}{\sim} \mbox{Beta}\left(\nu_{1ijkr},\nu_{2ijkr}\right),
\label{eq:dp3}
\end{equation}
under $\bG_{0,ijk}$.
Coincidences among $\bP_{Mijk}=\left\{ \bp_{1ijk},\bp_{2ijk},\ldots,\bp_{Mijk}\right\}$, 
which occur with positive probability, is the property of the DP based mixture models that we exploit to learn about the actual
number of mixture components. 
%in (\ref{eq:mixture1}) falls below $M$, the maximum number of components, the
%mixing probabilities taking the form $M^*/M$, where $1\leq M^*\leq M$. See \ctn{Majumdar13},
%\ctn{Sabya11}, \ctn{Sabya12}, \ctn{Bhattacharya08}, for the details. 

The associated Polya urn distribution of $\bP_{Mijk}$ can be derived by marginalizing over $\bG_{ijk}$:
\begin{equation}
\left[\bp_{mijk}|\bP_{Mijk}\backslash \{\bp_{mijk}\}\right]
\sim\frac{\alpha_{ijk}}{\alpha_{ijk}+M-1}\bG_{0,ijk}\left(\bp_{mijk}\right)
+\frac{1}{\alpha_{ijk}+M-1}\sum_{m'\neq m=1}^M\delta_{\bp_{m'ijk}}\left(\bp_{mijk}\right),
\label{eq:polya}
\end{equation}
where $\delta_{\bp_{m'ijk}}(\cdot)$ denotes point mass at $\bp_{m'ijk}$.
This scheme is useful
for constructing an efficient Gibbs sampling strategy for simulating the mixtures conditional
on the other parameters, embedded in a parallel MCMC strategy that we devise, bypassing the infinite-dimensional random measure $\bG_{ijk}$.
%The property of coincidences among the parameter vectors is clearly preserved by the Polya urn scheme.

%It is important to remark that the mixtures associated with different triplets $(i,j,k)$, are independent.
%This entails that, after coincidences among the mixture components, 
Coincidences among the mixture components associate the triplets $(i,j,k)$ to 
different mixtures with varying number of components. Indeed, 
the genotype distributions of any two individuals $i$ and $i'$ arising from a given sub-population with the same 
gene indexed by $j$ but with different case-control status, are likely to be different, so that
$(i,j,k=0)$ and $(i',j,k=1)$ may correspond to different mixtures. 
Also, for any two genes indexed by $j$ and $j'$,
$(i,j,k)$ and $(i,j',k)$ may correspond to different mixtures because of differences in the distribution of
genotypes of genes $j$ and $j'$ for the $i$-th individual. %from any given sub-population.
Furthermore, for any two individuals indexed by $i$ and $i'$, $(i,j,k)$ and $(i',j,k)$ are likely to be associated
with different mixtures because the genotype distribution of the $j$-th gene may be affected by different environmental exposures $\bE_i$ and $\bE_{i'}$. Thus, it seems that the Dirichlet process based mixtures
realistically take account of the various genotypic sub-populations and the number of such sub-populations the data arise from.

The above ideas are similar in essence to those in BB, but note that in their case, since
the environmental effect $\bE_i$ is not considered,
the mixtures were with respect to $(j,k)$ only, not with respect to $(i,j,k)$ as in our current scenario
influenced by $\bE_i$. 

%Following \ctn{Majumdar13}, \ctn{Sabya12}, \ctn{Sabya11}, \ctn{Bhattacharya08}, we set $M=30$ in our applications.
%It follows from \ctn{Antoniak74} that the expected number of distinct parameter vectors in the set
%$\bp_{1ijk},\bp_{2ijk},\ldots,\bp_{Mijk}$ is approximately $\alpha_{ijk}\log\left(1+\frac{M}{\alpha_{ijk}}\right)$.
%When prior information regarding the true number of mixture components is lacking, 
%it may be reasonable to specify the expected number of distinct components
%to be close to half of the maximum number of components possible, namely, close to $M/2$. 
%With $M=30$, we fix $\alpha_{ijk}=10$, so that about $14$ distinct mixture components
%in (\ref{eq:mixture1}) are to be expected {\it a priori}. Apart from this choice, we also considered
%the possibilities $\alpha_{ijk}=1$, $\alpha_{ijk}\sim\mbox{Gamma}\left(0.1,0.1\right)$, that is, the gamma
%distribution with mean $1$ and variance $10$, and $\alpha_{ijk}\sim\mbox{Gamma}\left(1,0.1\right)$
%(so that the mean and variance are $10$ and $100$, respectively); however, the choice $\alpha_{ijk}=10$
%for all $(i,j,k)$ outperformed the other choices with regard to capturing the true number of mixture components.

Following BB, we set $M$, the maximum possible number of sub-populations to be $30$ and $\alpha_{ijk}=10$ in our applications. These choices are not
affected by the presence of environmental variables, and performed adequately in our Bayesian analyses.

\subsection{{\bf Modeling the complex dependence structure with appropriate modeling of the parameters of $\bG_{0,ijk}$}}
\label{subsec:dependence_structure}

We specify the dependence structure between the genes and the environment by primarily seeing to it that the
environment may act upon gene-gene interaction without affecting the 
marginal distributions of the genotypes of the individual genes. However, we also  
take into account the fact that in some cases the environmental variables may cause changes in the distributions of the genotypes. %We accomplish all these by suitably modeling
%the parameters of $\bG_{0,ijk}$ using matrix-normal prior.% Briefly, we model the Beta parameters $\nu_{1ijkr}$ and $\nu_{2ijkr}$
%to incorporate the effect due to $\bE$ and the dependence between the SNPs and the genes
%by specifying suitable dependence structures between the parameters of $\nu_{1ijkr}$ and $\nu_{2ijkr}$
%through matrix-normal distributions. Importantly, a relevant matrix-normal structure 
%helps suitably express the influence of the environmental variables on gene-gene interactions.
%Specific details follow.
Modelling the parameters of the expected probability measure $\bG_{0,ijk}$ through a relevant hierarchical 
matrix-normal prior helps us incorporate the complex G$\times$E, G$\times$G and also the SNP$\times$SNP effects appropriately.

\subsubsection{{\bf Modeling the parameters of $\bG_{0,ijk}$}}
\label{subsubsec:model_G_0}
%The Beta parameters $\nu_{1ijkr}$ and $\nu_{2ijkr}$ of (\ref{eq:dp3}) are modeled under the consideration 
%that the SNPs within a gene may not be independently distributed due to linkage disequilibrium and 
%there may exist a complex dependence structure between different genes. 
%Specifically, for $r=1,\ldots,L_j$, and for every $(i,j,k)$, 
We model $\nu_{1ijkr}$ and $\nu_{2ijkr}$, for each loci $r=1,\ldots,L_j$, in $j$-th gene, of every individual $i$, having case or control status $k$, that is for every $(i,j,k)$, as the following:
\begin{align}
\nu_{1ijkr}&=\exp\left(u_{jr}+\lambda_{ijk}+\mu_{jk}+\bbeta'_{jk}\bE_i\right);\label{eq:nu_1}\\
\nu_{2ijkr}&=\exp\left(v_{jr}+\lambda_{ijk}+\mu_{jk}+\bbeta'_{jk}\bE_i\right).\label{eq:nu_2}
\end{align}
%Note that $\lambda_{ijk}$ is shared by every locus of the $j$-th gene and
%$k$-th case-control status. 
The complex dependence structure that may exist between the SNPs 
within a gene and between the genes has been incorporated in our model by the parameters 
$u_{jr}$ , $v_{jr}$ and $\lambda_{ijk}$, $\mu_{jk}$ %,$\bbeta_{jk}$
 respectively (see BB for details). 
%has been discussed in details in the subsequent sections. 
Here $\bE_{i}$ is the $d$-dimensional 
vector of continuous environmental variables for the $i$th individual. The model can be easily extended to 
include categorical environmental variables along with the continuous ones.

Note that, non-null $\bbeta_{jk}$ indicates significant marginal effect of the environmental variable $\bE$ on the $j$-th gene. 
In Section \ref{subsubsec:matrix_normal} we introduce a modeling strategy that accounts for the complex phenomenon 
through which gene-gene interaction gets modified under the environmental effect, even though the marginal effects
of the genes remain unchanged.

\subsubsection{{\bf Matrix normal prior for $\lambda_{ijk}$'s}}
\label{subsubsec:matrix_normal}

Let $\blambda=(\blambda_1,\ldots,\blambda_J)$, 
where $\blambda_j=(\lambda_{1j0},\ldots,\lambda_{n_0j0},\lambda_{1j1},\ldots,\lambda_{n_1j1})$, for $j=1,\ldots,J$.
Note that $\lambda_{ijk}$ is shared by every locus of the $j$-th gene of the individual indexed by $(i,k)$.

We consider the following model for $\blambda$:
\begin{equation}
\blambda\sim %N\left(\bzero,\bC\right), 
\mathcal N\left(\bxi,\bA\otimes\tilde\bSigma\right),
\label{eq:mn1}
\end{equation}
%noting that $j$th row of $\bLambda$ corresponds to the $j$th gene of all the ($N_0$+$N_1$) individuals and the $k_1$th column of $\Lambda$ corresponds to all $J$ genes of the $k_1$th individual, belonging either to the casses or the control group, we can say that the right covariance matrix $\tilde\bSigma$ siginifies the interaction effects between the $j$ genes whereas, the left covariance matrix $A$ reflects the variance-covariance between the distributions of cases and controls, under independence which is expected to be close to an identity matrix. 
where $\bA$ is the $J\times J$ left covariance matrix, indicating gene-gene interaction in the
absence of environmental effect, and
$\tilde\bSigma=\bSigma+\phi\mathcal E$ is the right covariance matrix under the effect of the environmental variable $E$. Here $\phi\geq 0$, $\bSigma$ is some positive definite matrix,
and the $(i,j)$-th element of the positive definite matrix $\mathcal E$, associated with the environmental
variable $\bE$, is given by
\begin{equation}
\mathcal E_{ij}=\exp\left(-b\|\bE_i-\bE_j\|^2\right),
\label{eq:E}
\end{equation}
where $b>0$ is a smoothness parameter.

Note that $\phi=0$ indicates absence of environmental effects on gene-gene interaction.
It is quite important to observe that, because of the above Gaussian assumption, 
even for non-zero $\phi$, which points towards indirect effect of environmental factors on the epigenome, triggering genetic interactions, the marginal genotypic distributions associated with the $J$ genes
of our model remain unaffected by $\bE$. %As already discussed
%in Section \ref{subsec:statistical_definition}, the phenomenon through which the environmental factors called mutagens directly %affect the genotypic distributions is regarded as mutation, which  is a rare phenomenon.% In any case, our model accounts for %both the %phenemonena, through (\ref{eq:nu_1}), 
%(\ref{eq:nu_2}) and via (\ref{eq:mn1}), (\ref{eq:E}).

For convenience, we represent the $JN$-dimensional vector $\blambda$ as a $J\times N$ matrix $\bLambda$, 
%so that (\ref{eq:mn1})  represented as a matrix normal distribution with mean matrix $\bmu^{J\times N}$, 
%$\bzero^{J\times 2}$,
%left covariance matrix $\bA$ and right covariance matrix $\tilde\bSigma$, having probability density function
which has the following probability density function:
\begin{equation}
\pi(\bLambda)=\frac{\exp\left[-tr\left\{\tilde\bSigma^{-1}\left(\bLambda-\bxi\right)^T\bA^{-1}
\left(\bLambda-\bxi\right)\right\}\right]}
{\left(2\pi\right)^J\left|\bA\right|^N\left|\bLambda\right|^J}.
\label{eq:pi_Lambda}
\end{equation}
%We note that the $k$-th column of $\bLambda$, which we denote by $\bLambda^{col,k}$, follows the multivariate 
%normal distribution:
It follows that
\begin{equation}
\bLambda^{col,k}\sim \mathcal N_J\left(\bxi^{col,k},\tilde\sigma_{kk}\bA\right),
\label{eq:mvn_col}
\end{equation}
where $\bLambda^{col,k}$ and $\bxi^{col,k}$ are the $k$-th columns of $\bLambda$ and $\bxi$, respectively.
The covariance matrix between $\bLambda^{col,k_1}$ and $\bLambda^{col,k_2}$ is given by
\begin{equation}
cov\left(\bLambda^{col,k_1},\bLambda^{col,k_2}\right)=\tilde\sigma_{k_1k_2}\bA,
\label{eq:mvn_cov}
\end{equation}
where $\tilde\sigma_{k_1k_2}$ denotes the $(k_1,k_2)$-the element of $\tilde\bSigma$. 
%
%Similarly, the $j$-th row of $\bLambda$, which we denote by $\bLambda^{row,j}$, has the following
%multivariate normal distribution:
Also,
\begin{equation}
\bLambda^{row,j}\sim \mathcal N_{N}\left(\bxi^{row,j},a_{jj}\tilde\bSigma\right),
\label{eq:mvn_row}
\end{equation}
where $\bLambda^{row,j}$ and $\bxi^{row,j}$ are the $j$-th rows of $\bLambda$ and $\bxi$, respectively.
Further,
\begin{equation}
cov\left(\bLambda^{row,j_1},\bLambda^{row,j_2}\right)=a_{j_1j_2}\tilde\bSigma.
\label{eq:mvn_cov2}
\end{equation}
In our applications, following BB, we choose $\bxi=\bzero$.

To summarize, the matrix-normal prior imposes a dependence structure between the genes 
through the gene-gene interaction matrix $\bA$, and $\tilde\bSigma$ features the direct or indirect effect of the environmental factors, on the epigenome of the individuals.
The randomness associated with the matrix-normal prior on $\bLambda$ incorporates 
dependence between the SNPs within a gene. 

Further discussion regarding the effect of environmental variables on gene-gene interaction is provided in Section S-1
of the supplement.

\subsubsection{{\bf Priors for $u_{jr}$ and $v_{jr}$}}
\label{subsubsec:u_v_prior}

%Note that, for fixed $k$, $u_{jr}+\lambda_{jk}$ may be interpreted as the effect of the
%$r$-th SNP of the $j$-th gene. Now if $u_{jr}$ and $v_{jr}$ are allowed to depend upon gene $j$, then SNP $r$ 
%would be receive less weightage than the corresponding gene, since $\lambda_{jk}$ depends upon the $j$-th gene as well.
%This is likely to suppress detection of the DPL of the genes.

%Hence, in (\ref{eq:nu_1}) and (\ref{eq:nu_2}), 
We follow BB in setting,
for $j=1,\ldots,J$, 
$u_{jr'}=u_{r'}$ and $v_{jr'}=v_{r'}$ for $r'=1,\ldots,L$, where
$L=\max\{L_j;~j=1,\ldots,J\}$, and assuming
for $r'=1,\ldots,L$,
\begin{align}
u_{r'} &\stackrel{iid}{\sim} N(0,1);\label{eq:u_r}\\
v_{r'} &\stackrel{iid}{\sim} N(0,1).\label{eq:v_r}
\end{align}

See BB for the details regarding the choice of $u_{jr}$ and $v_{jr}$.

%Apart from facilitating identification of the DPL, the above modeling strategy reduces the number of parameters to be
%updated for model-fitting. Moreover, since the genes share many $u_{r'}$ and $v_{r'}$ which are random
%this creates additional dependence between the genes. 
%The standard normal prior turned out to be quite adequate. In fact, since these parameters are in addition
%with $\lambda_{jk}$ and since we assume a hierarchical matrix-normal prior for $\blambda$ (to be descirbed below),
%Gaussian priors on $u_{r'}$ and $v_{r'}$ with other means and variances are unlikely to yield significantly 
%different results, thus pointing towards in-built prior robustness in our modeling strategy.

\subsubsection{{\bf Priors on $\mu_{jk}$, $\bbeta_{jk}$, $\bA$, $\bSigma$, $b$ and $\phi$}}
\label{subsubsec:prior_alpha_beta}

We put the following hierarchical priors on $\bmu=(\mu_{jk};~j=1,\ldots,J;~k=0,1)$
and $\bbeta=(\bbeta_{\ell};~\ell=1,\ldots,D)$, where $\bbeta_{\ell}=(\beta_{\ell jk};~j=1,\ldots,J;~k=0,1)$: 
\begin{align}
\bmu&\sim \mathcal N\left(\bzero,\bA_{\alpha}\otimes\bSigma_{\alpha}\right)\label{eq:alpha_prior}\\
\bbeta_{\ell}&\stackrel{iid}{\sim} 
\mathcal N\left(\bzero,\bA_{\beta}\otimes\bSigma_{\beta}\right);~\ell=1,\ldots,D.\label{eq:beta_prior}
\end{align}
For priors on $\bA_{\alpha}$, $\bA_{\beta}$, $\bSigma_{\alpha}$ and $\bSigma_{\beta}$, we first consider
their respective Cholesky decompositions:
$\bA_{\alpha}=\bC_{\alpha}\bC_{\alpha}'$, $\bA_{\beta}=\bC_{\beta}\bC_{\beta}'$, 
$\bSigma_{\alpha}=\bD_{\beta}\bD_{\beta}'$ and $\bSigma_{\beta}=\bD_{\beta}\bD_{\beta}'$.
We assume that the diagonal elements of the above Cholesky factors are $iid$ $Gamma(0.01,0.01)$, that is,
gamma distribution with mean 1 and variance 100. We assume the non-zero off-diagonal elements of the Cholesky
factors to be $iid$ $\mathcal N(0,10^2)$.

Using the same Cholesky decomposition idea, we assume that the off-diagonal elements of the Cholesky factors
of $\bA$ and $\bSigma$ to be $iid$ $\mathcal N(0,10^2)$, and the diagonal elements to be $iid$ $Gamma(0.01,0.01)$. 

We put log-normal priors on $b$ and $\phi$, so that both
$\log (b)$ and $\log (\phi)$ are normally distributed with mean zero and variance 100.

%\section{{\bf Model fitting through a parallel MCMC algorithm}}
%\label{sec:computation}

Recall that the mixtures associated with gene $j\in\{1,\ldots,J\}$, and individual
$i\in\{1,\ldots,N_k\}$ and case-control status $k\in\{0,1\}$,
are conditionally independent of each other, given the interaction parameters.
This allows us to update the mixture components in separate parallel processors, conditionally on the
interaction parameters. Once the mixture components are updated, we update the interaction parameters 
using a specialized form of TMCMC, in a single processor. A schematic representation of our model and the parallel processing algorithm
is provided in Figures \ref{fig:schematic1}. Details of our parallel processing algorithm are provided
in Section S-2 of the supplement.

\begin{figure}%[htp]
\centering
\includegraphics[width=10cm,height=10cm]{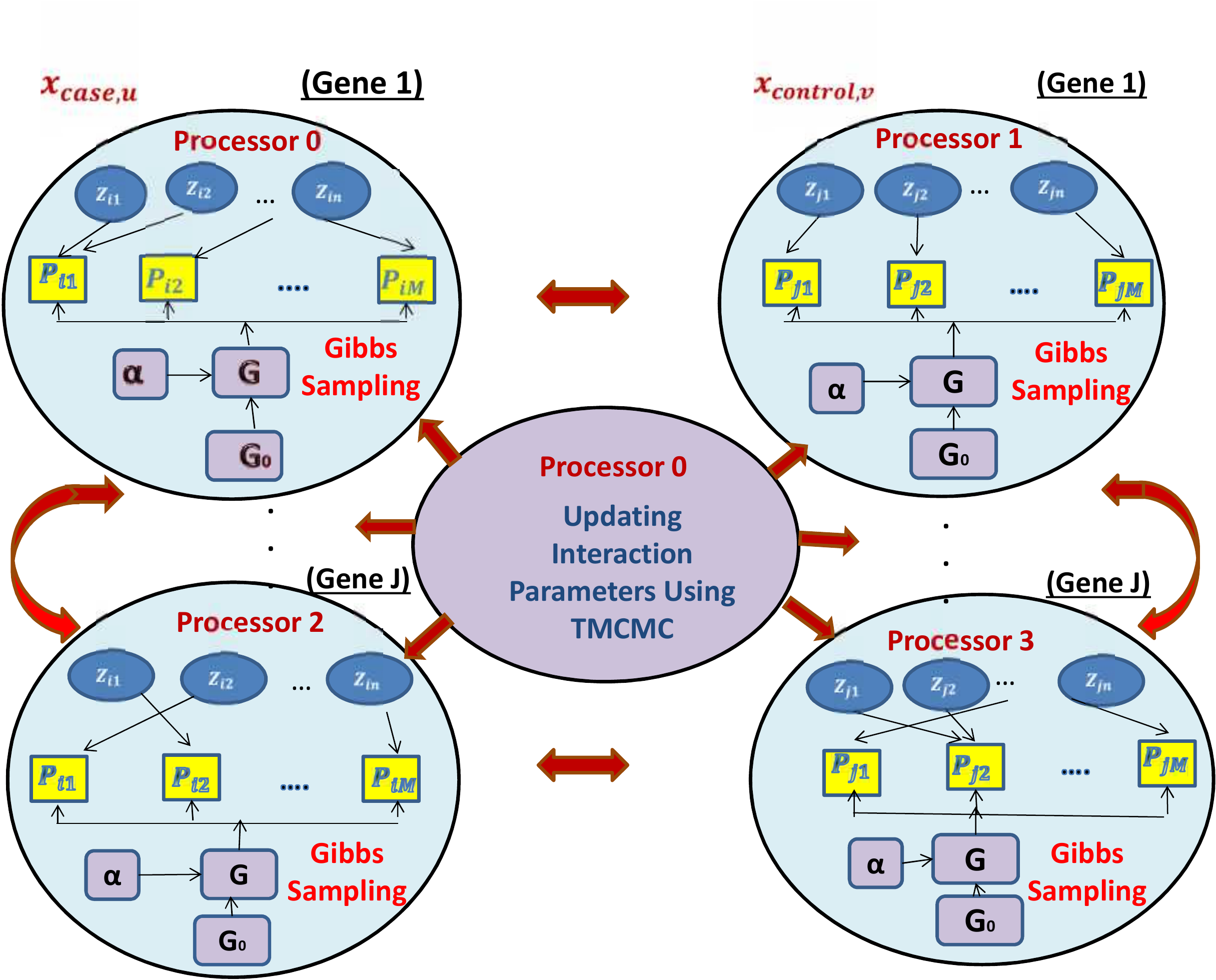}
\caption{{\bf Schematic diagram for our model and parallel processing idea:} The arrows in the diagram
represent dependence between the variables. The ranks of the processors updating the sets of parameters in parallel 
using Gibbs sampling are also
shown. Once the other parameters are updated in parallel, the interaction parameters are updated using TMCMC 
by the processor with rank zero.}
\label{fig:schematic1}
\end{figure}

\pagebreak

\section{{\bf Detection of the roles of environment, genes and their interactions in case-control studies}}
\label{sec:detection}

\subsection{{\bf Formulation of appropriate Bayesian hypothesis testing procedures}}
\label{subsec:test_formulation}

%Recall the definition of $\bar{w}_{ijk}$ from (\ref{eq:w_bar1}). 
%%Let $\bar{W}_{ijk}$ denote the random variable associated with the observation $\bar{w}_{ijk}$.
%Let $h_{0j}$ and $h_{1j}$ denote the distributions of $\bar{w}_{ijk=0}$ and $\bar{w}_{ijk=1}$, respectively.
%Clearly, $h_{0j}$ and $h_{1j}$ are also $M$-component mixtures, where the $m$-th mixture component of the respective
%distributions is characterized by $\bp_{mjk=0}$ and $\bp_{mjk=1}$, haing the same mixing probability $1/M$.
%%If, for case ($k=1$) and control ($k=0$), the distributions of the form (\ref{eq:mixture1}) are

%If $h_{0j}$ and $h_{1j}$ are not significantly different, then it is plausible to conclude that the role of genes is not
%significant in the case-control study. As such,
In order to investigate if genes have any effect on case-control, it is pertinent to test
\begin{equation}
H_{01}: h_{0j}=h_{1j}; ~j=1,\ldots,J,
%H_0:\sum_{m=1}^M\pi_{m jk=0}\prod_{r=1}^{L_j}f\left(\cdot\vert p^r_{m jk=0}\right)
%=\sum_{m=1}^M\pi_{m jk=1}\prod_{r=1}^{L_j}f\left(\cdot\vert p^r_{m jk=1}\right);~j=1,\ldots,J,
\label{eq:H_0}
\end{equation}
versus
\begin{equation}
H_{11}:\mbox{not}~H_{01},
\label{eq:H_1}
\end{equation}
where
\begin{align}
h_{0j}(\cdot)&=\prod_{i=1}^{N_0}\left\{\sum_{m=1}^M\pi_{m ijk=0}
\prod_{r=1}^{L_j}f\left(\cdot\vert p^r_{m ijk=0}\right)\right\};\label{eq:h_0}\\
h_{1j}(\cdot)&=\prod_{i=1}^{N_1}\left\{\sum_{m=1}^M\pi_{m ijk=1}
\prod_{r=1}^{L_j}f\left(\cdot\vert p^r_{m ijk=1}\right)\right\}.\label{eq:h_1}
\end{align}
We shall also test, for $\ell=1,\ldots,D$; $j=1,\ldots,J$, and $k=0,1$:
\begin{equation}
H_{02}:\beta_{\ell j k}=0~\mbox{versus}~H_{12}:\beta_{\ell j k}\neq 0,
\label{eq:beta_test}
\end{equation}
and
\begin{equation}
H_{03}:\phi=0~\mbox{versus}~H_{13}:\phi\neq 0.
\label{eq:phi_test}
\end{equation}
The cases that can possibly arise and the respective conclusions are the following:
\begin{itemize}
\item 
If $\underset{1\leq j\leq J}{\max}~d(h_{0j},h_{1j})$ is significantly small with
high posterior probability, then $H_{01}$ is to be accepted. 
If $h_{0j}$ and $h_{1j}$ are not significantly different, then it is plausible 
to conclude that the $j$-th gene is not marginally significant in the case-control study.

\item Suppose that $H_{01}$ is accepted (so that genes have no significant role) and 
that $\beta_{\ell jk}$ is significant, at least for some $\ell$, $j$ and $k$, but $\phi$ is insignificant.
This may be interpreted as the environmental variable $\bE$ having some altering effect on the $j$-th gene, that doesn't affect the disease status. If $\phi$ turns out to be significant, then this would additionally imply that
the environmental variable $\bE$ influences gene-gene interaction, but not in a way that causes the disease.
%Significance of $\phi$ and at least one $\beta_{\ell jk}$ would indicate that 
%both mutation and gene-gene interaction have been influenced by $\bE$, but not in a way that triggers the disease.

\item If $H_{01}$ is rejected, indicating that the genes have significant roles to play in causing the disease,
but none of the $\beta_{\ell jk}$ or $\phi$ turn out to be significant, then only genes, not $\bE$, are responsible
for causing the disease. In that case, the disease may be thought to be of purely genetic in nature.

\item Suppose $H_{01}$ is rejected, $\beta_{\ell j0}$ and $\beta_{\ell j1}$ turn out to be significant, 
but that $H_{0\ell j}:\beta_{\ell j0}=\beta_{\ell j1}$ is accepted.Then although $\bE$ is insignificant with respect to the marginal effect of gene $j$, it affects the disease status by triggering gene-gene interaction in some genes if $\phi$ turns out to be significant.

\item If $H_{01}$ is rejected, $\beta_{\ell jk}$ is significant for some $\ell$, $j$, $k$, and $\phi$ is insignificant,
then the presence of $\bE$ has altering effect on some genes, which, in turn, cause the disease. In this case, since
$\phi$ is insignificant, $\bE$ does not seem to influence gene-gene interaction.

\item If $H_{01}$ is rejected, $\beta_{\ell jk}$ is insignificant for all $\ell$, $j$, $k$, but 
$\phi$ is significant, then significant effect of $\bE$ on altering the marginal effect of genes is to be ruled out, and one may conclude that the underlying cause of the disease is gene-gene interaction, which has been adversely affected by the environmental variable.

\item If $H_{01}$ is rejected, $\beta_{\ell jk}$ is significant for some $\ell$, $j$, $k$, and $\phi$ is also significant,
then the environmental variable has possibly significantly affected both the marginal and also gene-gene interaction adversely
to cause the disease.

\end{itemize}

\subsection{{\bf Hypothesis testing based on clustering modes}}
\label{subsec:clustering}

%Ideas on clusterings of the mixture distributions $h_{0j}$ and $h_{1j}$ provides us with a novel
%and computationally efficient procedure of testing $H_0$. 
%Briefly, we assess discrepancies between the two mixture distributions
%%implied by $k=0$ and $k=1$ 
%$h_{0j}$ and $h_{1j}$

For $k=0,1$, let $i_k$ denote the index of the ``central" clusterings of
$\bP_{Mijk}=\left\{ \bp_{1ijk},\bp_{2ijk},\ldots,\bp_{Mijk}\right\}$, $i=1,\ldots,N_k$.
The concept of central clustering has been introduced by \ctn{Sabya11}.
%we shall briefly discuss
%the main ideas in Section S-3 of the supplement. %\ref{subsubsec:clustering_metric_mode}.
Significant divergence between the two clusterings 
of $\bP_{Mi_0jk=0}=\left\{ \bp_{1i_0jk=0},\bp_{2i_0jk=0},\ldots,\bp_{Mi_0jk=0}\right\}$
and $\bP_{Mi_1jk=1}=\left\{ \bp_{1i_1jk=1},\bp_{2i_1jk=1},\ldots,\bp_{Mi_1jk=1}\right\}$, for $j=1,\ldots,J$.
%Significantly large divergence between the two clusterings 
clearly indicates
that the $j$-th gene is marginally significant.
%leads to rejection of $H_{0j}$.
%An appropriate metric for studying divergence between clusterings is described next. %in Section \ref{subsec:metric}.
%In Section S-3 of the supplement we discuss suitable measures of divergence. 
%These ideas require a suitable divergence measure to compare any two clusterings, which we discuss
%in Section S-3 of the supplement, also
%providing a briefing on the central clustering idea. 
%In our case, we shall obtain $i_0$ and $i_1$, the indices of the central clusterings associated with
%$\bP_{Mijk=0};~i=1,\ldots,N_0$ and $\bP_{Mijk=1};~i=1,\ldots,N_1$, respectively, obtained by the above
%method. 
Once $i_0$ and $i_1$ are determined,
we shall consider the clustering distance between $\bP_{Mi_0jk=0}$ and $\bP_{Mi_1jk=1}$, denoted by
$\hat d\left(\bP_{Mi_0jk=0},\bP_{Mi_1jk=1}\right)$, as a suitable measure of divergence. 
We shall be particularly interested in
\begin{equation}
d^*=\max_{1\leq j\leq J}\hat d\left(\bP_{Mi_0jk=0},\bP_{Mi_1jk=1}\right);
\label{eq:d_star}
\end{equation}
In Section S-3 of the supplement we include a brief discussion of the aforementioned methodology.
%In Section 
%\ref{subsec:clustering_metric_mode}, we provide a brief discussion on our clustering metric and central clustering.

BB point out that 
although significantly large divergence between clusterings %of $\bP_{Mjk=0}$ and $\bP_{Mjk=1}$ 
indicate rejection of the null hypothesis, %$H_0$,
insignificant clustering distance %between $\bP_{Mjk=0}$ and $\bP_{Mjk=1}$ 
need not necessarily 
provide strong enough evidence in favour of the null. In other words, even if the clustering distance is insignificant, 
it is important to check if the parameter vectors being compared are significantly different. 
In this regard, BB propose an appropriate divergence measure based on Euclidean distances of the logit transformations
of the minor allele frequencies. The necessary ideas in our current context are discussed in Section S-3.1 of the supplement.
In our case, in order to compute the Euclidean distance, we first compute the averages 
$\bar{p}_{mijk}=\sum_{r=1}^{L_j}p_{m,ijkr}/L_j$, then consider their logit transformations
$\mbox{logit}\left(\bar{p}_{mijk}\right)=\log\left\{\bar{p}_{mijk}/(1-\bar{p}_{mijk})\right\}$. 
Then, we compute the Euclidean distance between the vectors
$$\mbox{logit}\left(\bar{\bP}_{Mi_0jk=0}\right)=\left\{\mbox{logit}\left(\bar{p}_{1i_0jk=0}\right),
\mbox{logit}\left(\bar{p}_{2i_0jk=0}\right),
\ldots, \mbox{logit}\left(\bar{p}_{Mi_0jk=0}\right)\right\}$$
and 
$$\mbox{logit}\left(\bar{\bP}_{Mi_1jk=1}\right)=\left\{\mbox{logit}\left(\bar{p}_{1i_1jk=1}\right),
\mbox{logit}\left(\bar{p}_{2i_1jk=1}\right),
\ldots, \mbox{logit}\left(\bar{p}_{Mi_1jk=1}\right)\right\}.$$
We denote the Euclidean distance associated with the $j$-th gene by 
$$d_{E,j}=d_{E,j}\left(\mbox{logit}\left(\bar{\bP}_{Mi_0jk=0}\right),
\mbox{logit}\left(\bar{\bP}_{Mi_1jk=1}\right)\right),$$ and denote 
$\underset{1\leq j\leq J}{\max}~d_{E,j}$ by $d^*_E$.

\subsection{{\bf Formal Bayesian hypothesis testing procedure integrating the above developments}}
\label{subsec:testing}

In our problem, we need to test the following for reasonably small choices of $\varepsilon$'s: 
%$\varepsilon_1,\ldots,\varepsilon_4$:
\begin{equation}
H_{0,d^*}:~d^*< \varepsilon_{d^*}\hspace{2mm}\mbox{versus}\hspace{2mm}H_{1,d^*}:~d^*\geq\varepsilon_{d^*};
\label{eq:hypothesis_d_star}
\end{equation}
%where
%\begin{equation}
%d^*=\max_{1\leq j\leq J}\hat d\left(\bP_{Mi_0jk=0},\bP_{Mi_1jk=1}\right);
%\label{eq:d_star}
%\end{equation}
\begin{equation}
H_{0,d^*_E}:~d^*_E< \varepsilon_{d^*_E}\hspace{2mm}\mbox{versus}\hspace{2mm}H_{1,d^*_E}:~d^*_E\geq\varepsilon_{d^*_E};
\label{eq:hypothesis_d_star_E}
\end{equation}
\begin{equation}
H_{0,\beta_{\ell jk}}:~\left|\beta_{\ell jk}\right|< \varepsilon_{\ell jk}\hspace{2mm}
\mbox{versus}\hspace{2mm}H_{1,\beta_{\ell jk}}:~\left|\beta_{\ell jk}\right|\geq\varepsilon_{\ell jk},
%~\mbox{for}~\ell=1,\ldots,D;~j=1,\ldots,J;~k=0,1;
\label{eq:hypothesis_beta}
\end{equation}
$$\mbox{for}~\ell=1,\ldots,D;~j=1,\ldots,J;~k=0,1;$$
\begin{equation}
H_{0,\phi}:~\phi< \varepsilon_{\phi}\hspace{2mm}
\mbox{versus}\hspace{2mm}H_{1,\phi}:~\phi\geq\varepsilon_{\phi}.
\label{eq:hypothesis_phi}
\end{equation}
If $H_0$ is rejected in (\ref{eq:hypothesis_d_star}) or in (\ref{eq:hypothesis_d_star_E}), we could also
test if the $j$-th gene is influential by testing, for $j=1,\ldots,J$, 
$H_{0,\hat d_j}:~\hat d_j< \varepsilon_{\hat d_j}\hspace{2mm}\mbox{versus}\hspace{2mm}
H_{1,\hat d_j}:~\hat d_j\geq\varepsilon_{\hat d_j}$,
where $\hat d_j=\hat d\left(\bP_{Mi_0jk=0},\bP_{Mi_1jk=0}\right)$; we could also test
$H_{0,d_{E,j}}:~d_{E,j}< \varepsilon_{d_{E,j}}\hspace{2mm}\mbox{versus}\hspace{2mm}H_{1,d_{E,j}}:~d_{E,j}\geq\varepsilon_{d_{E,j}}$.

%If, on the other hand, $H_{0j}$ is accepted, then it is possible that the $j$-th gene is not
%individually influential, but some interaction effect involving the $j$-th gene may be significant. To check which interactions
%are significant (we may check this even if $H_{0j}$ is rejected, since the $j$-th gene may be marginally
%significant as well as interactive with the other genes), one may conduct the tests 

To test if gene-gene interactions are significant, one may test, following BB,
$H_{0,j,j^*}:~\left|\bA_{jj^*}\right|<\varepsilon_{A_{jj^*}}$ 
versus $H_{1,j,j^*}:~\left|\bA_{jj^*}\right|\geq\varepsilon_{A_{jj^*}}$,
for $j^*\neq j$,  
%\begin{equation}
%\rho_{jj^*} = \frac{\bA_{jj^*}}{\sqrt{\bA_{jj}\bA_{j^*j^*}}};
%\label{eq:rho}
%\end{equation}
$\bA_{jj^*}$ being the $(j,j^*)$-th
element of $\bA$.
If $H_{1,j,j^*}$ is accepted for some (or many) $j^*\neq j$, then this would indicate significant interaction 
between the $j^*$-th and the $j$-th genes. 

\begin{figure}%[htp]
\centering
\includegraphics[width=10cm,height=10cm]{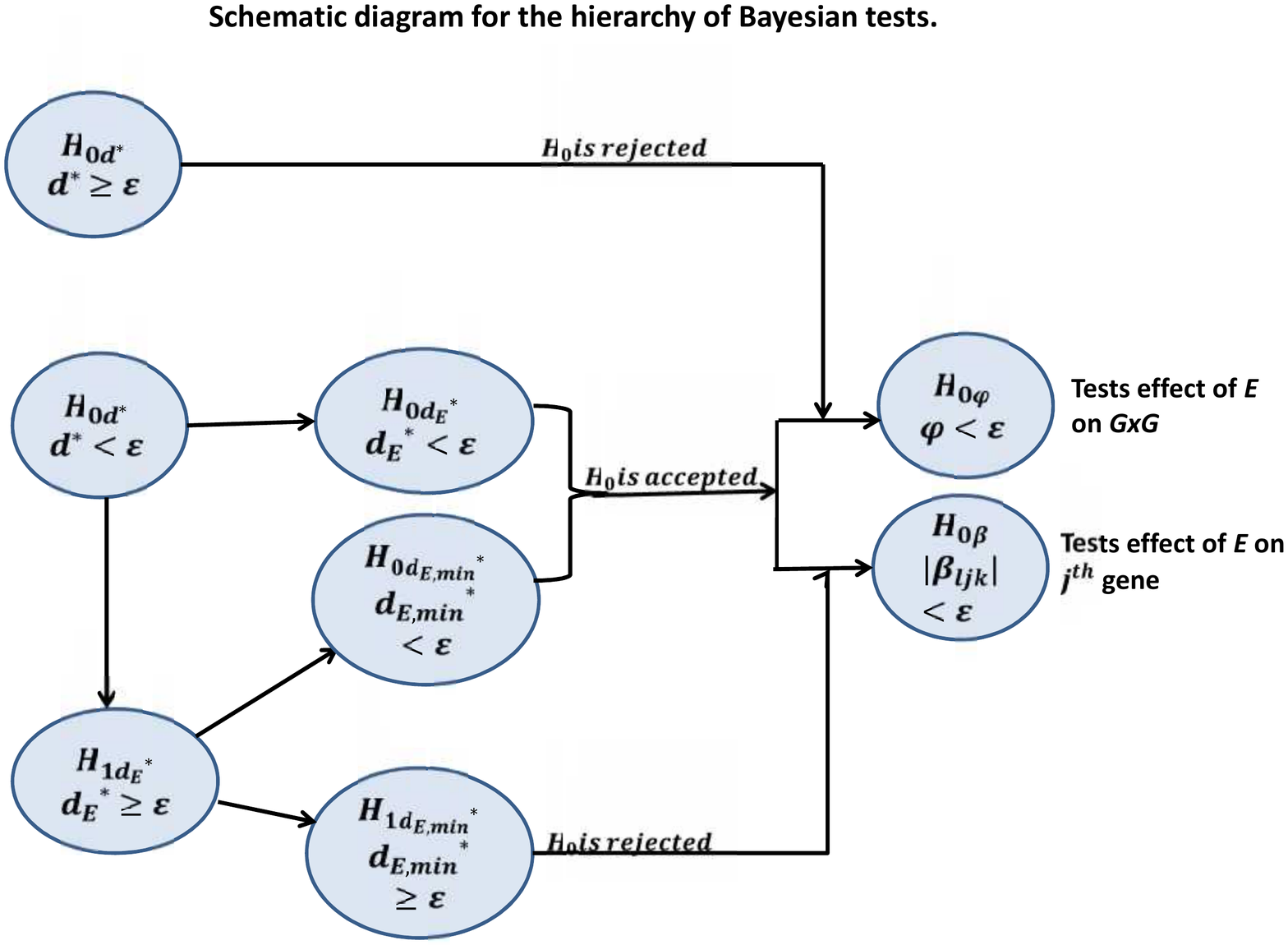}
\caption{{\bf Schematic diagram for our Bayesian testing idea.}}
\label{fig:schematic3}
\end{figure}

As argued in BB, here also it is easily seen that our testing procecure is equivalent to Bayesian multiple
testing procedures that minimize the Bayes risk of additive ``0-1" and ``$0-1-c$" loss functions (see BB for
the details; see also \ctn{Berger85}). Since it is well-known that Bayesian multiple tetsing methods automatically
provide multiplicity control through the inherent hierarchy (see, for example, \ctn{Scott10}), 
separate error control is not necessary. 
A brief, schematic representation of the hierarchy of the hypothesis tests is shown in 
Figure \ref{fig:schematic3}.

Our choices of the $\varepsilon$'s are based on the idea of null model introduced in BB.
In a nutshell, we first specify an appropriate null model, which, for example, is the same model as ours but 
with $\bA$ and $\tilde\bSigma$ set to identity matrices
to reflect the null hypotheses of ``no interaction" and the same mixture distributions under cases and
controls for each gene for no genetic effect. From the null model thus specified, we then generate case-control 
genotype data and fit our general Bayesian model to this ``null data" and set $\varepsilon$ to be the $55$-th percentile
of the relevant posterior distribution. The rationale and details of this procedure are provided in BB (particularly
in Section S-7 of their supplement)

\section{{\bf Simulation studies}}
\label{sec:simulation_study}

For simulation studies, we first generate biologically realistic genotype data sets under stratified population with known
G$\times$G and G$\times$E set ups from the GENS2 software of \ctn{Pinelli12}.  
%To this data, we then apply our model and methodologies in an effort to detect gene-environment
%interaction effects that are present in the data. 
We consider simulation studies 
in $5$ different true model set-ups: (a) presence of gene-gene and gene-environment interaction,
(b) absence of genetic or gene-environmental interaction effect,
(c) absence of genetic and gene-gene interaction effects but presence of environmental effect,
(d) presence of genetic and gene-gene interaction effects but absence of environmental effect,
and 
(e) independent and additive genetic and environmental effects.

As we demonstrate, our model and methodologies successfully identify the marginal effects of the genes, 
along with the G$\times$G and G$\times$E, and the number of sub-populations. Details are provided in 
Section S-4 of the supplement.

\section{{\bf Application of our model and methodologies to a real, case-control dataset on Myocardial Infarction}}
\label{sec:realdata}

MI (more commonly, heart attack), has been subjected to much investigation for detecting the underlying genetic causes, the possible environmental factors and their interactions. 
Application of our ideas to a case-control genotype dataset on early-onset of myocardial infarction (MI) from MI Gen study, obtained from the dbGaP database
({\bf http://www.ncbi.nlm.nih.gov/gap}), led to some interesting insights into gene-environment and gene-gene interactions on incorporating sex as the environmental factor.

\subsection{{\bf Data description}}
\label{subsec:myo_data}
The MI Gen data obtained from dbGaP consists of observations on presence/absence of
minor alleles at $727478$ SNP markers associated with 22 autosomes and the sex chromosomes of $2967$ cases of early-onset myocardial infarction, $3075$ age and sex matched controls. The average age at the time of MI was 41 years among the male cases and 47 years among the female cases. 
The data also consists of the sex information of the individuals, which we incorporate in our Bayesian model. 
The data broadly represents a mixture of four sub-populations: Caucasian, Han Chinese,
Japanese and Yoruban. SNPs were mapped on to the corresponding genes using the  Ensembl human genome database ({\bf http://www.ensembl.org/}). However, technical glitches prevented us from
obtaining information on the genes associated with all the markers. As such, we could categorize
$446765$ markers out of $727478$ with respect to $37233$ genes.

For our analysis, we considered a set of SNPs that are found to be individually associated with different cardiovascular end points like LDL cholesterol, smoking, blood pressure, body mass etc. in various GWA studies published in NHGRI catalogue and augmented this set further with another set of SNPs found to be marginally associated with MI in the MIGen study (see \ctn{LucasG12}). Our study also includes SNPs that are reported to be associated with MI in various other studies; see \ctn{Erdmann10}, \ctn{LuQi11} and \ctn{Wang04}. In all, we obtained 271 SNPs.
Unfortunately, only 33 of them turned out to be common to the SNPs of our original MI dataset on genotypes, which has been mapped on to the genes using the Ensembl human genome database.
However, we included in our study all the SNPs associated with the genes containing the 33 common SNPs. Specifically, our study involves the genotypic information on 32 genes covering 1251 loci, including
the 33  previously identified loci for $200$ individuals. We chose this relatively small number of individuals
to ensure computational feasibility. However, even this data set, along with our model and prior, yielded results
that are not only compatible with, but also complement the results established in the literature.

Categorization of the case-control genotype data into the four sub-populations, each of which are likely
to represent several further and rather varied sub-populations genetically, implies that  
the maximum number of mixture components must be fixed at some value much higher than $4$. As before, we set $M=30$
and $\alpha_{jk}=10$ for every $(j,k)$, to facilitate data-driven inference. 
%Interestingly, the distributions of the number of distinct components for $\alpha_{jk}=1.5$
%(so that the prior mean and variance are approximately $5$) were not significantly different from
%those of $\alpha_{jk}=10$, indicating prior robustness. 

We chose a similar set-up for the null model. That is, we chose the same number of genes and the
same number of loci for each gene, the same number of cases and controls, the same value $M=30$, but
$\alpha_{jk}=1.5$ for every $(j,k)$, as in our simulation studies. 
%As in the simulation studies, this entails that about $5$ components are to be expected 
%{\it a priori} and {\it a posteriori} under the null model for each $(j,k)$ pair.
We use the same priors as in the real data set-up except that we set
$\bA$ and $\bSigma$ to be identity matrices to ensure that the genetic interaction is not present 
and set the same mixture distribution under cases and controls for each gene to ensure the absence of genetic effects. 
%For details see Section 4.1.2 of the supplement.   

\subsection{\bf Remarks on incorporation of the sex variable in our model}
\label{subsec:sex_incorporation}
In our case, $\bE_i=E_i$, a one-dimensional binary variable, where $E_i=1$ if the $i$-th individual is male and
$E_i=0$ if female. Hence, $\bbeta_{jk}=\beta_{jk}$ is a scalar quantity. In (\ref{eq:nu_1}) and (\ref{eq:nu_2})
we considered the environmental variable to be continuous, but remarked that the model can be easily extended
to include categorical variables. Indeed, in this case the exponentials of (\ref{eq:nu_1}) and (\ref{eq:nu_2})
can be thought of as binary regressions with sex as the covariate. 

As regards $\mathcal E_{ij}$ of (\ref{eq:E}), we first consider $a_0+a_1 E_i$ as a binary regression, and then
write
\begin{equation}
\mathcal E_{ij}=\exp\left(-\|(a_0+a_1E_i)-(a_0+a_1E_j)\|^2\right)=\exp\left[-a^2_1(E_i-E_j)^2\right],
\label{eq:E_binary}
\end{equation}
with $b=a^2_1$ being the smoothness parameter. Observe that for the same sex, $\mathcal E_{ij}=1$
while for different sex, $\mathcal E_{ij}=\exp(-b)<1$.

\subsection{{\bf Remarks on model implementation}}
\label{subsec:myo_implementation}

We first obtain the number of parameters to be updated by TMCMC in our case; other unknowns associated
with the mixtures, to be updated using Gibbs steps in parallel.
Note that in our case, the interaction matrix $\bA$ is of order $32\times 32=1024$, and the associated Cholesky decomposition
then consists of $33\times 16=528$ parameters.
Also, $\blambda$ is a $NJ=200\times 2=400$-dimensional random vector and $\bSigma$ is of order $N\times N=200\times 200$,
so that its Cholesky decomposition consists of $201\times 100=20100$ parameters. Furthermore,
$\left\{(u_r,v_r):r=1,\ldots,L\right\}$, where $L=207$, consists of $2\times 207=414$ parameters,
$\bmu$ and $\bbeta$ consist of $64$ parameters each, and there are two more parameters $b$ and $\phi$. So,
in all, there are $21572$ parameters to be updated simultaneously in a single block using TMCMC.

We implemented our parallel MCMC algorithm detailed in S-2 of the supplement
on a VMware consisting of $50$ double-threaded, $64$-bit physical cores, each running at $2493.990$ MHz.
In spite of the large number of parameters associated with the interaction part, our mixture of additive 
and additive-multiplicative TMCMC still ensured reasonable performance.

Our parallel MCMC algorithm takes about $11$ days to yield $100,000$ iterations in our aforementioned
VMware machine. We discard the first $50,000$ iterations as burn-in.
%Implementation of the null model took about $19$ hours and $21$ minutes to yield $30,000$ iterations in the same machine.
%The reduced time compared to the real data implementation is because of lesser number of mixture components
%under the null model.
% Recall that gene-gene interactions in the non-null model influence the posterior distributions
%of the number of components, while this is not the case for the null model which lacks gene-gene interaction.
%This issue seems to be responsible for the discrepancy between the posteriors of the number of distinct components
%in the null and non-null models.
%
Informal convergence diagnostics such as trace plots exhibited adequate mixing properties
of our parallel algorithm.

\subsection{{\bf Results of the real data analysis}}
\label{subsec:realdata_results}

\subsubsection{{\bf Effect of the sex variable}}
\label{subsubsec:sex_effect}

\begin{figure}%[htp]
\centering
\includegraphics[width=15cm,height=6cm]{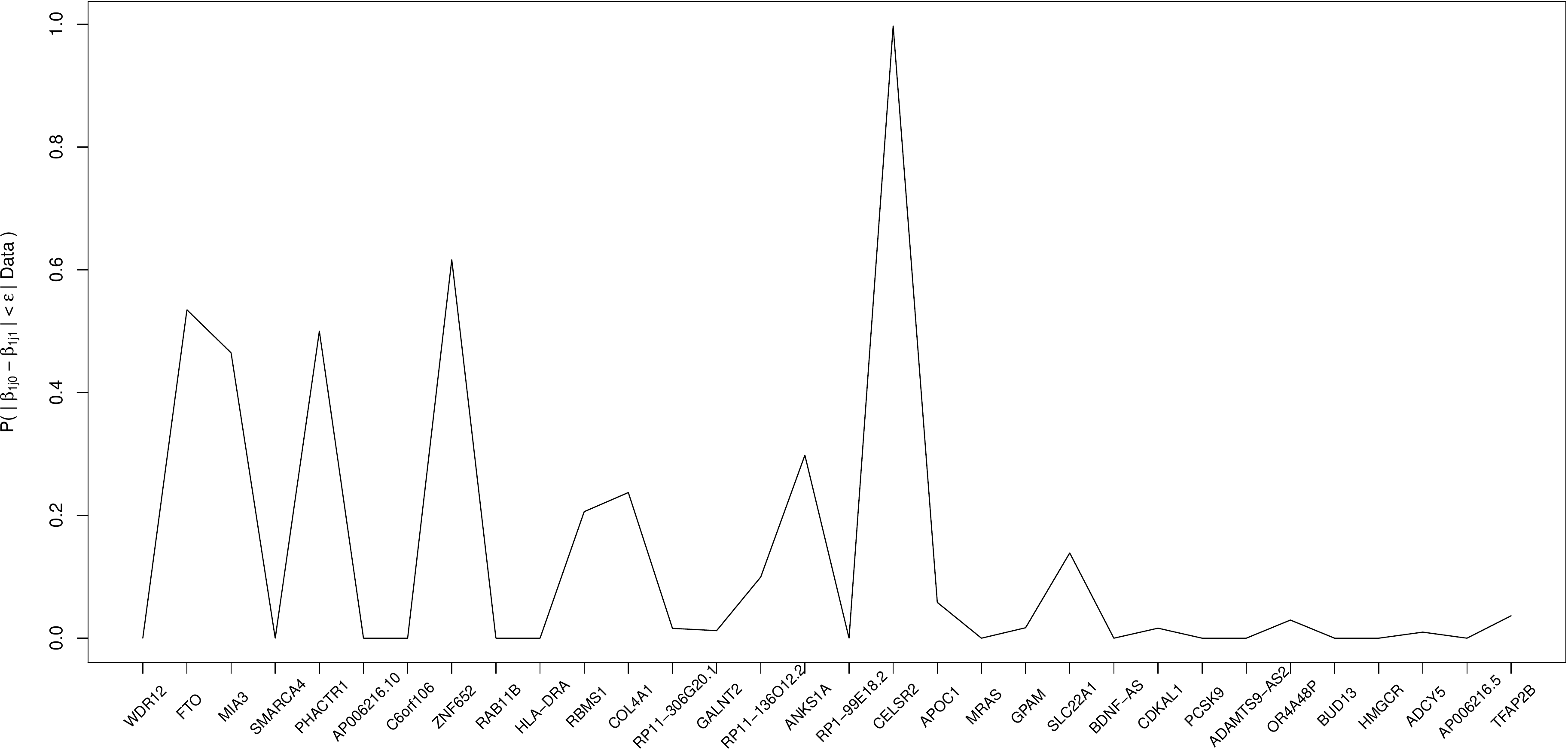}
\caption{{\bf Index plots of posterior probabilities of no environmental effect with respect to 
$|\beta_{1j0}-\beta_{1j1}|<\varepsilon$,
for $j=1,\ldots,32$.}} 
\label{fig:beta_probs}
\end{figure}

It turned out that $\varepsilon_{\phi}=1.043069$ and $P(\phi<\varepsilon_{\phi}|\mbox{Data})\approx 1$, so that
$\phi$ is clearly insignificant, indicating no differential effect of sex on the genetic interactions. The posterior probabilities 
$P(|\beta_{1j1}-\beta_{1j0}|<\varepsilon|\mbox{Data})$ are shown in Figure \ref{fig:beta_probs}. 
As before, $\varepsilon$ is the $55$-th percentile of the posterior distribution of $|\beta_{1j1}-\beta_{1j0}|$ under the
null model.
Under the 0-1 loss function, the above posterior probability exceeding $0.5$ indicates significant environmental effect 
on the $j$th gene. From the figure it is interesting to note that there is significant differential effect due 
to sex on the marginal effects of several genes although sex does not affect the genetic interactions significantly.

\subsubsection{{\bf Influence of genes and gene-gene interactions on MI based on our study}}
\label{subsubsec:influential_genes}

Our Bayesian hypotheses testing using the clustering metric yielded $P\left(d^*<\epsilon_1|\mbox{Data}\right)\approx 0.35202$
while that with the Euclidean distance we obtained $P\left(d^*_E<\epsilon_2|\mbox{Data}\right)\approx 0.51078$.
In other words, it seems rather debatable whether or not the genes have significant overall effect on MI.
This is in sharp contrast with the results obtained by BB where both clustering metric and Euclidean distance
confirmed significant overall genetic influence on MI.  
However, both the posterior probabilities are substantially large, practically indicating that the genes are
not very significant.
%The resolution of this dilemma seems to lie in incorporation
%of sex in our current model, which accounts for most of the variation between case and control. 

As far as testing of significance of the individual genes are concerned, it turned out that
under the clustering metric, except genes $SMARCA4$, $RBMS1$, $COL4A1$, $RP11-306G20.1$, $MRAS$, $SLC22A1$, $CDKAL1$, 
$PCSK9$, $ADAMTS9-AS2$, and $AP006216.5$, the rest turned out to be significant, while with respect to the
Euclidean metric the only insignificant genes are $AP006216.10$, $CELSR2$, $MRAS$, $PCSK9$, $OR4A48P$ and $BUD13$.
The posterior probabilities of the null hypotheses (of no significant genetic influence) are shown in 
Figure S-3 of the supplement. %\ref{fig:null_hypotheses}.
%\begin{figure}%[htp]
%\centering
%\subfigure[Posterior probability of no genetic effect with respect to clustering metric.]{ \label{fig:clustering_hypotheses}
%\includegraphics[width=15cm,height=6cm]{plots_realdata/postprob_clustering-crop.pdf}}\\
%\vspace{4mm}
%\subfigure[Posterior probability of no genetic effect with respect to Euclidean metric.]{ \label{fig:euclidean_hypotheses} 
%\includegraphics[width=15cm,height=6cm]{plots_realdata/postprob_euclidean-crop.pdf}}
%\caption{{\bf Posterior probabilities of no individual genetic influence:} 
%Index plots of the posterior probabilities of the null hypotheses for (a) clustering metric
%and (b) Euclidean metric, for the $32$ genes.}
%\label{fig:null_hypotheses}
%\end{figure}
%Interestingly, with
%respect to the Euclidean metric, all the five posterior probablities of the null hypotheses associated
%with the aforementioned 5 genes, %AP006216.10, AP006216.5, APOC1 and OR4A48P and AP006216.5, 
%turned out to be empirically zero.
%Thus, even though the clustering metric accepts 5 null hypotheses, the
%confirmation tests with the Euclidean distances suggest rejection of all of them. We hence conclude
%that all the genes considered in the study have significant effect on MI. This is in keeping with the fact that the genes considered in our study were found to be associated with different cardiovascular endpoints in various GWA studies or have been confirmed to play important roles in causing MI in earlier studies.
The figure reveals that the posterior probabilities of no significant genetic influence, although generally did not 
cross $0.5$, are not adequately small to reflect very strong evidence against the null hypotheses. This is consistent
with the result on overall genetic significance that we obtained.

The actual gene-gene correlations based on medians of the posterior covariances,
are shown in Figure S-4 of the supplement. %\ref{fig:ggi_plots2}. 
The color intensities correspond to the absolute values of the
correlations. Consistent with the figure, all the tests on interaction turned out to support the hypotheses
of no interaction.
%\begin{figure}%[htp]
%\centering
%\subfigure[Colorplot of actual posterior gene-gene interaction.]{ \label{fig:ggi_plot} 
%\includegraphics[width=16cm,height=16cm]{plots_realdata/ggi_plot-crop.pdf}}
%\caption{{\bf Gene-gene interaction plot:} Actual gene-gene interactions
%based on medians of the absolute values of the posterior covariances.}
%\label{fig:ggi_plots2}
%\end{figure}

%Thus, it seems that sex has greater influence on MI compared to genes. 
Thus, individual genes have impact on MI but not gene-gene interactions. Moreover, the relatively weak evidences 
against the null suggest that external
factors, in our case sex, may be playing a bigger role in explaining case-control with respect to MI. 
As such, given our data set of size $200$
with $77$ cases, the empirical conditional probability of a male given case is $0.3766234$, while
the empirical conditional probability of a male given control is $0.504065$, indicating that with respect to our data,
females seem to be more at risk compared to males. Coherency of Bayesian models in general is instrumental 
in reflecting this information in our inference in the way of downplaying the genes, suggesting at the same time
that the only external factor, namely, sex, must have more important effect.

A detailed investigation of the disease predisposing loci detected by our model and methods, and the role of
SNP-SNP interactions behind such disease predisposing loci, is carried out in Section S-5 of the supplement,
and a discussion on the posterior distribution of the number of distinct mixture components is provided in Section S-6
of the supplement.

\subsection{{\bf Discussion of our Bayesian methods and GWAS in light of our findings}}
\label{subsec:discussion}

Our results of Bayesian analysis of the MI data set demonstrate that sex plays more significant role than the
genes in triggering the disease, and in particular, do not support gene-gene interaction. In these regards,
our results significantly differ from those obtained by BB, who do not consider the sex variable in their model. 
Since as per our inference sex seems to be far more influential compared to the genes with respect to MI,
there is internal consistency of our more general gene-gene and gene-environment interaction model
with the gene-gene interaction model of BB.
It is important to note that \ctn{Lucas12} analyzed the same MI dataset
using logistic regression and reached the same conclusion as ours that there is no significant gene-gene interaction.
Since two completely different methods of analyses are in such strong agreement, it is pertinent to
presume that the data contains enough information on the lack of gene-gene interaction. However,
as we demonstrated, SNP-SNP correlations have important roles to play in determining the DPLs. 
These are responsible for suppression of the SNPs considered influential in the literature
by implicit induction of negative correlations between Euclidean distances between cases and controls 
for the associated SNPs. Thus, even though the genes did not turn out to be as significant, it is clear
that sophisticated nonparametric modeling of gene-gene and SNP-SNP interactions is of utmost importance.

\section{{\bf Summary and conclusion}}
\label{sec:conclusion}
In this paper, we have extended the Bayesian semiparametric gene-gene interaction model of BB 
to realistically include the case of gene-environment interactions. Careful attention has been paid to the
fact that in the absence of mutation, the environmental variable does not affect the marginal genotypic
distributions, in spite of influencing gene-gene interaction. Needless to mention, our model considers
dependence between SNPs as well to account for LD effects, in addition to gene-gene, gene-environment
and dependencies between individuals. Besides, our model, via Dirichlet processes, facilitates learning
about the number of genotypic sub-populations associated with the individuals and the genes, while
accounting for the environmental effect at the same time. 
%As such, when the environmental effects are
%absent, our model reduces to the gene-gene interaction model of BB.

We extend the Bayesian hypotheses testing methods introduced in BB to enable
test for significances of marginal genetic and environmental effects, gene-gene interactions,
effect of environment on gene-gene interaction and mutational effect. 
The basis for our tests are extensions of the clustering metric based tests proposed by BB
to account for the environmental variables, in conjunction with the tests based on Euclidean metric.
We recommended careful application of our tests based on the clustering metric, followed by re-confirmation
with respect to the Euclidean metric.

On the Bayesian computational side, we propose a powerful parallel processing algorithm
that takes advantage of the conditional independence structures built within our model through
the Dirichlet process based mixture framework for parallelisation, and is complemented by the efficiency of TTMCMC,
which updates the interaction parameters within a single processor.

We validate our model and methodologies with applications to biologically realistic datasets generated
from under $5$ different set-ups characterized by different combinations and structures associated
with gene-gene and gene-environment interactions. Adequate performance of our model and
methods are demonstrated in every situation. Additionally, our ideas correctly captured the true
number of genetic sub-populations in each case, and attempted to capture the DPL adequately
even in the face of highly complex dependence structures. %unlike those in \ctn{Bhattacharya16}.

We apply our model and methods to the MI Gen data set also studied by BB and because
of inclusion of the sex variable, succeeded in obtaining results that are quite compatible with
those reported in the literature. Although the gene-gene interactions turned out to be
insignificant, the SNP-SNP correlations associated with case-control Euclidean distances 
facilitated understanding the mismatch of our DPL with those reported in the literature as
having significant impact on MI. Interestingly, our Bayesian approach allowed us 
obtain insightful results even with a sample consisting of only $200$ individuals, showing the importance
of building sophisticated models and prior structures, and efficient computational methods and technologies.

\newpage

\renewcommand\thefigure{S-\arabic{figure}}
\renewcommand\thetable{S-\arabic{table}}
\renewcommand\thesection{S-\arabic{section}}

\setcounter{section}{0}
\setcounter{figure}{0}
\setcounter{table}{0}

\begin{center}
{\bf \Large Supplementary Material}
\end{center}

\section{{\bf Further discussion regarding the effect of environmental variables on gene-gene interaction}}
\label{sec:E_effect}

It is important to elucidate how our above modeling strategy accounts for the G$\times$G and G$\times$E. %effect
%of the environmental variables on the gene-gene interaction matrix, given that the gene-gene interactions are expected
%to be different for different individuals exposed to different effects of the environmental variable. For instance,
%the gene-gene interactions of a smoker may be expected to be significantly different from those of a non-smoker.

Recall that in our model, $\bA$ represents the gene-gene interaction matrix in the absence of environmental variables,
and has essentially the same interpretation as that of BB. When there is no significant environmental effect on the genes, it is pertinent to test for the significance of the elements of $\bA$  
%see (\ref{eq:mvn_col}) and (\ref{eq:mvn_cov}), 
(see (2.14) and (2.15) of our main manuscript), ignoring the multiplicative constants, 
to learn about gene-gene interactions. %Indeed, when environmental variables are not relevant, 
%then the only difference with respect to the gene-gene covariance structures of the $\blambda$'s of the models of BB and ours is
%with respect to the multiplicative constants of $\bA$ that are elements of the respective $\bSigma$ matrices; for BB's model
%$\bSigma$ stands for $2\times 2$ case-control matrix, while in our case it is a more detailed $N\times N$ matrix 
%representing dependence between all $N$ individuals.
%The elements of the $\bSigma$ matrices, which represent the
%variabilities and covariances between the relevant components of $\blambda$ in both the models, encapsulate the
%influences of various other factors, such as various environmental variables that are not suspected as influential
%with respect to gene-gene interactions and not explicitly incorporated in the models. 
%Hence, when no environmental variable is suspected to be particularly important,

However, when $\bE_i$ affects gene-gene interactions of individual $i$, then it follows from (2.15) %(\ref{eq:mvn_cov}) 
of our main manuscript
that the relevant gene-gene covariance matrix for individual $i$ is $\tilde\sigma_{ii}\bA$, which
involves the effect of $\bE_i$ through $\tilde\sigma_{ii}$. %Here testing for significance of the elements of 

\section{{\bf A parallel MCMC algorithm for model fitting}}
\label{sec:computation}

Recall that the mixtures associated with gene $j\in\{1,\ldots,J\}$, and individual
$i\in\{1,\ldots,N_k\}$ and case-control status $k\in\{0,1\}$,
are conditionally independent of each other, given the interaction parameters.
This allows us to update the mixture components in separate parallel processors, conditionally on the
interaction parameters. Once the mixture components are updated, we update the interaction parameters 
using a specialized form of TMCMC, in a single processor. The details of updating the mixture components 
in parallel are as follows.

\begin{itemize}

%\subsection{{\bf Updating the allocation variables using Gibbs steps}}
%\label{subsec:fullcond_z}

\item[(1)] Split the triplets $\left\{(i,j,k):~i=1,\ldots,N_k;~j=1,\ldots,J;~k=0,1\right\}$ in the available
parallel processors. For our convenience, we split the triplets sequentially into 
$$\mathcal T_1=\left\{(i,j,0):~i=1,\ldots,N_0;~j=1,\ldots,J\right\}$$ and 
$$\mathcal T_2=\left\{(i,j,1):~i=1,\ldots,N_1;~j=1,\ldots,J\right\};$$
we then parallelise updation of the mixtures associated with $\mathcal T_1$, followed by those of
$\mathcal T_2$. 

\item[(2)] During each MCMC iteration, for each $(i,j,k)$ 
in each available parallel processor, do the following 
\begin{enumerate}
%\item[{\bf Update allocation variables}]
\item[(i)]
%For $i=1,\ldots,N_k$, 
Update the allocation variables $z_{ijk}$ by simulating
from the full conditional distribution of $z_{ijk}$, given by
\begin{equation}
[z_{ijk}=m|\cdots]\propto\pi_{mijk}\prod_{r=1}^{L_j}f\left(\bx_{ijkr}|p_{mijkr}\right);
%{\sum_{m'=1}^M\pi_{m'jk}\prod_{r=1}^{L_j}f\left(\bx_{ijkr}|p_{m'jkr}\right)}
\label{eq:fullcond_z}
\end{equation}
for $m=1,\ldots,M$.

%\subsection{{\bf Updating the configuration indicators associated with the mixtures using Gibbs steps}}
%\label{subsec:fullcond_c}

%\item[{\bf Updating the configuration indicators associated with the mixtures using Gibbs steps}]
\item[(ii)]

%For each $(j,k)$, 
Let $\left\{\bp^*_{1ijk},\ldots,\bp^*_{\tau_{ijk} ijk}\right\}$ denote the distinct elements
in $\bP_{Mijk}=\left\{\bp_{1ijk},\ldots,\bp_{Mijk}\right\}$. Also let $\bC_{ijk}=\left\{c_{1ijk},\ldots,c_{Mijk}\right\}$
denote the configuration vector, where $c_{mijk}=\ell$ if and only if $\bp_{mijk}=\bp^*_{\ell ijk}$.

Now let $\tau^{(m)}_{ijk}$ denote the number of distinct elements in 
$\bP_{-Mijkm}=\bP\backslash\left\{\bp_{mijk}\right\}$ 
and let ${\bp^m}^*_{\ell}=\left\{{p^{m}}^*_{\ell ijkr};~r=1,\ldots,L_j\right\};~\ell=1,\ldots,\tau^{(m)}_{ijk}$ 
denote the distinct parameter vectors. Further, let ${\bp^m}^*_{\ell}$
occur $M_{\ell m}$ times. 

Then update $c_{mijk}$ using Gibbs steps, where the full conditional distribution of $c_{mijk}$ is given by
\begin{equation}
[c_{mijk}=\ell|\cdots]\propto\left\{\begin{array}{ccc}q^*_{\ell, mijk} & \mbox{if} & \ell=1,\ldots,\tau^{(m)}_{ijk};\\
q_{0,mijk} & \mbox{if} & \ell=\tau^{(m)}_{ijk}+1,\end{array}\right.
\label{eq:fullcond_c}
\end{equation}
where
\begin{align}
q_{0,mijk} &=\alpha_{ijk}\prod_{r=1}^{L_j}\frac{\beta\left(n_{1mijkr}+\nu_{1ijkr},n_{2mijkr}+\nu_{2ijkr}\right)}
{\beta\left(\nu_{1ijkr},\nu_{2ijkr}\right)};
\label{eq:q0}\\
q^*_{\ell, mjk} &=M_{\ell m}\prod_{r=1}^{L_j}\left\{{p^{m}}^*_{\ell jkr}\right\}^{n_{1mijkr}}
\left\{1-{p^{m}}^*_{\ell ijkr}\right\}^{n_{2mijkr}}.
\label{eq:q1}
\end{align}
In (\ref{eq:q0}) and (\ref{eq:q1}), $n_{1mijkr}$ and $n_{2mijkr}$ denote the number of $``a"$ and $``A"$ alleles,
respectively, at the $r$-th locus of the $j$-th gene of the $i$-th individual, associated with the $m$-th mixture component.
In other words, 
%$n_{1mjr}=\sum_{i:z_{ijk}=m}\left(x^1_{ijkr}+x^2_{ijkr}\right)$ and
%$n_{2mjr}=\sum_{i:z_{ijk}=m}\left\{2-\left(x^1_{ijkr}+x^2_{ijkr}\right)\right\}$.
$n_{1mijkr}=x^1_{ijkr}+x^2_{ijkr}$ and
$n_{2mjr}=2-\left(x^1_{ijkr}+x^2_{ijkr}\right)$.
The function $\beta(\cdot,\cdot)$ in the above equations is the Beta function such that for any $s_1>0, s_2>0$,
$\beta(s_1,s_2)=\frac{\Gamma(s_1)\Gamma(s_2)}{\Gamma(s_1+s_2)}$; $\Gamma(\cdot)$ being the Gamma function.

%\subsection{{\bf Updating the parameters of the distinct components associated with the mixtures using Gibbs steps}}
%\label{subsec:fullcond_distinct}

%\item[{\bf Update the parameters of the distinct components associated with the mixtures using Gibbs steps}]
\item[(iii)]

Let ${n}^*_{1\ell ijkr}=\sum_{m:c_{mijk}=\ell}n_{1mijkr}$ and  ${n}^*_{2\ell ijkr}=\sum_{m:c_{mijk}=\ell}n_{2mijkr}$.
Then, for $\ell=1,\ldots,\tau_{jk}$; $r=1,\ldots,L_j$; $j=1,\ldots,J$ and $k=0,1$, 
update ${p}^*_{\ell ijkr}$ by simulating from its full conditional distribution,
given by
\begin{equation}
[{p}^*_{\ell ijkr}|\cdots]\sim \mbox{Beta}\left({n}^*_{1\ell ijkr}+\nu_{1ijkr},{n}^*_{2\ell ijkr}+\nu_{2ijkr}\right).
\label{eq:fullcond_p}
\end{equation}
\end{enumerate}
\item[(3)] During each MCMC iteration, update the interaction parameters $\left\{(u_{r'},v_{r'});~r'=1,\ldots,L\right\}$,
$\bLambda$, $\bA$ and $\bSigma$, $\bA_{\alpha}$, $\bA_{\beta}$, $\bSigma_{\alpha}$, $\bSigma_{\beta}$, $b$, and $\phi$ 
in a single processor using TMCMC, conditionally on the remaining 
parameters. As in BB, we update these parameters using a mixture of additive and additive-multiplicative
TMCMC, exploiting the Cholesky factorizations of the positive definite matrices, 
and updating only the non-zero elements of the respective lower triangular matrices.
%The details of updating the interaction parameters are provided in Section \ref{subsec:interaction_update}.
\end{itemize}

\section{{\bf Clustering metric, clustering mode, and divergence measures based on Euclidean distance}}
\label{sec:clustering_metric_mode}

Let $C_1$ and $C_2$ denote two possible clusterings of some dataset. Let ${K_1}$ and $K_2$ denote the number of
clusters of clusterings $C_1$ and $C_2$ respectively, and let $\tilde n_{{i}{j}}$ denote the number of units belonging to the 
${i}$-th cluster of $C_1$ and ${j}$-th cluster of $C_2$,
and $\tilde n_{00}=\sum\sum \tilde n_{ij}$ is the total number of units. 
Following \ctn{Sabya11} and BB we consider the following divergence between $C_1$ and $C_2$,
which has been conjectured to be a metric by \ctn{Sabya11}: 
\begin{equation}
\hat d(C_1,C_2)=\max\left\{\bar d(C_1,C_2),\bar d(C_2,C_1)\right\},
\label{eq:approx_final}
\end{equation}
where
\begin{eqnarray}
\bar d(C_1,C_2)&=&\left\{\tilde n_{00}-\sum_{{i}=1}^{K_1}
\max_{1\leq j\leq K_2}\tilde n_{{i}{j}}\right\}\bigg/{\tilde n}_{00}\label{eq:approx1}\\
&=&1-\frac{\sum_{{i}=1}^{K_1}\underset{1\leq j\leq K_2}{\max}
{\tilde n}_{{i}{j}}}{{\tilde n}_{00}}.\label{eq:approx2}
\end{eqnarray}

Let $\mathcal C$ denote the set of all possible clusterings of some dataset.
Motivated by the definition of mode in the case of parametric distributions, \ctn{Sabya11} define that clustering 
$C^*$ as ``central," which, for a given small $\epsilon>0$, satisfies the following equation:
\begin{equation}
P\left(\left\{C\in\mathcal C:\hat d(C^*,C)<\epsilon\right\}\right)
=\sup_{C'\in\mathcal C} P\left(\left\{C\in\mathcal C:\hat d(C',C)<\epsilon\right\}\right).
\label{eq:central_cluster}
\end{equation}
Note that $C^*$ is the global mode of the posterior distribution of clustering as $\epsilon\rightarrow 0$. 
Thus, for a sufficiently small $\epsilon>0$, the probability of an $\epsilon$-neighborhood of an 
arbitrary clustering $C'$, of the form $\left\{C:\hat d(C',C)<\epsilon\right\}$,
is highest when $C'=C^*$, the central clustering. 

In a set of clusterings $\left\{C^{(\ell)}:\ell=1,\ldots,L\right\}$, 
\ctn{Sabya11} define that clustering $C^{(j)}$ as ``approximately central," which, for a given small $\epsilon>0$, satisfies 
the following equation:
\begin{equation}
C^{(j)}=\arg\max_{1\leq \ell\leq L}\frac{1}{L}\#\left\{C^{(k)};1\leq k\leq L:\hat d(C^{(\ell)},C^{(k)})<\epsilon\right\}.
\label{eq:empirical_central_cluster}
\end{equation}
The central clustering $C^{(j)}$ is easily computable, given $\epsilon>0$ and a suitable metric $d$. Also,
by the ergodic theorem, as $L\rightarrow\infty$ the empirical central clustering $C^{(j)}$ converges almost surely to the exact
central clustering $C^*$.

\begin{figure}%[htp]
\centering
\includegraphics[width=10cm,height=10cm]{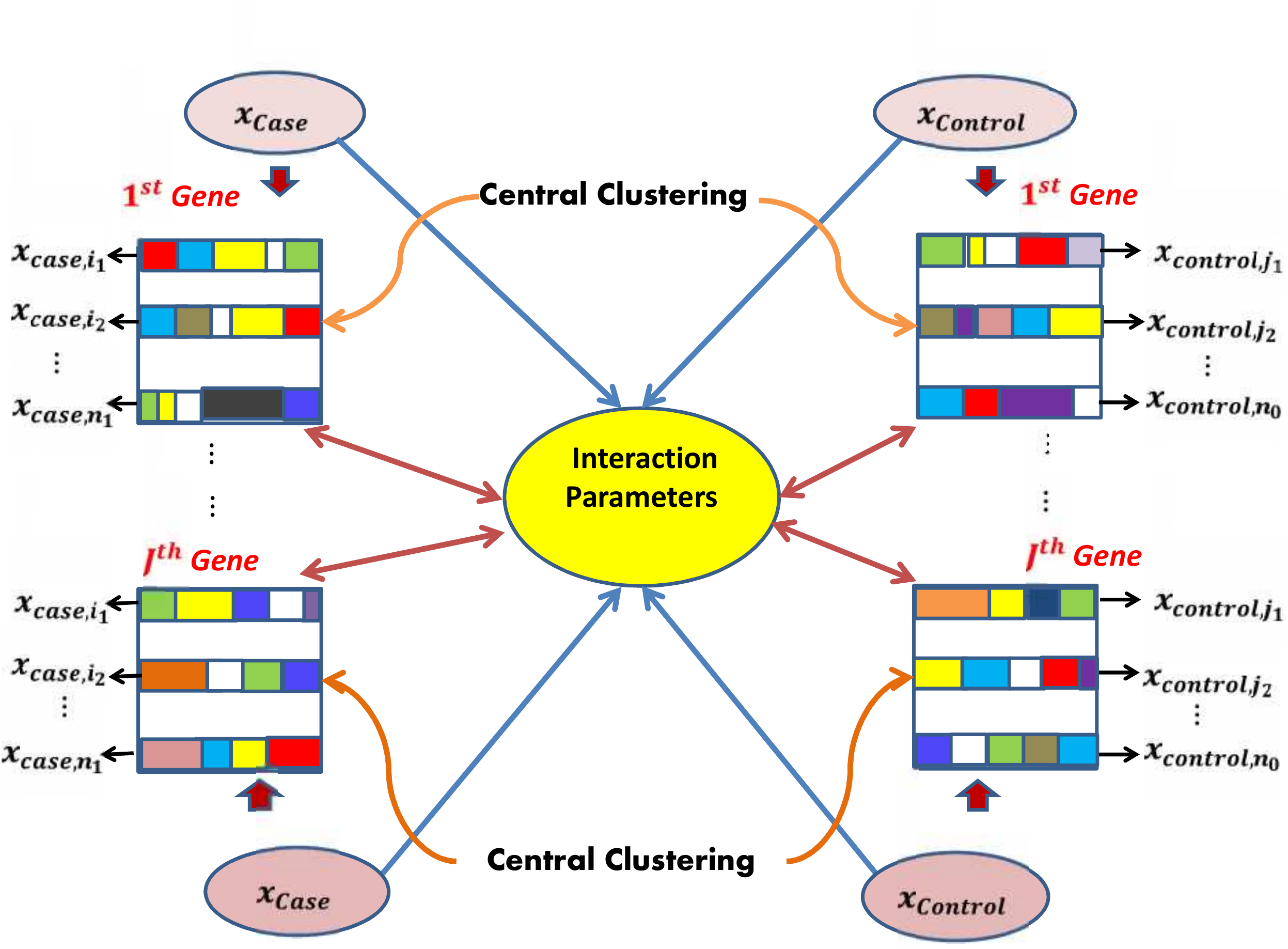}
\caption{{\bf Schematic diagram for our model and testing of hypothesis based on the ideas of central clustering.}}
\label{fig:schematic2}
\end{figure}

In our case, we shall obtain $i_0$ and $i_1$, the indices of the central clusterings associated with
$\bP_{Mijk=0};~i=1,\ldots,N_0$ and $\bP_{Mijk=1};~i=1,\ldots,N_1$, respectively, obtained by the above
method. Once $i_0$ and $i_1$ are determined,
we shall consider clustering distances between $\bP_{Mi_0jk=0}$ and $\bP_{Mi_1jk=1}$, denoted by
$\hat d\left(\bP_{Mi_0jk=0},\bP_{Mi_1jk=1}\right)$. We shall be particularly interested in (\ref{eq:d_star}).
%\begin{equation}
%d^*=\max_{1\leq j\leq J}\hat d\left(\bP_{Mi_0jk=0},\bP_{Mi_1jk=1}\right);
%\label{eq:d_star}
%\end{equation}
A schematic representation of our model and hypothesis testing based on the ideas of central clustering is shown 
in Figure \ref{fig:schematic2}.

\subsection{{\bf Shortcoming of the clustering metric for hypothesis testing and a divergence
measure based on Euclidean distance}}
\label{subsec:clustering_shortcoming}

%BB point out that 
%although significantly large divergence between clusterings of $\bP_{Mjk=0}$ and $\bP_{Mjk=1}$ indicate rejection of $H_0$,
%insignificant clustering distance between $\bP_{Mjk=0}$ and $\bP_{Mjk=1}$ need not necessarily 
%provide strong enough evidence in favour of $H_0$. In other words, even if the clustering distance is insignificant, 
%it is important to check if the parameter vectors being compared are significantly different. 

BB note that when two clusterings are the same, %(at least not significantly different under $H_0$), 
minimizing
the Euclidean distance over all possible permutations of the clusters, provides a sensible measure
of divergence. 
In other words, for any two vectors $\bv^{(1)}=\left(v^{(1)}_1,\ldots,v^{(1)}_K\right)$ and 
$\bv^{(2)}=\left(v^{(2)}_1,\ldots,v^{(2)}_K\right)$ 
in $K$-dimensional Euclidean space, where $K>1$,
BB propose the following divergence measure:
\begin{equation}
d_{E,\mbox{min}}\left(\bv^{(1)},\bv^{(2)}\right)=\min_{j_1,\ldots,j_K}\sqrt{\sum_{i=1}^K\left(v^{(1)}_i-v^{(2)}_{j_i}\right)^2}, 
\label{eq:d_min}
\end{equation}
the minimization being over all possible permutations $(j_1,j_2,\ldots,j_K)$ of $(1,2,\ldots,K)$.
The above divergence is non-negative, symmetric in that 
$d_{E,\mbox{min}}\left(\bv^{(1)},\bv^{(2)}\right)=d_{E,\mbox{min}}\left(\bv^{(2)},\bv^{(1)}\right)$, 
satisfies the property
$d_{E,\mbox{min}}\left(\bv^{(1)},\bv^{(2)}\right)=0$ if and only if $\bv^{(1)}=\bv^{(2)}$,
and is 
invariant with respect to permutations of the clusters. 

Since computation of $d_{E,\mbox{min}}$ involves minimization over all possible permutations, great computational
burden will be incurred. BB devise a strategy based on the simple Euclidean distance 
$d_E$ (which does not require minimization over permutations), which can often avoid such computational burden. 
The idea is that, if the null hypothesis is accepted with respect to $d_E$, 
then this clearly implies acceptance of the null with respect to $d_{E,\mbox{min}}$, so that
minimization over permutations is completely avoided.
If, on the other hand, the null is rejected when tested with $d_E$, then one must re-test the null 
using $d_{E,\mbox{min}}$, which would involve dealing with permutations. In our case
we compute the Euclidean distance after giving the logit transformation to the minor allele frequences.
The details are provided in Section 3.2 of the main manuscript.
The method of selecting appropriate $\varepsilon$'s for the hypotheses tests are provided in Section S-4
of the supplement.

\section{{\bf Simulation studies}}
\label{sec:simulation_study_supp}

\subsection{{\bf First simulation study: presence of gene-gene and gene-environment interaction}}
\label{subsec:first_simulation_study}

\subsubsection{{\bf Data description}}
\label{subsubsec:data_description}
As in BB we consider two genetic factors as allowed by GENS2
and simulated 5 data sets with gene-gene and gene-environment interaction with a one-dimensional environmental
variable, associated with 5 sub-populations. 
%The data sets consist of disease status, gender, environmental exposures and genotypes for each individual.
One of the genes consists of 1084 SNPs and another has 1206 SNPs, with 
one DPL at each gene. There are 113 individuals in each of the 5 data sets, from which
we selected a total of 100 individuals without replacement with probabilities 
assigned to the 5 data sets being $(0.1, 0.4, 0.2, 0.15, 0.15)$. 
%That is, we chose one of the 
%5 data sets with these probabilities and selected a row randomly from the chosen data set; we repeated
%this procedure 100 times without replacing the rows. In our final data set thus obtained, there
Our final dataset consists of 46 cases and 54 controls.
Since, in our case, the environmental variable is one-dimensional, $d=1$.

\subsubsection{{\bf Model implementation}}
\label{subsubsec:model_implementation}
We implemented our parallel MCMC algorithm on i7 processors by splitting the mixture
updating mechanisms in 8 parallel processors, and updating the interaction parameters in a single processor.
Our code is written in C in conjunction with the Message Passing Interface (MPI) protocol for parallelisation.

The total time taken to implement $100,000$ MCMC iterations, where the first $50,000$ are discarded as burn-in,
is approximately 4 days. We assessed convergence informally with trace plots, which indicated 
adequate mixing properties of our algorithm.

\subsubsection{{\bf Specifications of the thresholds $\varepsilon$'s using null distributions}}
\label{subsubsec:threshold}

Following the method outlined in Section 3.3.1 of our main manuscript, %\ref{subsubsec:e_choice}, 
setting $M$, the maximum number
of distinct components to be 30, and $\alpha_{ijk}=1.5$ following BB, we obtain
%$\varepsilon_{d^*}=0.633$, $\varepsilon_{\hat d_1}=0.6$, $\varepsilon_{\hat d_2}=0.6$,
%$\varepsilon_{d^*_E}=17.891$, $\varepsilon_{d^*_{E,1}}=17.370$, $\varepsilon_{d^*_{E,2}}=16.015$,
%$\varepsilon_{\beta_{110}}=1.792$, $\varepsilon_{\beta_{120}}=0.496$, 
%$\varepsilon_{\beta_{111}}=0.149$, $\varepsilon_{\beta_{121}}=0.218$,
%$\varepsilon_{\phi}=1.993$.

$\varepsilon_{d^*}=0.633$, $\varepsilon_{\hat d_1}=0.6$, $\varepsilon_{\hat d_2}=0.6$,
$\varepsilon_{d^*_E}=18.000$, $\varepsilon_{d^*_{E,1}}=17.483$, $\varepsilon_{d^*_{E,2}}=17.249$,
$\varepsilon_{\beta_{110}}=0.570$, $\varepsilon_{\beta_{120}}=0.665$, 
$\varepsilon_{\beta_{111}}=1.819$, $\varepsilon_{\beta_{121}}=1.106$,
$\varepsilon_{\phi}=0.658$, $\varepsilon_{A_{12}}=\varepsilon_{A_{21}}=0.200$.

\subsubsection{{\bf Results of fitting our model}}
\label{subsubsec:results_first_simulation_study}

%Figure \ref{fig:ggi_metric_plots} displays the posterior distributions of 
%$d^*=\underset{j=1,2}{\max}~\hat d\left(\bP_{30,j,0},\bP_{30,j,1}\right)$, 
%$\hat d_1=\hat d\left(\bP_{30,1,0},\bP_{30,1,1}\right)$ and 
%$\hat d_2=\hat d\left(\bP_{30,2,0},\bP_{30,2,1}\right)$, respectively.
%The diagrams show that in all the three cases, regions that are significantly bounded away from zero
%have high posterior probabilities compared to those closer to zero.
%%the region to the right of $0.2$ has significantly higher posterior 
%%probability compared to the left of $0.2$. 
%%, while the maximum value is around $0.6$. 
%For the purpose of formal Bayesian
%hypothesis, following the discussion in Section \ref{subsubsec:threshold}, we
%set $\varepsilon=0.233$. 
%%to be $1/3$ of the maximum value $0.6$, that is,
%%we specify $\varepsilon=0.2$. 

The posterior probabilities
$P\left(d^*<\varepsilon_{d^*}|\mbox{Data}\right)$, $P\left(\hat d_1<\varepsilon_{\hat d_1}|\mbox{Data}\right)$
and $P\left(\hat d_2<\varepsilon_{\hat d_2}|\mbox{Data}\right)$
empirically obtained from $50,000$ MCMC samples,
turned out to be $0.358$, $0.334$ and $0.336$, respectively. 
Hence, $H_{0,d^*}$, $H_{0,\hat d_1}$ and $H_{0,\hat d_2}$ are rejected, suggesting
the influence of significant genetic effects in the case-control study. Moreover,
$P\left(d^*_E<\varepsilon_{d^*_E}|\mbox{Data}\right)$, $P\left(\hat d_{E,1}<\varepsilon_{\hat d_{E,1}}|\mbox{Data}\right)$
and $P\left(\hat d_{E,2}<\varepsilon_{\hat d_{E,2}}|\mbox{Data}\right)$ are given, approximately, by
$0.460$, $0.916$ and $0.361$, respectively. That is, even though $H_{0,\hat d^*_{E,1}}$ is to be accepted,
there is not enough evidence to suggest acceptance of $H_{0,\hat d^*_{E,2}}$ and $H_{0,\hat d^*_E}$. 
Thus, with respect to the ``0-1" loss, the test with respect to the Euclidean-based
metric is consistent with the test with respect to the clustering metric.   

To check the influence of the environmental 
variable on the genes we compute the posterior probabilities 
$P\left(|\beta_{1jk}|<\varepsilon_{\beta_{1jk}}|\mbox{Data}\right)$, for $j=1,2$ and $k=0,1$.
The probabilities turned out to be $0.759$, $0.253$, $1.000$ and $1.000$, respectively, showing that
$\beta_{111}$ is significant. That is, the environmental variable has a significant effect on gene $j=1$. 
Now if gene-gene interaction is found to be significant, then the interaction of the environment and gene 1 would seem to have affected  
gene $j=2$ as well, so that both $H_{0,\hat d_1}$ and $H_{0,\hat d_2}$ are rejected. 
Hence, we now investigate the significance of gene-gene interaction.

%Figure \ref{fig:ggi_A} shows the posterior distributions of the elements of $\bA$. Note that the
%posterior of $\bA_{12}$ ($=\bA_{21}$), which may be thought of as the gene-gene interaction parameter,
%has a long left tail supported on negative values. This seems to suggest significant
%gene-gene interaction.
%Using the procedure detailed in Section \ref{subsubsec:threshold}, we obtain $\varepsilon=0.166$. %0.09.
%The relevant empirical posterior probability is given by 
%$\mbox{P}\left(\left|\bA_{12}\right|<0.09|\mbox{Data}\right)\approx 0.169$, clearly pointing
As regards $\phi$, the corresponding posterior
probability turned out to be $0.982$, indicating its insignificance. Noting that the model of \ctn{Pinelli12}
does not have provision for any interaction terms related to our matrix $\mathcal E$, $\phi=0$
is the true hypothesis.
The relevant posterior probability of $\bA_{12}$ ($=\bA_{21}$) is given by
$\mbox{P}\left(\left|\bA_{12}\right|<\varepsilon_{A_{12}}|\mbox{Data}\right)\approx 0.463$, which implies
statistically significant gene-gene interaction under the ``$0-1$" loss.

This seems to confirm the roles of genes, influenced by the environmental variable in the simulated case-control study.

\begin{figure}%[htp]
\centering
\subfigure[Posterior of $\tau_{110}$.]{ \label{fig:ggi_comp_prob1}
\includegraphics[width=7cm,height=7cm]{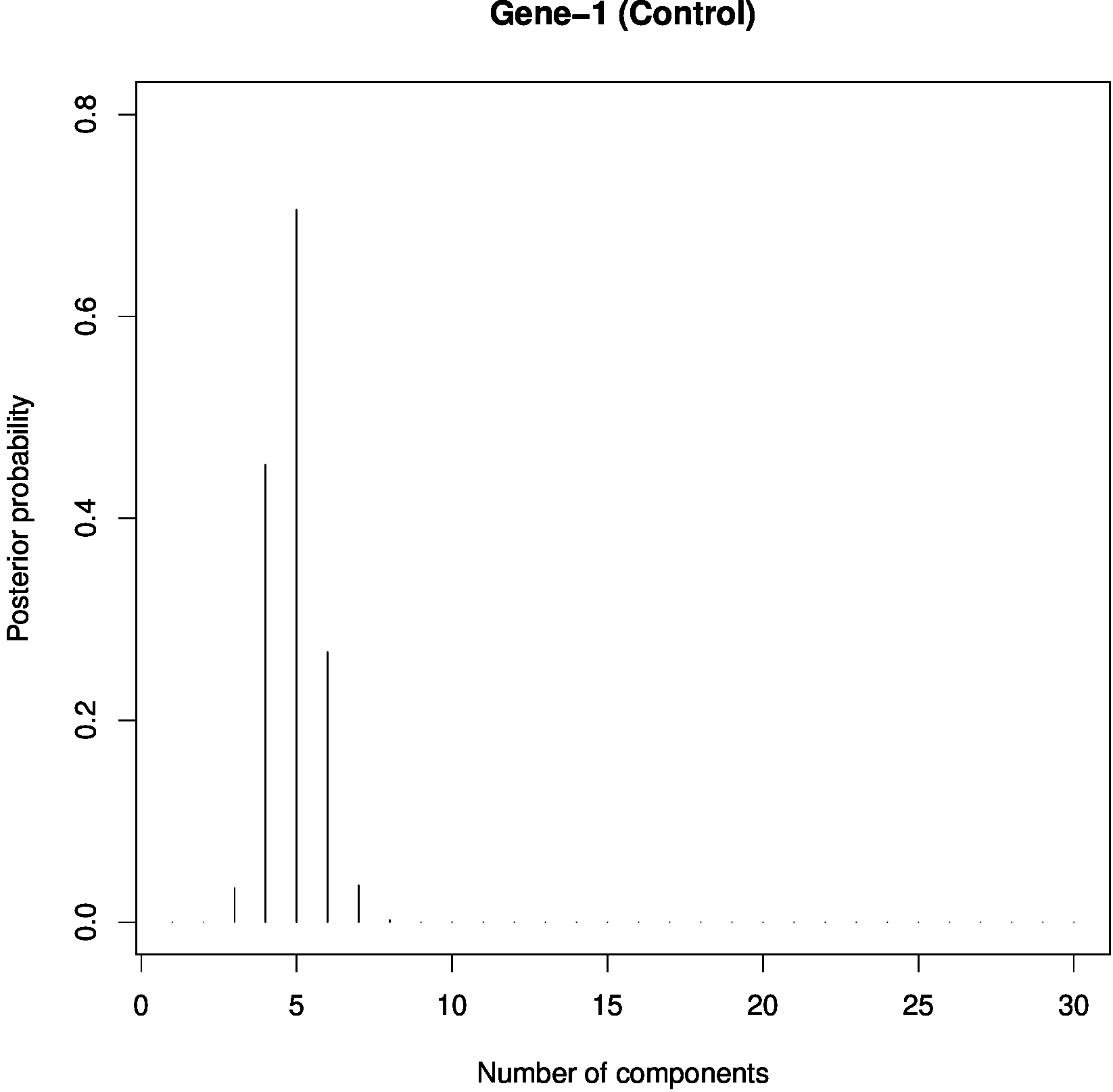}}
\hspace{2mm}
\subfigure[Posterior of $\tau_{211}$.]{ \label{fig:ggi_comp_prob2}
\includegraphics[width=7cm,height=7cm]{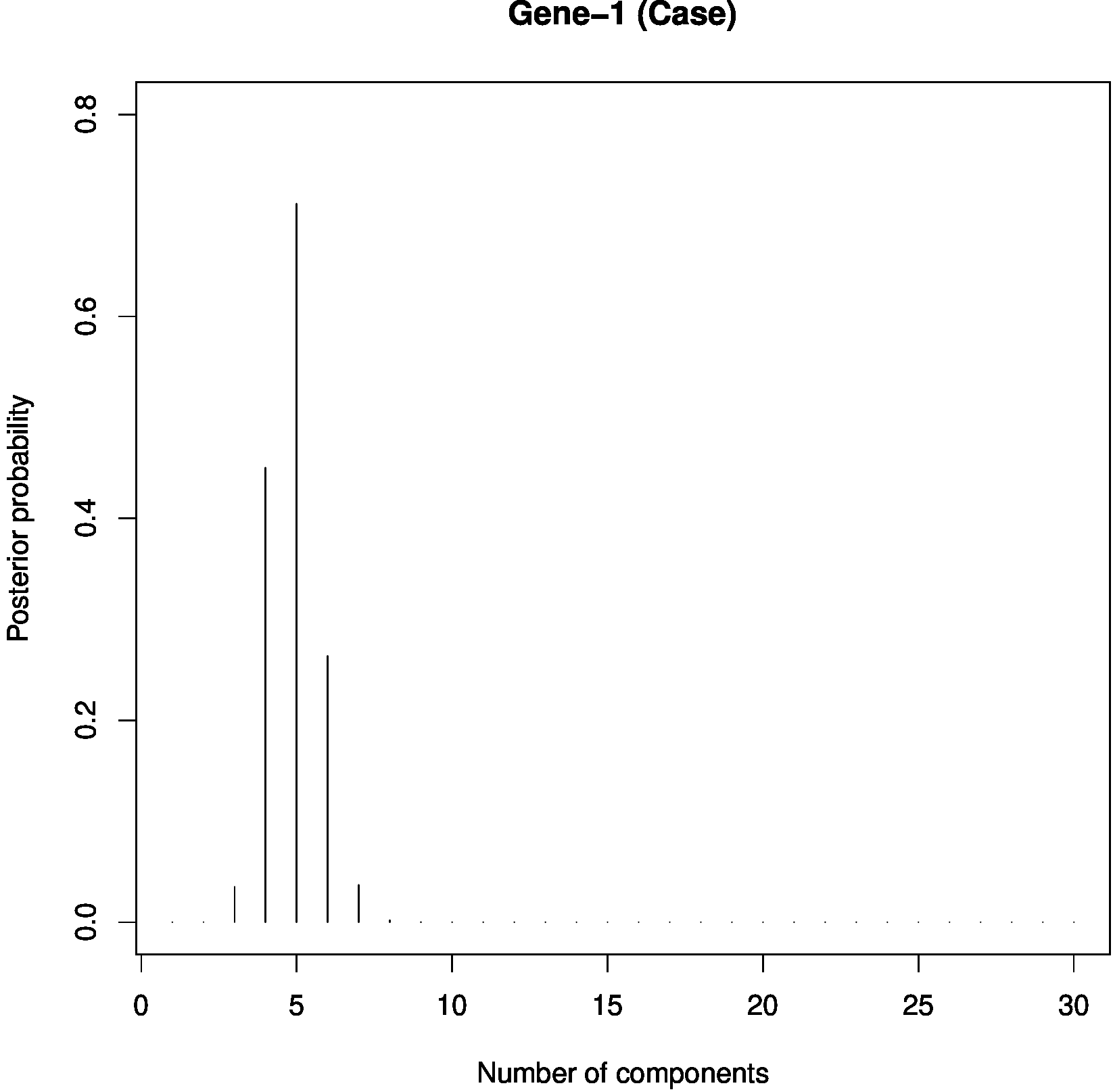}}
\hspace{2mm}
\subfigure[Posterior of $\tau_{120}$.]{ \label{fig:ggi_comp_prob3}
\includegraphics[width=7cm,height=7cm]{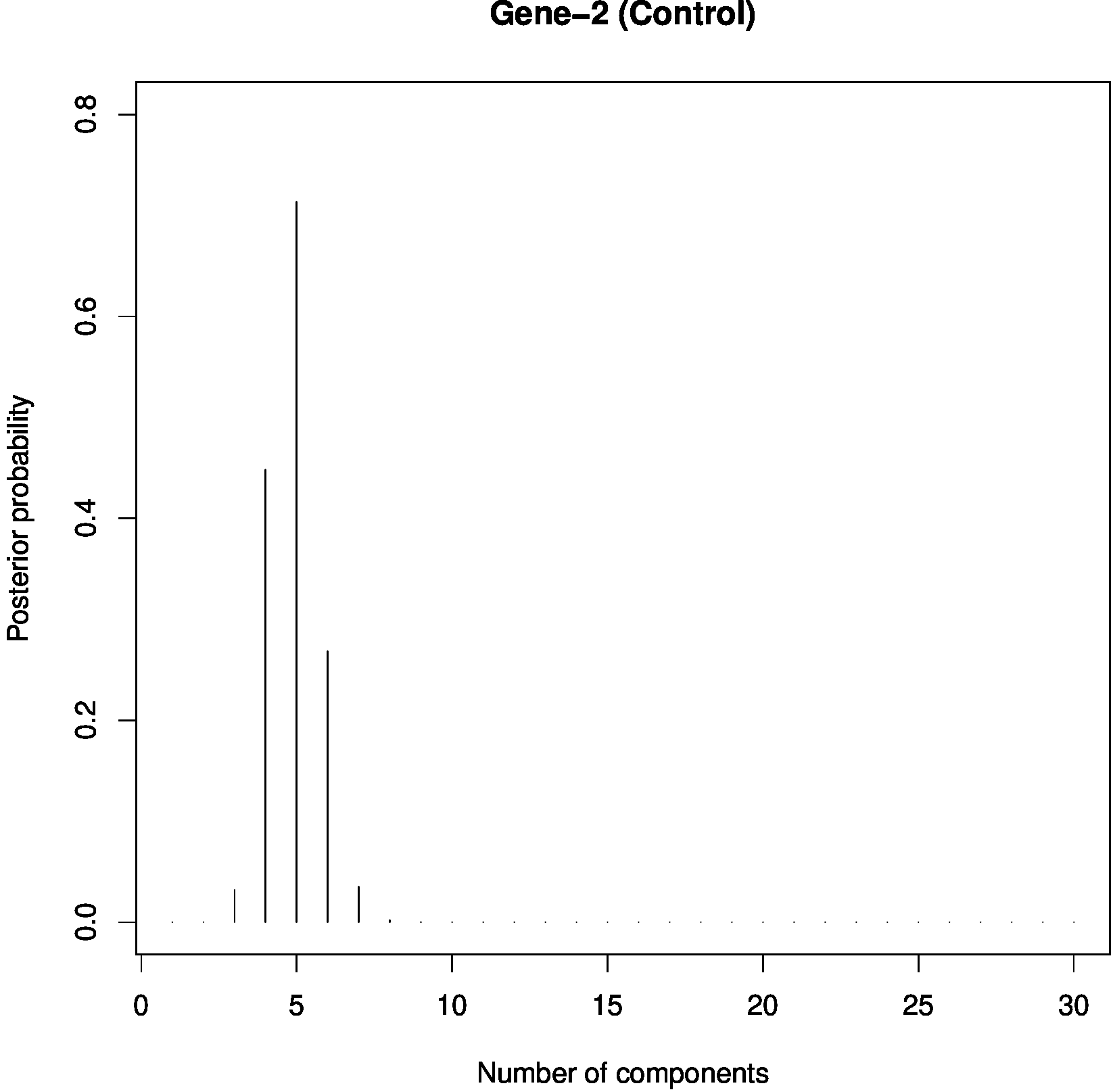}}
\hspace{2mm}
\subfigure[Posterior of $\tau_{221}$.]{ \label{fig:ggi_comp_prob4}
\includegraphics[width=7cm,height=7cm]{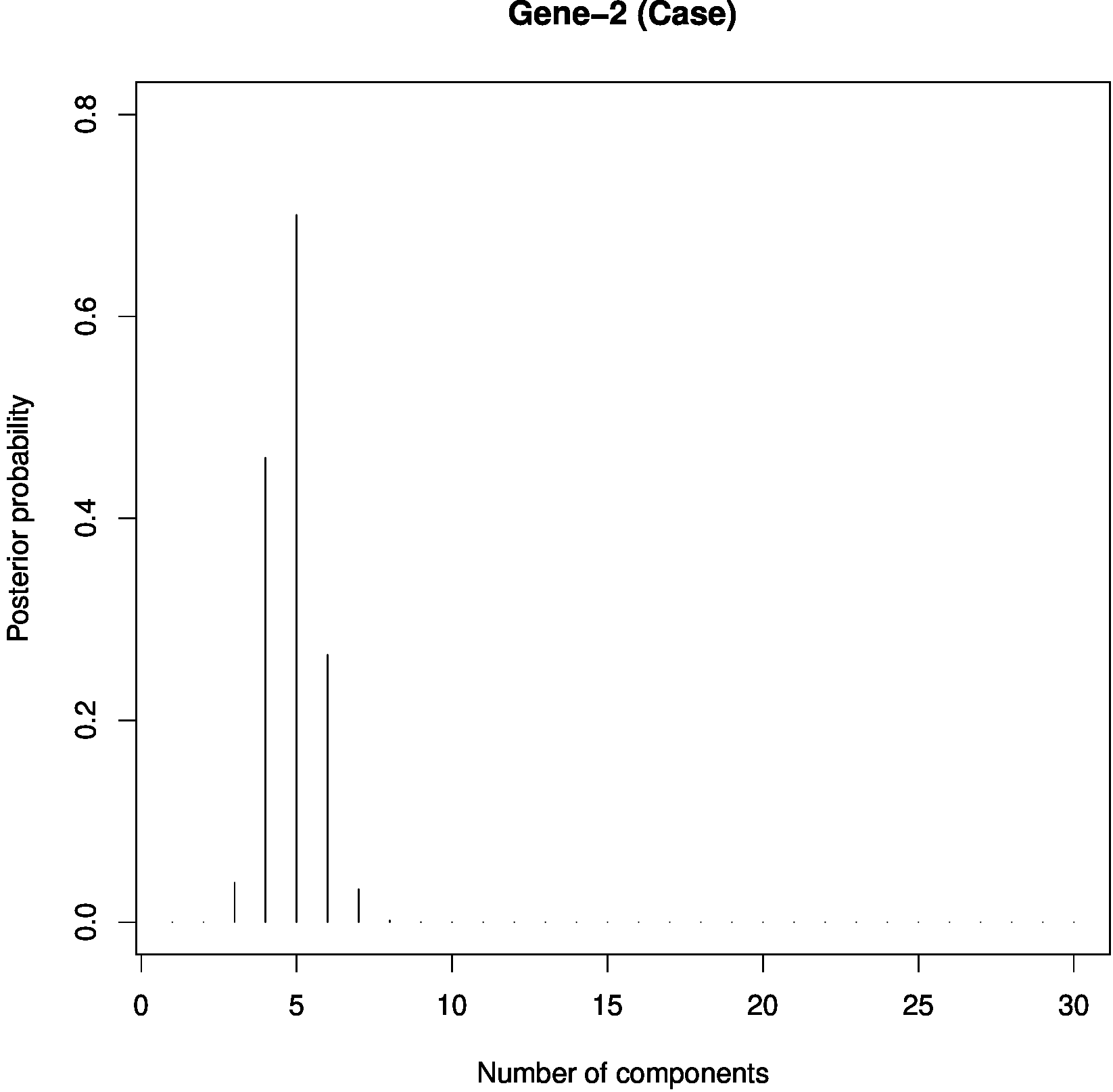}}
\caption{{\bf Gene-gene and gene-environment interaction:} Posterior distributions of the number of distinct components.}
%for each pair ($j,k$); $j=1,2$; $k=0,1$.}
\label{fig:ggi_comp}
\end{figure}

Finally, the true number of sub-populations has been identified correctly by our model and methods,
even though we set the maximum number of components $M$ to be $30$. All the posteriors related to the number of components have correctly concentrated around 5, the true number of components, a few of which are shown in Figure \ref{fig:ggi_comp}.

\subsubsection{{\bf Detection of DPL}}
\label{subsubsec:dpl}

The correct positions of the DPL, provided by GENS2, are $rs13266634$ and $rs7903146$, 
for the first and second gene respectively. Due to the LD effects implied by the correlated structured of our model
the actual DPL need not be easy to locate. For the gene-gene interaction model of BB 
it has been possible to identify a relatively small set of loci which included the actual DPLs. Our current 
gene-gene and gene-environment interaction model is, however, much more structured due to incorporation
of gene-environment dependence in addition to gene-gene dependence. In particular, since 
$\nu_{1ijkr}$ and $\nu_{2ijkr}$ %given by (\ref{eq:nu_1}) and (\ref{eq:nu_2}) 
consist of $\lambda_{ijk}$, $\mu_{jk}$ and $\bbeta_{jk}$ that are shared by every locus of 
the $j$-th gene of the individual denoted by $(i,k)$,
and because $\beta_{121}$ is significant in our example, this induces further dependence   
between the loci of the second gene. Because of gene-gene interaction, this also implicitly induces
further dependence between the loci of the first gene.
Hence, it is rather challenging to locate the DPLs in the presence
of both gene-gene and gene-environmental interactions.

But in spite of the difficulties, it has been possible to segregate the DPL, albeit
not as precisely as in BB. Borrowing the idea of BB, and letting 
$\bp^r_{ijk}=\left\{p_{mijkr}:m=1,\ldots,M\right\}$, we declare the $r$-th locus of the $j$-th 
gene as disease pre-disposing if, for the $r$-th locus, the Euclidean distance
$d^r_j\left(\mbox{logit}\left(\bp^r_{i_0jk=0}\right),\mbox{logit}\left(\bp^r_{i_1jk=1}\right)\right)$, 
between $\mbox{logit}\left(\bp^r_{i_0jk=0}\right)$ and $\mbox{logit}\left(\bp^r_{i_1jk=1}\right)$, is significantly larger than 
$d^{r'}_j\left(\bp^{r'}_{i_0jk=0},\bp^{r'}_{i_1jk=1}\right)$, for $r'\neq r$.  
We adopt the graphical method as BB.

\begin{figure}%[htp]
\centering
\subfigure[Index plot for the first gene]{ \label{fig:index_plot_gene1}
\includegraphics[width=8cm,height=8cm]{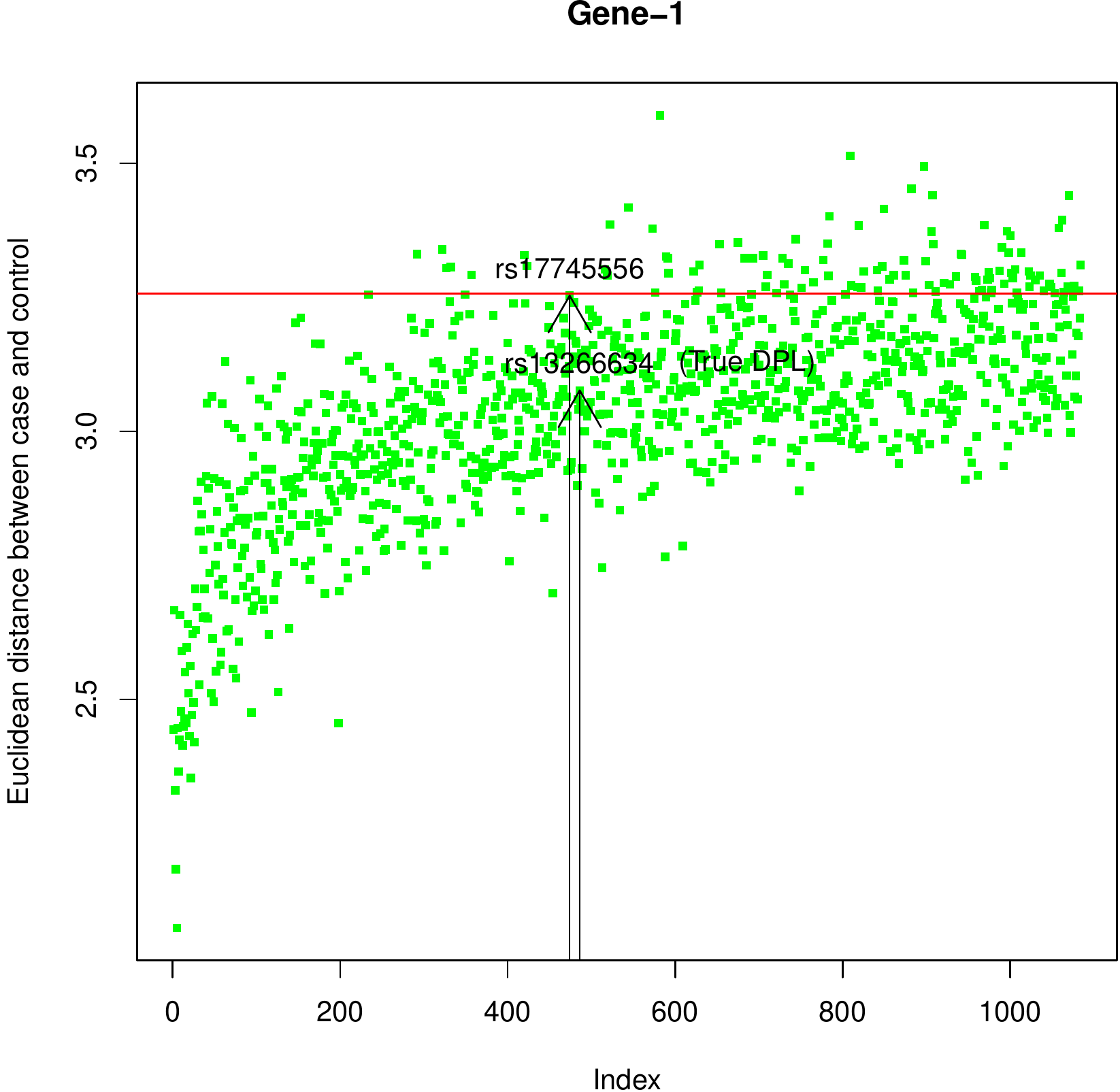}}
%\vspace{2mm}
\subfigure[Index plot for the second gene.]{ \label{fig:index_plot_gene2}
\includegraphics[width=8cm,height=8cm]{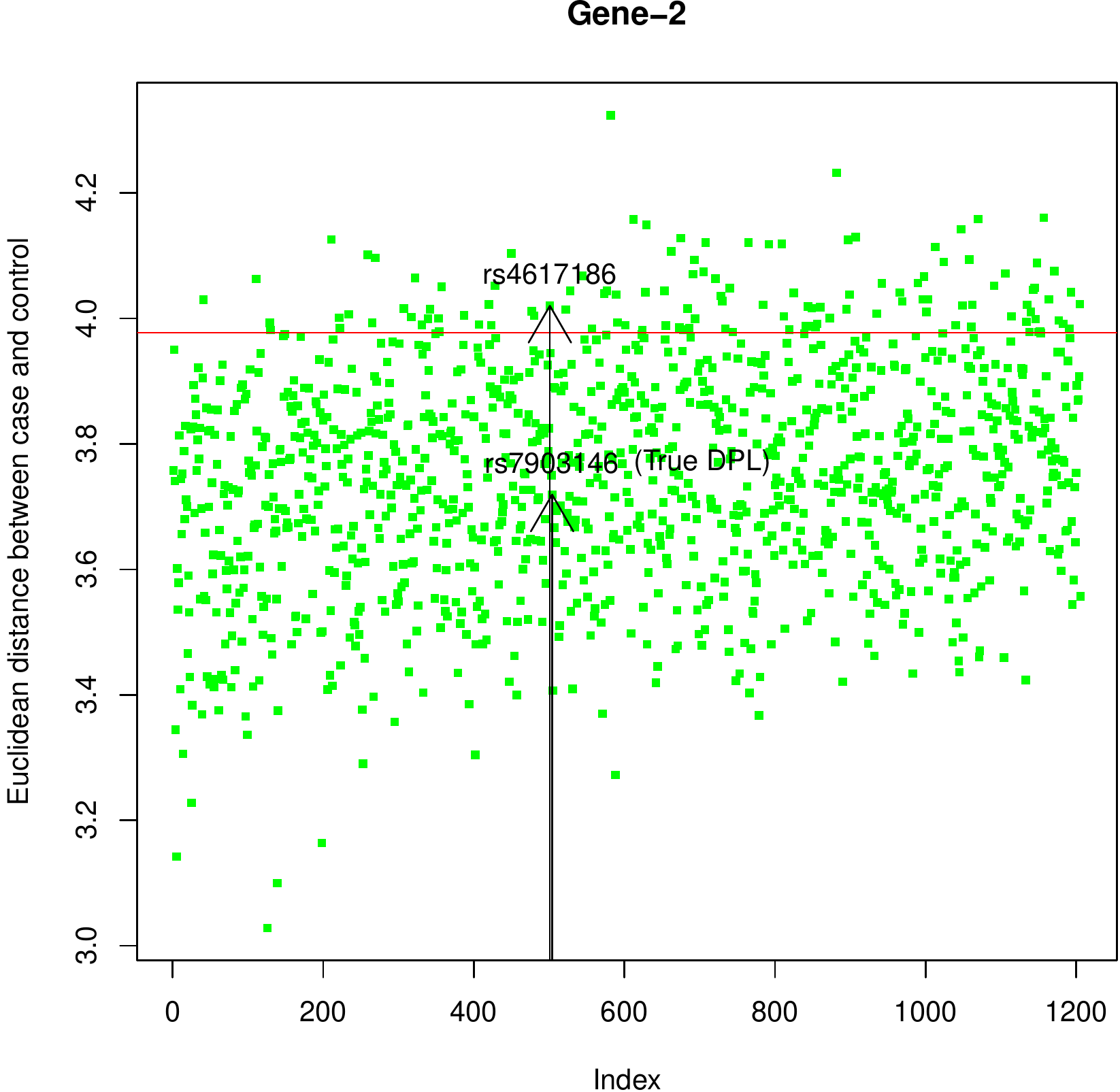}}
\caption{{\bf Index plots:} Plots of the Euclidean distances 
$\left\{d^r_j\left(\bp^r_{i_0jk=0},\bp^r_{i_1jk=1}\right);~r=1,\ldots,L_j\right\}$
against the indices of the loci, for $j=1$ (panel (a)) and $j=2$ (panel (b)).}
\label{fig:index_plots}
\end{figure}
The red, horizontal lines in the panels of Figure \ref{fig:index_plots} 
represent the cut-off value such that the points above the
horizontal line are those with the highest $10\%$ Euclidean distances. 
The true DPLs and the SNPs which are the nearest neighbors of the true DPLs with Euclidean distances on or above the
red, horizontal line are shown in the figures. It is interesting to note that even though the Euclidean distances
of the true DPLs fall below the red, horizontal line (due to LD effects), they are quite close to SNPs that 
cross the 10\% horizontal line. Thus, examination of the close neighbors of the SNPs whose Euclidean distances
are high, would reveal the actual DPLs.

\subsection{{\bf Second simulation study: no genetic or environmental effect}}
\label{subsec:second_simulation_study}

Here we use the same case-control genotype data set as used by BB in their second simulation study
where genetic effects are absent, 
consisting of 49 cases and 51 controls and 5 sub-populations with the mixing proportions
$(0.1, 0.4, 0.2, 0.15, 0.15)$. We use the same environmental data set generated in our first simulation study described in
Section \ref{subsec:first_simulation_study}, which is unrelated to this genotype data.

%For specification of the thresholds $\varepsilon$'s, we employ the same method proposed in Section 
%\ref{subsubsec:threshold}. %but here we also set $\bA_{01}=\bA_{10}=0$.

%Following the method outlined in Section \ref{subsubsec:e_choice}, setting $M$, the maximum number
%of distinct components to be 30, and $\alpha_{ijk}=1.5$ following \ctn{Bhattacharya16}, we obtain
%$\varepsilon_{d^*}=0.633$, $\varepsilon_{\hat d_1}=0.6$, $\varepsilon_{\hat d_2}=0.6$,
%$\varepsilon_{d^*_E}=19.365$, $\varepsilon_{d^*_{E,1}}=14.859$, $\varepsilon_{d^*_{E,2}}=19.310$,
%$\varepsilon_{110}=0.087$, $\varepsilon_{120}=0.364$, $\varepsilon_{111}=0.294$, $\varepsilon_{121}=0.169$,
%$\varepsilon_{\phi}=0.704$.

Here we obtain, from $50,000$ MCMC samples,
$P\left(d^*<\varepsilon_{d^*}|\mbox{Data}\right)\approx 0.359$, 
$P\left(\hat d_1<\varepsilon_{\hat d_1}|\mbox{Data}\right)\approx 0.337$
and $P\left(\hat d_2<\varepsilon_{\hat d_2}|\mbox{Data}\right)\approx 0.334$.
Thus, even though neither genes nor environment are responsible for the case-control status under the true,
data-generating model of GENS2, 
still $H_{0,d^*}$, $H_{0,\hat d_1}$ and $H_{0,\hat d_2}$ are rejected. 
%suggesting
%the influence of genes in the case-control study. 

However, $P\left(d^*_E<\varepsilon_{d^*_E}|\mbox{Data}\right)\approx 0.761$, so that
$H_{0,d^*_E}$ is to be accepted. This also implies that there is no significant difference
between the mixture models $h_{0j}$ and $h_{1j}$ %(given by (\ref{eq:h_0}) and (\ref{eq:h_1})), 
for $j=1,2$. The apparent conflict between acceptance of  $H_{0,d^*_E}$ and rejection of $H_{0,d^*}$
can be clarified as follows.
%is due to issue that even if two distinct values of the parameter are close enough to have insignificant
%difference, they will be considered as representing two distinct clusters, and increase in the number
%of clusters increases the clustering distance. 
Since we are considering the distance between  
two central clusterings, and there is a non-negligible amount of uncertainty associated with the
central clustering because of relatively small sizes of case and control groups in these simulation
studies, the distance between the central clusterings turn out to be larger compared to
the gene-gene interaction studies carried out by BB, which did not involve
distances between central clusterings. In our situation, the number of clusters remained
around 5 as in the previous simulation study, but the clusters of two central clusterings 
associated with cases and controls turned out to have only a few common elements, thus contributing
towards relatively large distance. Also, the data sets generated by GENS2 provide somewhat lesser
information when fitted to our complex Bayesian nonparametric model, as compared to the data generated from
our null Bayesian nonparametric model itself. This problem is further aggravated since the central clusterings
themselves are subject to a (relatively large) degree of approximation, as discussed above. 
Consequently, the distance between central clusterings
associated with the null model and the null data is somewhat lesser than that associated with the
data simulated from GENS2. 

Hence, here one needs to exercise caution to reach
the right conclusion. Indeed, $P\left(d_{E,1}<\varepsilon_{d_{E,1}}|\mbox{Data}\right)\approx 0.713$
and $P\left(d_{E,2}<\varepsilon_{d_{E,2}}|\mbox{Data}\right)\approx 0.946$, also suggesting
acceptance of $H_{0,d_{E,1}}$ and $H_{0,d_{E,2}}$. Also recall that BB obtained,
for the same genotype data, the clear conclusion of acceptance of all three hypotheses
$H_{0,d^*}$, $H_{0,\hat d_1}$ and $H_{0,\hat d_2}$, with respect to both clustering and Euclidean distances
associated with their gene-gene interaction model.
Thus our results with respect to the Euclidean distance is consistent with the results obtained by BB. 
We conclude that genes are not responsible for the case-control outcome
in this study. Hence, the environmental variable has no negative influence on the genes in triggering
the disease.
Note that given the above conclusion, the tests involving $\beta_{\ell jk}$ and $\phi$ are rendered unimportant.
As before, our model has successfully captured 5 sub-populations.

Since we model the genotype data conditionally on the case-control status, rather than modelling
the case-control status directly as binary outcomes, it is not possible to infer from the above conclusion
that the environmental variable is irrelevant for the case-control outcome in this study. 
To test whether or not the environmental
variable is marginally influential, one may consider direct modelling of the case-control binary data
using, say, the logistic regression on the environment, and then test significance of the environmental variable,
independently of our Bayesian nonparametric model. 
Considering such a test, we obtain clear insignificance of the environmental variable.

\subsection{{\bf Third simulation study: absence of genetic and gene-gene interaction effects
but presence of environmental effect}}
\label{subsec:third_simulation_study}

In this study we consider a case-control genotype data set simulated from GENS2 where case-control status depends
only upon the environmental data. The number of cases turned out to be 47 among a total of 100 individuals.

We obtain $P\left(|\beta_{1jk}|<\varepsilon_{\beta_{110}}|\mbox{Data}\right)\approx 0.998$,
$P\left(|\beta_{1jk}|<\varepsilon_{\beta_{120}}|\mbox{Data}\right)\approx 1.000$,
$P\left(|\beta_{1jk}|<\varepsilon_{\beta_{111}}|\mbox{Data}\right)\approx 1.000$, and
$P\left(|\beta_{1jk}|<\varepsilon_{\beta_{121}}|\mbox{Data}\right)\approx 1.000$,
suggesting that all $\beta_{1jk}$ are insignificant. This indicates that the environmental variable does not
cause mutation of the genes.
But even though genes are not responsible for the case-control outcome in this study, rather counter-intuitively
we find that 
$P\left(d^*<\varepsilon_{d^*}|\mbox{Data}\right)\approx 0.359$, 
$P\left(\hat d_1<\varepsilon_{\hat d_1}|\mbox{Data}\right)\approx 0.336$
$P\left(\hat d_2<\varepsilon_{\hat d_2}|\mbox{Data}\right)\approx 0.332$,
$P\left(d^*_E<\varepsilon_{d^*_E}|\mbox{Data}\right)\approx 0.236$, 
$P\left(\hat d_{E,1}<\varepsilon_{\hat d_{E,1}}|\mbox{Data}\right)\approx 0.548$
and $P\left(\hat d_{E,2}<\varepsilon_{\hat d_{E,2}}|\mbox{Data}\right)\approx 0.298$,
all suggesting the relevance of genes in this experiment. Significant gene-gene interaction
is also indicated by
$P\left(\left|\bA_{12}\right|<\varepsilon_{A_{12}}|\mbox{Data}\right)\approx 0.450$.
%These led us to suspect that since the environmental variable plays the only role in this case-control study,
%it must have some impact on our model to which the data is fed. Since the $\beta_{\ell jk}$ turned
%out to be insignificant, so that the environmental variable does not force the genes to mutate,
%it must affect $\phi$ and make it significant to reflect its role in this study.
%As such, it turned out that
It also turned out, counter-intuitively, that
$P\left(\phi<\varepsilon_{\phi}|\mbox{Data}\right)\approx 0.351$, suggesting significant impact
of the environmental variable on gene-gene interaction. 
%This spurious gene-gene interaction 
%is responsible for making the genes seem significant. 

In an attempt to resolve this dilemma we again considered a logistic linear regression of the case-control status
on the environmental variable and the (summaries of the) genes, and obtained, using the Akaike Information Criterion (AIC), 
the model consisting of the
marginal effects of the environment and the second gene, as the best model. 
Thus, relevance of at least the second gene is also revealed by this simple logistic linear model.
%In fact, it turned out, rather counter-intuitively, that the second gene is significant while 
%the environmental variable is insignificant at 5\% level.  

Since gene-environment interaction is ruled out by the best logistic linear model, 
we re-implemented our model by setting $\phi=0$, so that the environmental
variable can not have any effect on gene-gene interaction. 
This can be interpreted as (data based) prior information obtained from the best logistic linear model.
With this prior information it then turned out that
$P\left(d^*<\varepsilon_{d^*}|\mbox{Data}\right)\approx 0.358$, 
$P\left(\hat d_1<\varepsilon_{\hat d_1}|\mbox{Data}\right)\approx 0.340$
$P\left(\hat d_2<\varepsilon_{\hat d_2}|\mbox{Data}\right)\approx 0.317$,
$P\left(d^*_E<\varepsilon_{d^*_E}|\mbox{Data}\right)\approx 0.658$, 
$P\left(\hat d_{E,1}<\varepsilon_{\hat d_{E,1}}|\mbox{Data}\right)\approx 0.653$
and $P\left(\hat d_{E,2}<\varepsilon_{\hat d_{E,2}}|\mbox{Data}\right)\approx 0.804$,
strongly suggesting that genes are not responsible for case-control status. And, as before,
all the $\beta_{\ell jk}$ turned out to be insignificant, demonstrating that the environmental variable
does not have any effect on the genes. 
Since the best logistic linear model includes the environmental variable one may conclude on this basis 
%If genes and the environmental variable are the only possible factors
%associated with the case-control outcome, then it is clear from our analyses 
that the environmental effect is the only factor responsible in this case-control experiment. Thus, our inference
is consistent with the true data-generating mechanism.
%
%Coefficients:
%             Estimate Std. Error z value Pr(>|z|)  
%	     (Intercept)  24.06452   19.85076   1.212   0.2254  
%	     x             0.01371    0.25153   0.055   0.9565  
%	     gene1         5.12431   18.24752   0.281   0.7788  
%	     gene2       -31.66786   14.81964  -2.137   0.0326 *
%	     ---
%	     Signif. codes:  0 ‘***’ 0.001 ‘**’ 0.01 ‘*’ 0.05 ‘.’ 0.1 ‘ ’ 1
%
%Coefficients:
%             Estimate Std. Error z value Pr(>|z|)  
%	     (Intercept)  28.01225   14.05171   1.994   0.0462 *
%	     x             0.03232    0.24261   0.133   0.8940  
%	     gene2       -31.44433   14.79148  -2.126   0.0335 *
%
%
%
%This suggests that marginally
%the environmental variable has no effect on case-control status. The only way the environmental variable
%can turn out to be significant is when logistic regression is applied on both genes and the environmental
%variable, so that dependencies between the regressor variables can make the environmental variable significant. 
%
%Hence, interestingly, the roles of genes and their interactions, along with their
%interactions with the environmental variable can not be ruled out in this experiment, even though
%the latter may be more responsible than the genes in causing the case-control outcomes. 

\subsection{{\bf Fourth simulation study: presence of genetic and gene-gene interaction effects
but absence of environmental effect}}
\label{subsec:fourth_simulation_study}

Here we use the same genotype data set as used by BB in their first simulation study
associated with genetic and gene-gene interaction effects, 
consisting of 41 cases and 59 controls and 5 sub-populations with the mixing proportions
$(0.1, 0.4, 0.2, 0.15, 0.15)$. We use the same environmental data set generated in our first simulation study described in
Section \ref{subsec:first_simulation_study}, which is unrelated to this case-control genotype data. 

Here we obtain
$P\left(d^*<\varepsilon_{d^*}|\mbox{Data}\right)\approx 0.362$, 
$P\left(\hat d_1<\varepsilon_{\hat d_1}|\mbox{Data}\right)\approx 0.336$
$P\left(\hat d_2<\varepsilon_{\hat d_2}|\mbox{Data}\right)\approx 0.337$,
$P\left(d^*_E<\varepsilon_{d^*_E}|\mbox{Data}\right)\approx 0.345$, 
$P\left(\hat d_{E,1}<\varepsilon_{\hat d_{E,1}}|\mbox{Data}\right)\approx 0.764$
and $P\left(\hat d_{E,2}<\varepsilon_{\hat d_{E,2}}|\mbox{Data}\right)\approx 0.317$,
so that importance of genes is correctly indicated by our tests.

As for the tests related to the environmental variable, we find
$P\left(|\beta_{1jk}|<\varepsilon_{\beta_{110}}|\mbox{Data}\right)\approx 0.633$,
$P\left(|\beta_{1jk}|<\varepsilon_{\beta_{120}}|\mbox{Data}\right)\approx 0.647$,
$P\left(|\beta_{1jk}|<\varepsilon_{\beta_{111}}|\mbox{Data}\right)\approx 1.000$, and
$P\left(|\beta_{1jk}|<\varepsilon_{\beta_{121}}|\mbox{Data}\right)\approx 1.000$,
meaning that all the $\beta_{\ell jk}$ are insignificant. That is, mutation is to be correctly ruled out.

That the environmental variable has no influence on gene-gene interaction is clear from the result
$P\left(\phi<\varepsilon_{\phi}|\mbox{Data}\right)\approx 0.640$, which correctly suggests
acceptance of the null hypotheses $\phi=0$.
Also, 
$P\left(\left|\bA_{12}\right|<\varepsilon_{A_{12}}|\mbox{Data}\right)\approx 0.423$, correctly suggesting
the presence of gene-gene interaction. 

To check if the environmental variable has no role to play in the case-control outcome of this study we again perform analyses
based on logistic regression and obtain insignificance of the environmental variable. 
%Here, based on AIC, we obtain the logistic linear model consisting of
%the second gene and the intercept as the best model. Hence, from the perspective of the best logistic linear model 
%one may rule out the environmental variable.
%Coefficients:
%            Estimate Std. Error z value Pr(>|z|)
%	    (Intercept)    9.899     12.354   0.801    0.423
%	    gene2        -11.331     13.641  -0.831    0.406
%
%	    (Dispersion parameter for binomial family taken to be 1)
%
%	        Null deviance: 135.37  on 99  degrees of freedom
%		Residual deviance: 134.67  on 98  degrees of freedom
%		AIC: 138.67

\subsection{{\bf Fifth simulation study: independent and additive genetic and environmental effects}}
\label{subsec:fifth_simulation_study}

Now we simulate a case-control genotype dataset from GENS2 where the genetic and environmental effects
are independent of each other and additive. Among 100 individuals obtained, there are 57 cases.

Note that in our Bayesian model there is no provision for additivity of genetic and environmental effects.
Hence this dataset is not expected to provide enough information to our Bayesian model to enable it capture
the true data-generating relationships between the genes and the environmental variable.
Here we obtain
$P\left(d^*_E<\varepsilon_{d^*_E}|\mbox{Data}\right)\approx 0.711$, 
$P\left(\hat d_{E,1}<\varepsilon_{\hat d_{E,1}}|\mbox{Data}\right)\approx 0.740$
and $P\left(\hat d_{E,2}<\varepsilon_{\hat d_{E,2}}|\mbox{Data}\right)\approx 0.816$,
indicating that the genes are unimportant in this study. 
All $\beta_{\ell jk}$ also turned out to be insignificant.
%but 
%$P\left(\left|\bA_{12}\right|<\varepsilon_{A_{12}}|\mbox{Data}\right)\approx 0.301$, and
However, $P\left(\phi<\varepsilon_{\phi}|\mbox{Data}\right)\approx 0.054$, suggesting that 
gene-gene interaction is influenced by the environmental variable.
Since genetic effect turned out to be insignificant, it is clear that gene-gene interaction
did not have substantial effect on the case-control data.  
%Re-implementing our model after setting $\phi=0$ again indicated that the genes are insignificant.

%Coefficients:
%            Estimate Std. Error z value Pr(>|z|)
%	    (Intercept)  24.0920    15.1302   1.592    0.111
%	    x            -0.2465     0.2162  -1.140    0.254
%	    gene1       -25.2022    17.2758  -1.459    0.145
%
%	    (Dispersion parameter for binomial family taken to be 1)
%
%	        Null deviance: 136.66  on 99  degrees of freedom
%		Residual deviance: 133.57  on 97  degrees of freedom
%		AIC: 139.57

On conducting independent logistic regression experiments as before we find that the 
best AIC-based model consists of the marginal effects of the environmental variable
and the first gene, along with an intercept, which is somewhat consistent with the actual
data-generating model.

%coefficient associated 
%with the environmental variable
%is approximately $-0.200$, and the associated $P$-value is $0.345$. Thus, once again, the marginal
%effect of the environmental variable seems to be insignificant; only considering the genes and the environmental
%variable simultaneously as regressors can render the latter significant. In such a case we anticipate
%that the environmental variable will have much stronger effect than the genes, as is reflected
%in the results obtained by our Bayesian model.

In summary, it seems that with respect to our Bayesian model, the additive effect has been almost wholly transformed
into the environmental effect, given that the provoked gene-gene interaction did not not affect the case-control data. 
From the practical perspective, it seems that the environmental
variable exerts much stronger influence in this case-control study compared to the genes and gene-gene interaction. 

\newpage

\begin{figure}%[htp]
\centering
\subfigure[Posterior probability of no genetic effect with respect to clustering metric.]{ \label{fig:clustering_hypotheses}
\includegraphics[width=15cm,height=6cm]{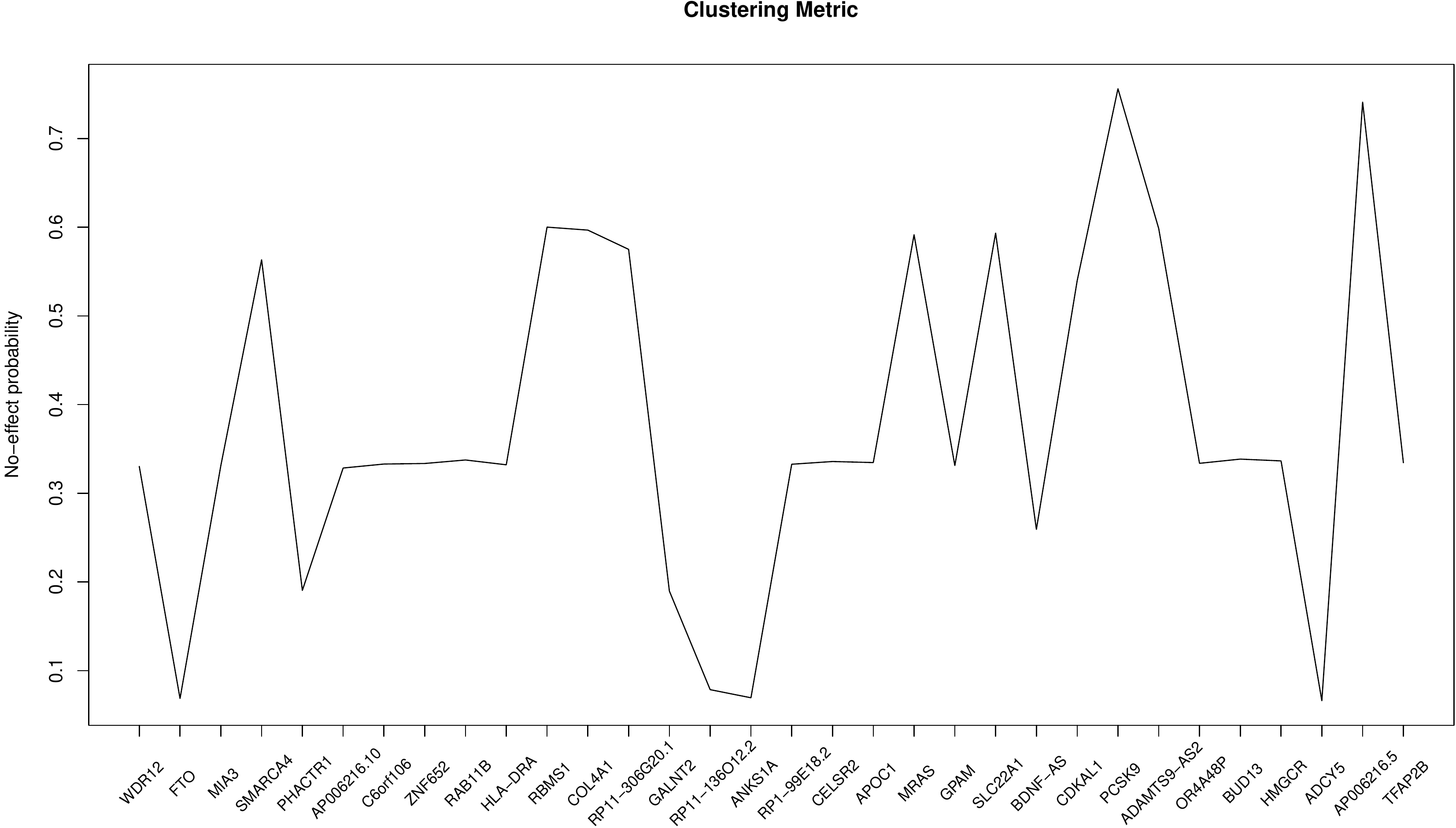}}\\
\vspace{4mm}
\subfigure[Posterior probability of no genetic effect with respect to Euclidean metric.]{ \label{fig:euclidean_hypotheses} 
\includegraphics[width=15cm,height=6cm]{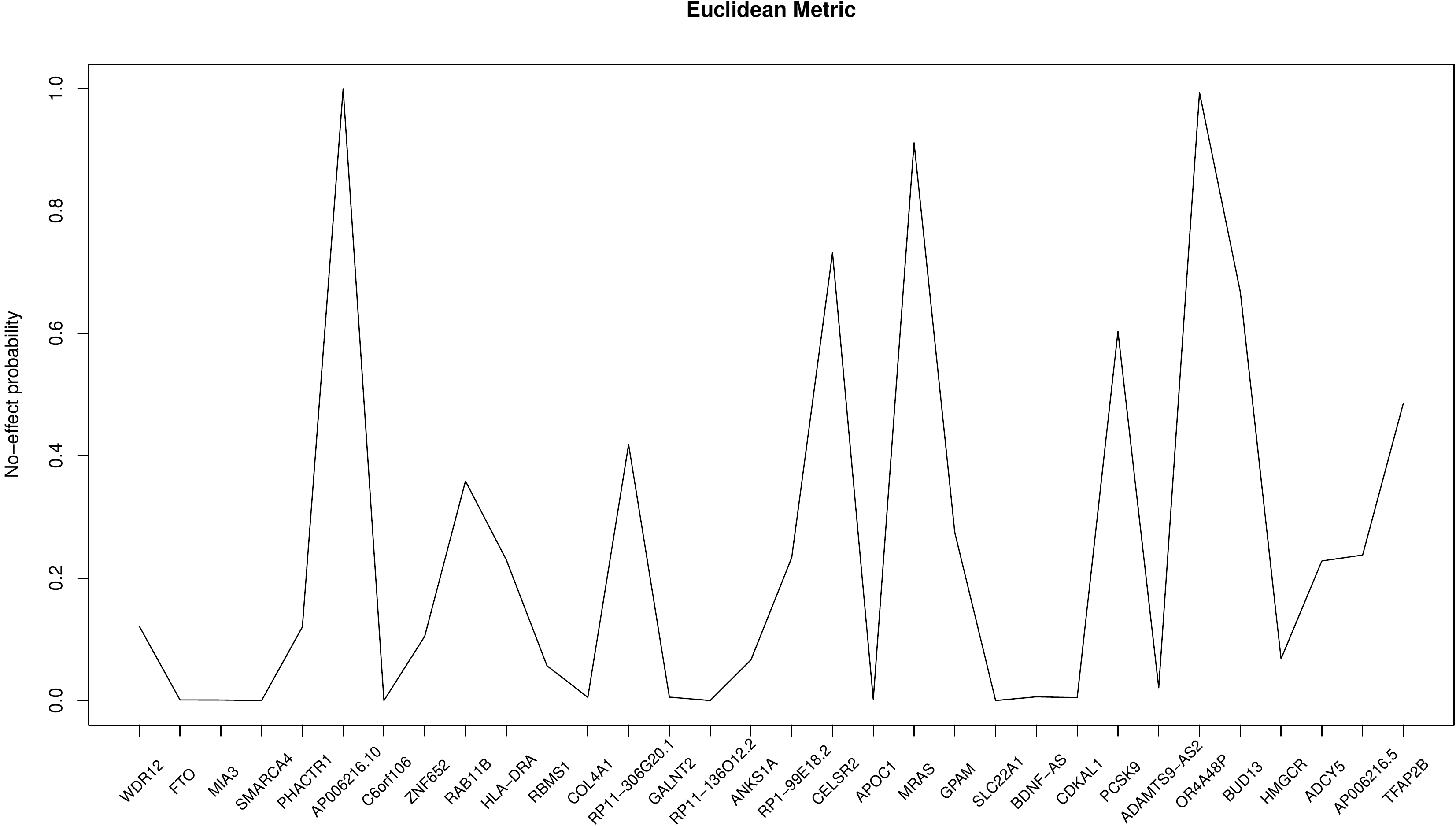}}
\caption{{\bf Posterior probabilities of no individual genetic influence in MI study:} 
Index plots of the posterior probabilities of the null hypotheses for (a) clustering metric
and (b) Euclidean metric, for the $32$ genes.}
\label{fig:null_hypotheses}
\end{figure}

\newpage

\begin{figure}%[htp]
\centering
\subfigure[Colorplot of actual posterior gene-gene interaction.]{ \label{fig:ggi_plot} 
\includegraphics[width=16cm,height=16cm]{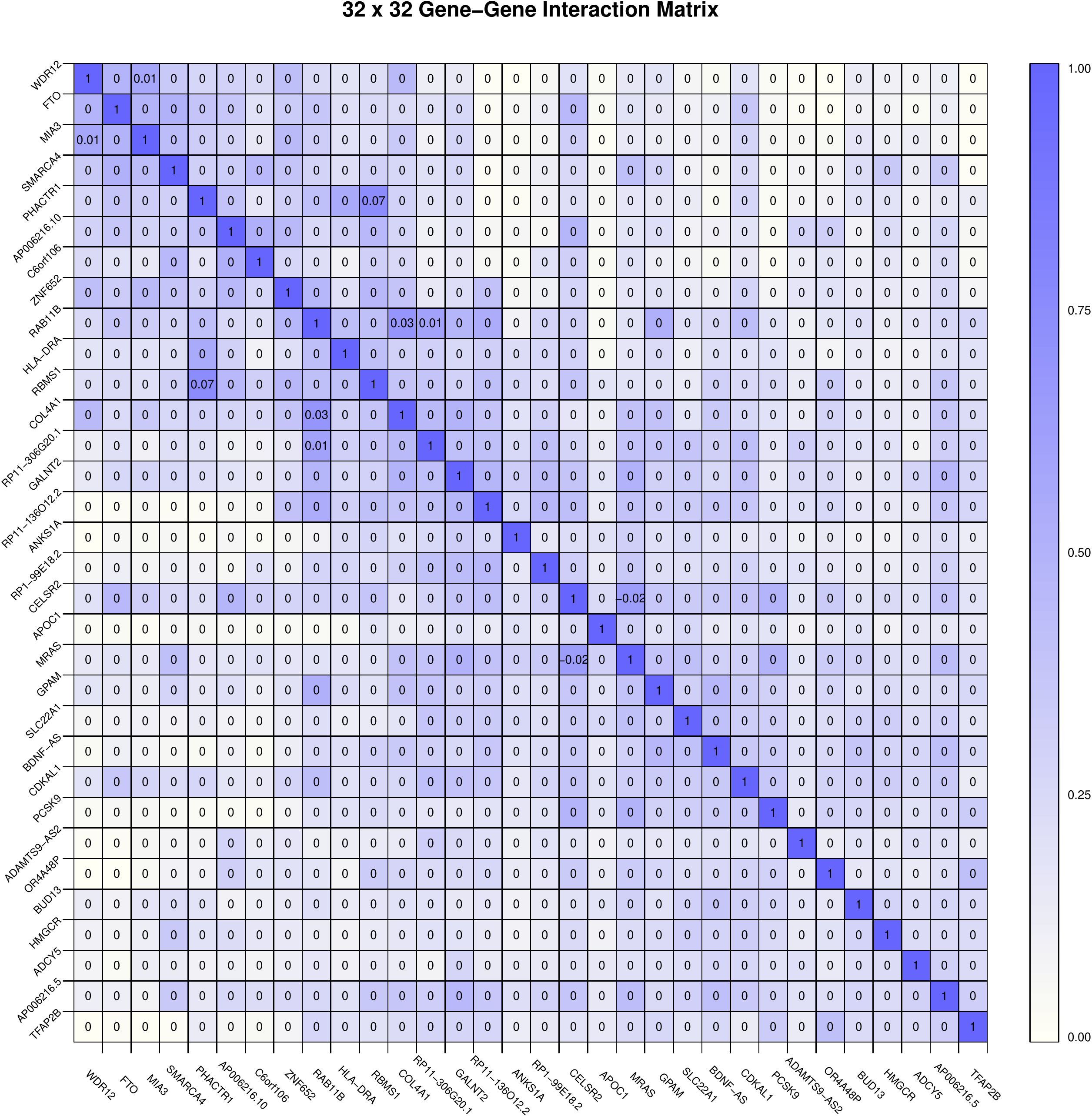}}
\caption{{\bf Gene-gene interaction plot in MI study:} Actual gene-gene interactions
based on medians of the absolute values of the posterior covariances.}
\label{fig:ggi_plots2}
\end{figure}

\newpage

\section{{\bf Disease predisposing loci detected by our Bayesian analysis}}
\label{subsec:our_DPL}
%\ctn{Erdmann10} enlists $11$ SNPs which have been flagged by GWAS investigations as having significant effects
%on MI. Such reportedly important markers, which are available in our dataset, 
%are $rs599839$, $rs6725887$, $rs9818870$, $rs12526453$, $rs2048327$, $rs3127599$, $rs7767084$, 
%$rs10755578$, $rs1333049$ and $rs2259816$. In our notation, these
%correspond to Gene-1 to Gene-9, with number of loci 
%$6$, $5$, $17$, $177$, $39$, $15$, $19$, $21$ and $8$, respectively. Note that $rs7767084$ and 
%$rs10755578$ represent the same gene. 
%We now report the DPLs that we obtain by our analysis
%and show that % the existing ones %For brevity of exposition, we consider 15 genes containing 
%important SNPs, 
%We now show that the most influential SNPs corresponding to the maximum Euclidean distance in each of the significant
%genes in our study, which we continue to refer to as the DPLs, are usually close to,
%and sometimes exactly the same as the SNPs, already flagged by the earlier studies as influential.

Figure \ref{fig:metric_medians} shows the index plots of the posterior medians of the 
clustering and Euclidean distances between case and control, with respect to the corresponding genes.
\addtocounter{figure}{4}
\begin{figure}%[htp]
\centering
\subfigure[Clustering metric medians.]{ \label{fig:clustering_medians}
\includegraphics[width=15cm,height=6cm]{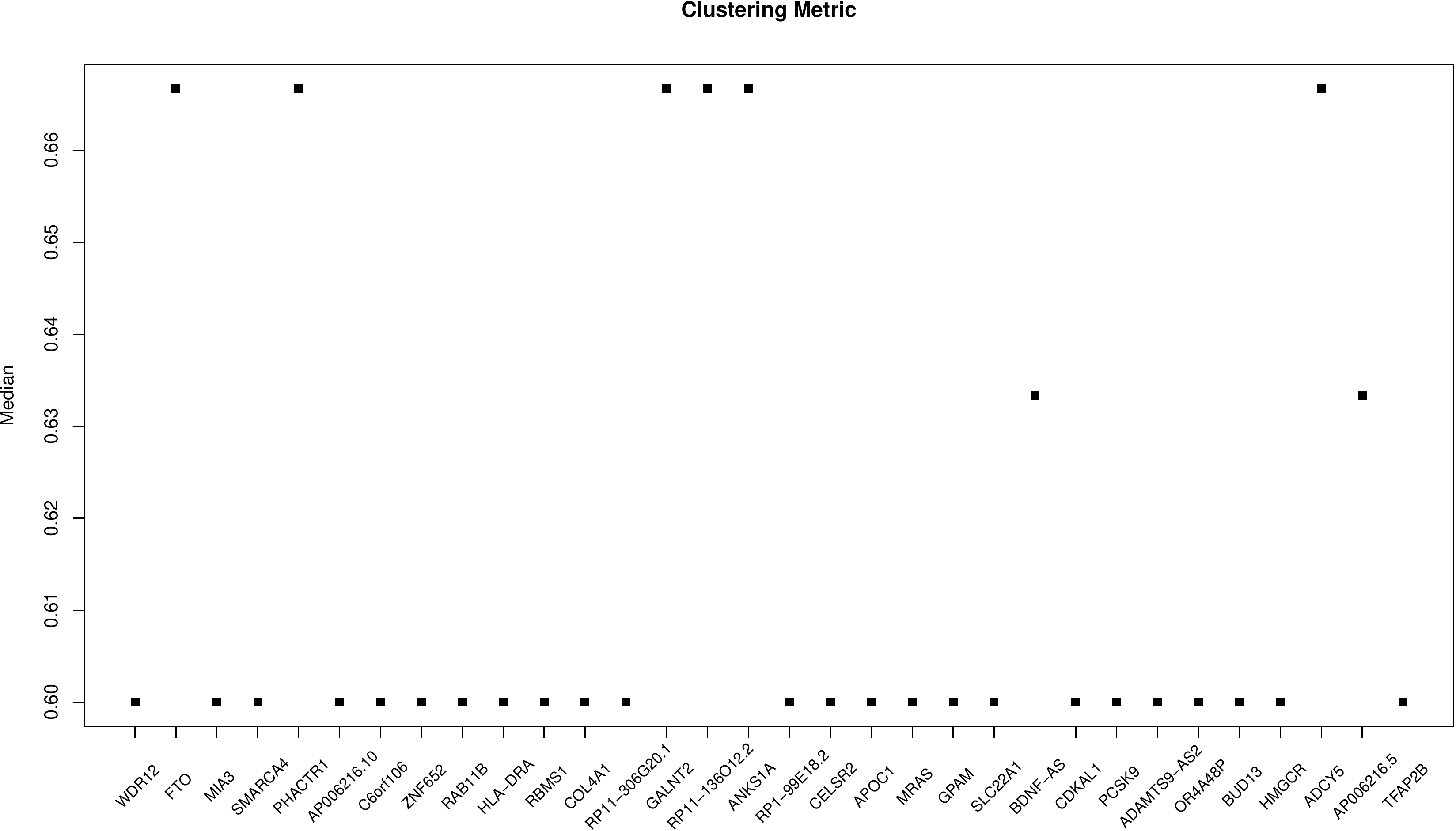}}\\
\vspace{4mm}
\subfigure[Euclidean metric medians.]{ \label{fig:euclidean_medians} 
\includegraphics[width=15cm,height=6cm]{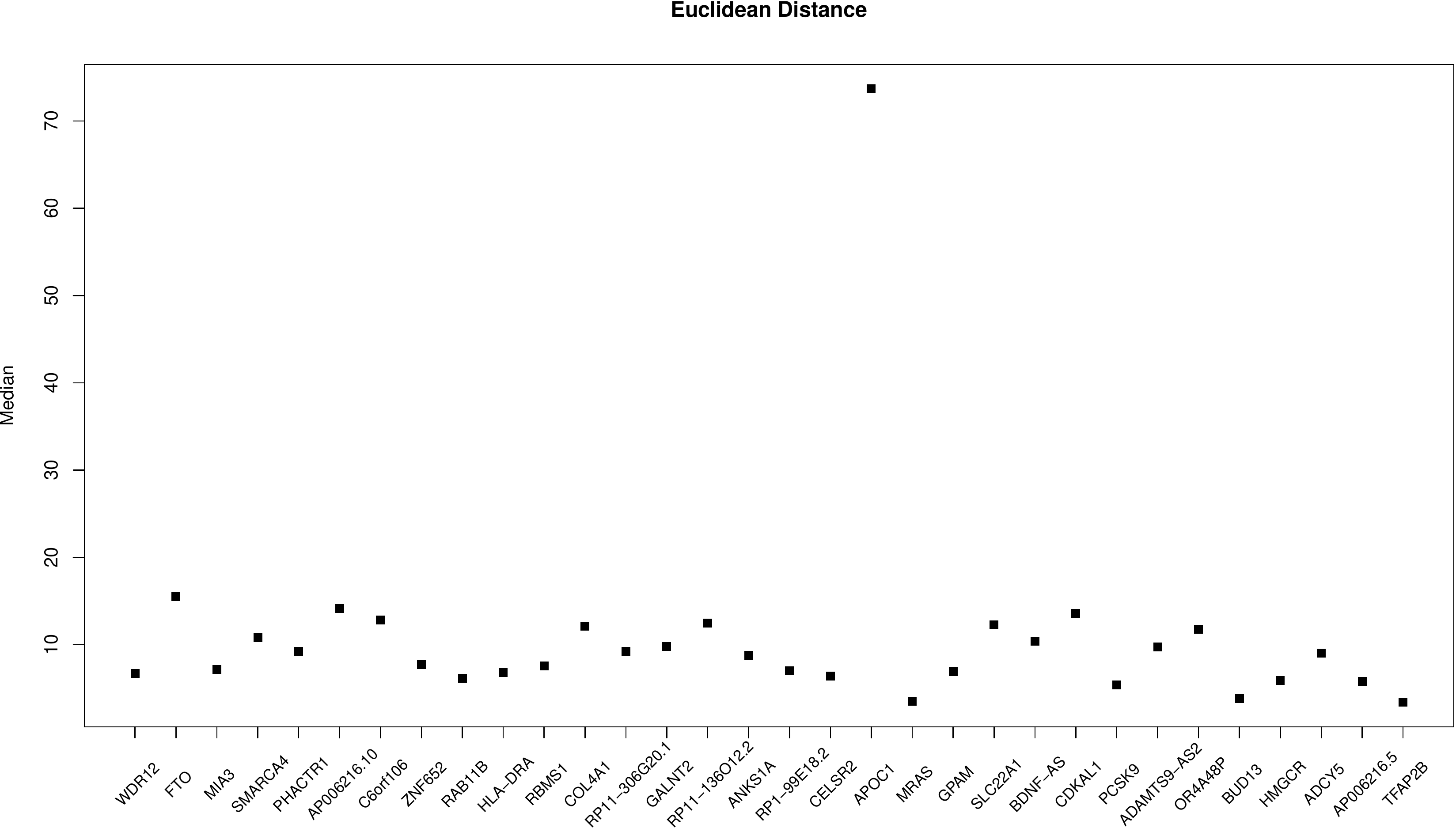}}
\caption{{\bf Posterior medians of the Euclidean distances:} 
Index plots of the posterior medians of the Euclidean distances
with respect to the $32$ genes.}
\label{fig:metric_medians}
\end{figure}
It is clear from the figure that in terms of the clustering metric, 
genes $FTO$, $PHACTR1$, $GALNT2$, $RP11-136O12.2$, $ANKS1A$ and $ADCY5$ 
exceed $0.66$, while $APOC1$ exceeds $70$ in terms of the Euclidean distance. The number of loci of these $7$ genes
are $137$, $177$, $89$, $45$, $54$, $95$ and $1$, respectively. 

\begin{figure}%[htp]
\centering
\subfigure[DPL of $FTO$.]{ \label{fig:DPL_gene_2}
\includegraphics[width=6cm,height=6cm]{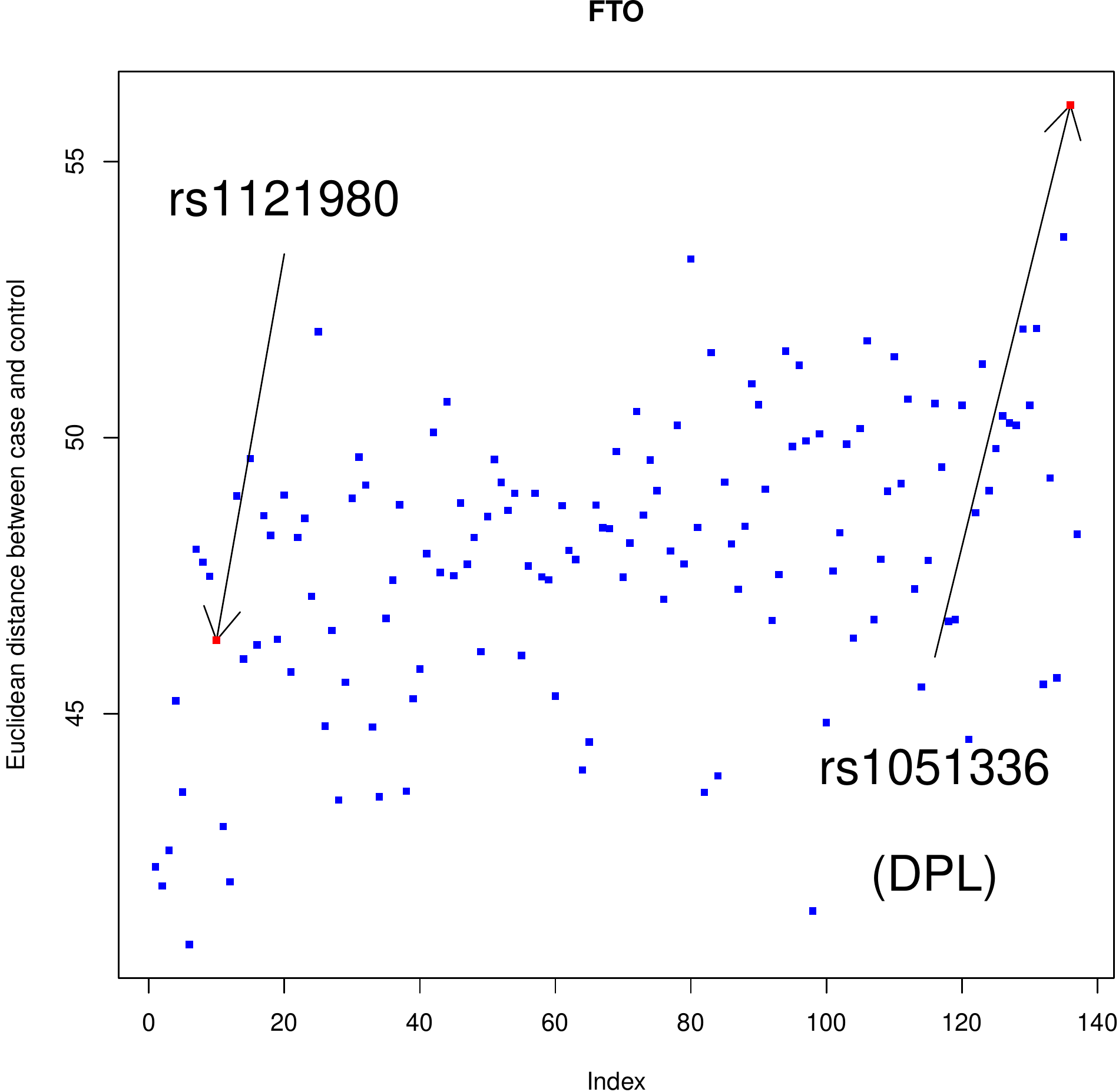}}
\hspace{2mm}
\subfigure[DPL of $PHACTR1$.]{ \label{fig:DPL_gene_5}
\includegraphics[width=6cm,height=6cm]{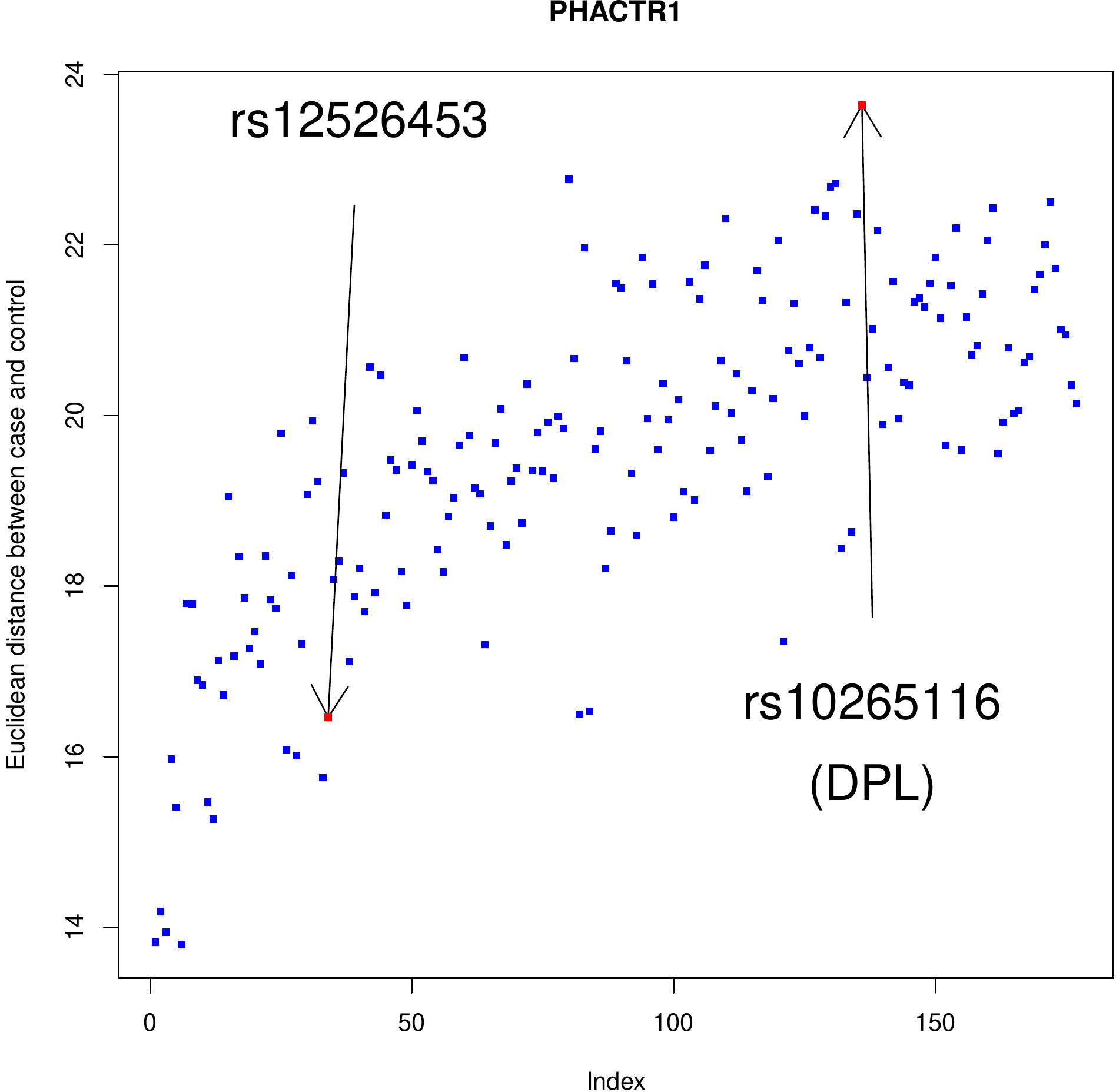}}\\
\vspace{2mm}
\subfigure[DPL of $GALNT2$.]{ \label{fig:DPL_gene_14}
\includegraphics[width=6cm,height=6cm]{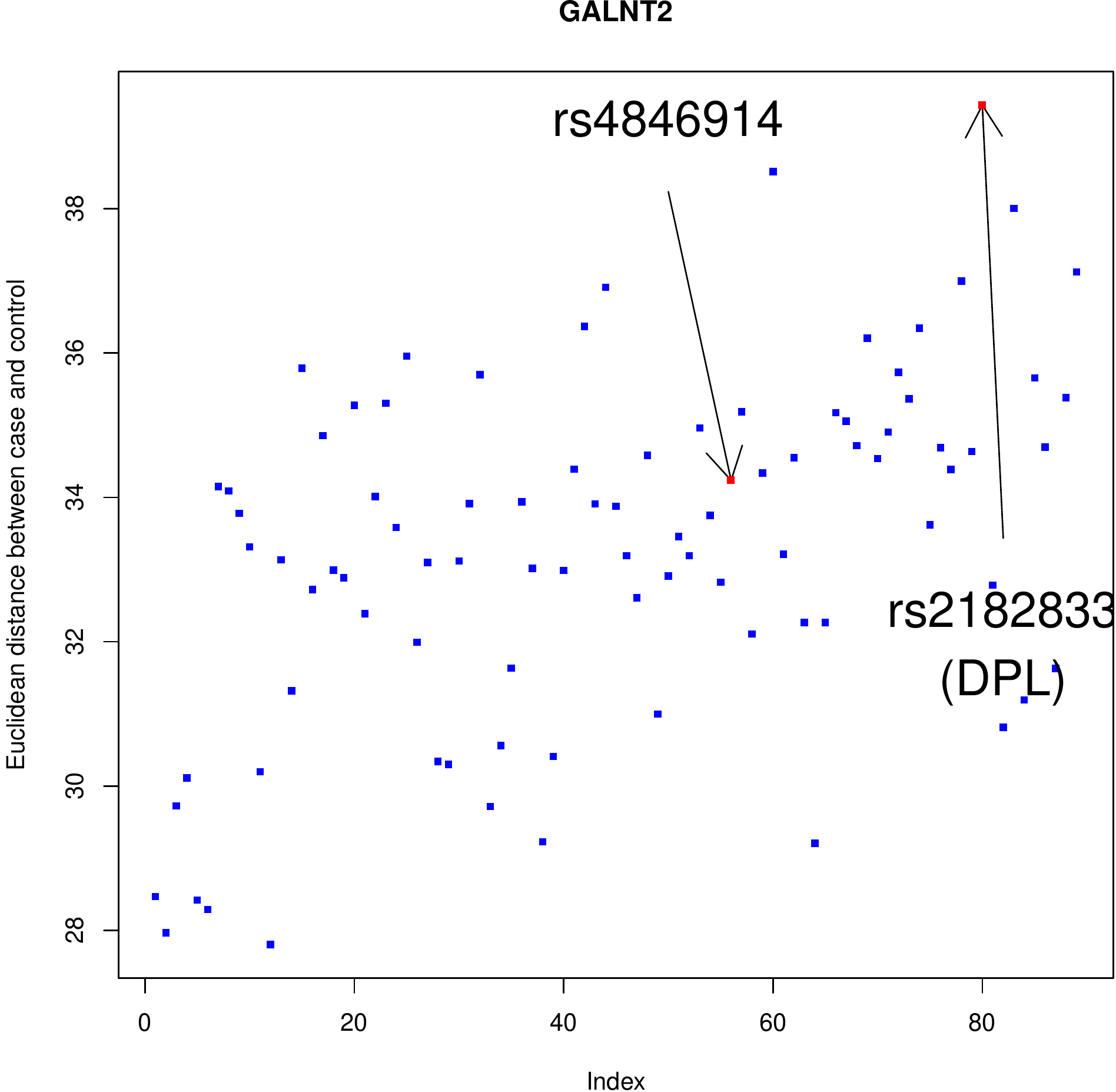}}
\hspace{2mm}
\subfigure[DPL of $RP11-136O12.2$.]{ \label{fig:DPL_gene_15}
\includegraphics[width=6cm,height=6cm]{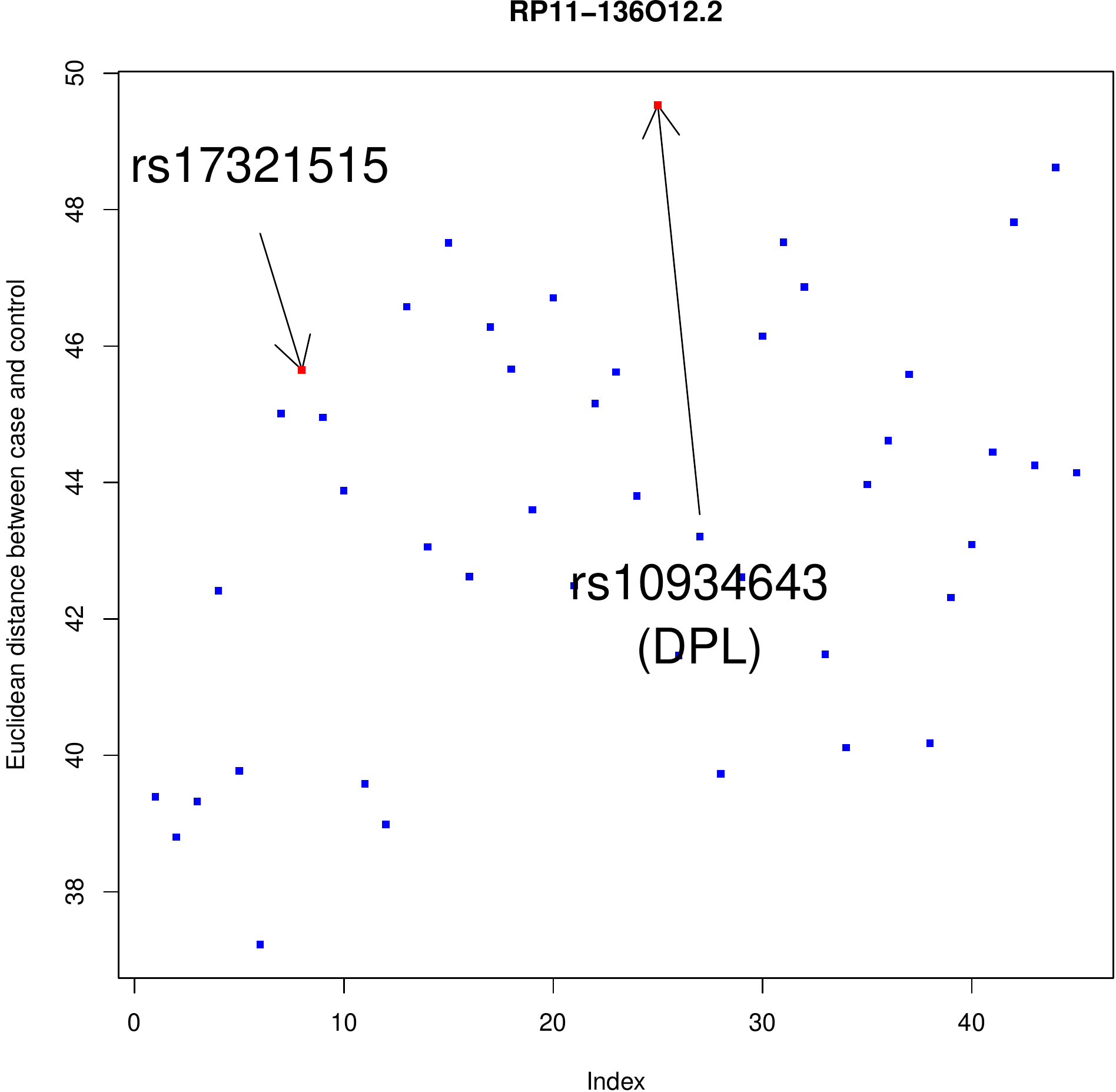}}
\caption{{\bf Disease predisposing loci of the genes influential with respect to the clustering metric:} 
Plots of the Euclidean distances 
$\left\{d^r_j\left(\mbox{logit}\bp^r_{jk=0},\mbox{logit}\bp^r_{jk=1}\right);~r=1,\ldots,L_j\right\}$
against the indices of the loci.} 
\label{fig:DPL_genes_clustering}
\end{figure}

\begin{figure}%[htp]
\centering
\subfigure[DPL of $ANKS1A$.]{ \label{fig:DPL_gene_16}
\includegraphics[width=6cm,height=6cm]{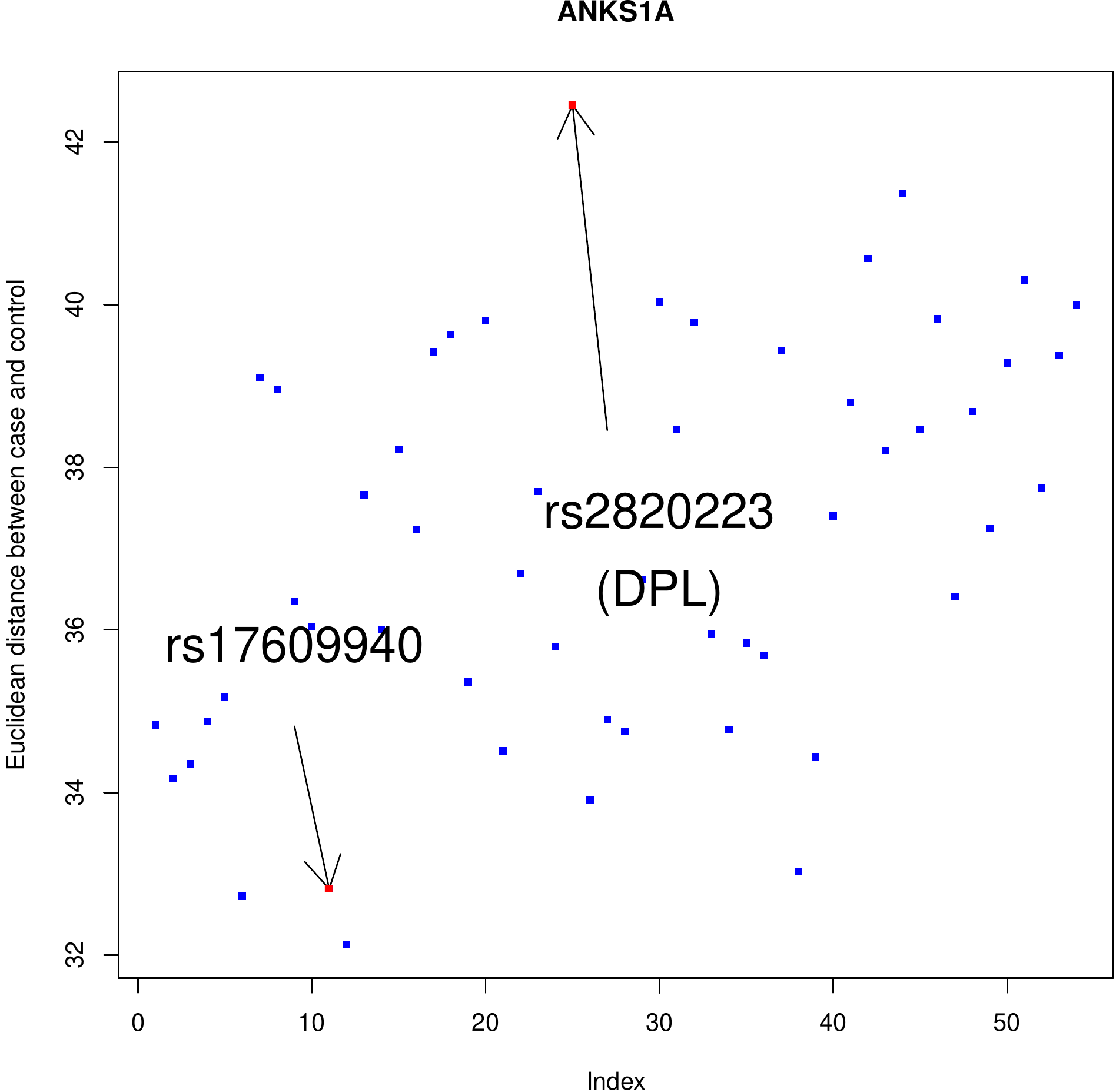}}
\hspace{2mm}
\subfigure[DPL of $ADCY5$.]{ \label{fig:DPL_gene_30}
\includegraphics[width=6cm,height=6cm]{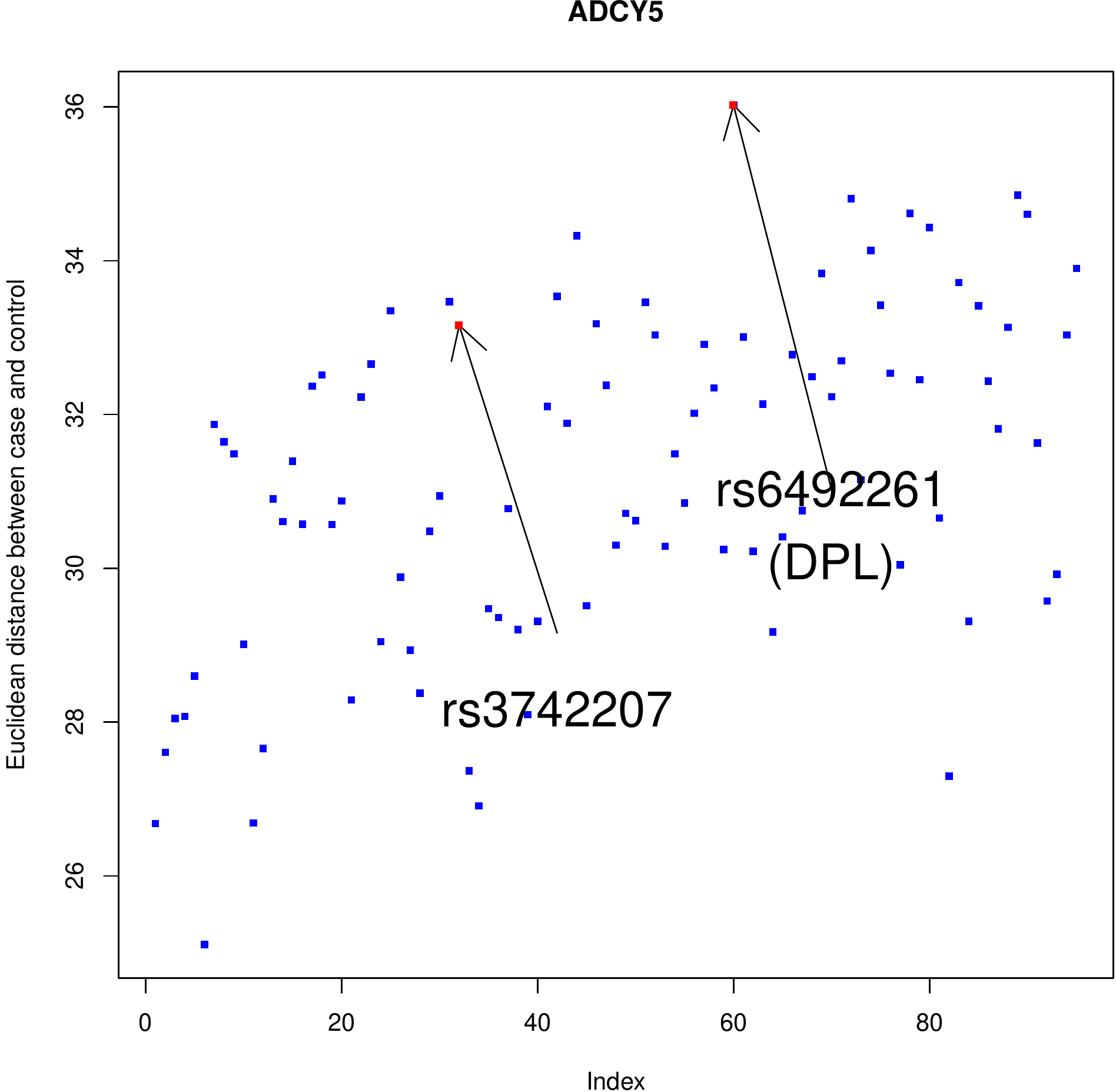}}
%\vspace{2mm}
%\subfigure[DPL of $APOC1$.]{ \label{fig:DPL_gene_19}
%\includegraphics[width=6cm,height=6cm]{plots_realdata/plot_DPL_gene_19-crop.pdf}}
\caption{{\bf Disease predisposing loci of genes influential with respect to the Euclidean metric:} 
Plots of the Euclidean distances 
$\left\{d^r_j\left(\mbox{logit}\bp^r_{jk=0},\mbox{logit}\bp^r_{jk=1}\right);~r=1,\ldots,L_j\right\}$
against the indices of the loci.}
\label{fig:DPL_genes_euclidean}
\end{figure}

After computing the averaged Euclidean distances 
$\left\{d^r_j\left(\mbox{logit}\left(\bp^r_{jk=0}\right),\mbox{logit}\left(\bp^r_{jk=1}\right)\right);
~r=1,\ldots,L_j\right\}$, of the loci in 
each such Gene-$j$, where the averages are taken over the TMCMC samples, we 
single out that SNP with maximum such distance and compare this SNP, which we continue to refer to as DPL, with
that SNP of Gene-$j$ which is reported in the literature as important. Our findings are reported in 
Figures \ref{fig:DPL_genes_clustering} and \ref{fig:DPL_genes_euclidean}. Since $APOC1$ consists of only one 
SNP ($rs4420638$), that SNP is clearly our DPL, and so this case does not present any new insight. As such, 
we do not display the corresponding diagram.

%In the next section we argue that SNP-SNP interactions, not gene-gene interactions, plays a vital role in explaining the discrepancies
%between our findings and the existing results based on previous studies.

\subsection{{\bf Role of SNP-SNP interactions behind our obtained DPLs}}
\label{subsec:ggi_contrast_influential}

%The results on interaction testing are depicted in Figure \ref{fig:ggi_plots1}. 

Figures \ref{fig:DPL_genes_clustering} and \ref{fig:DPL_genes_euclidean} show that most of the literature based
SNPs have turned out to be less influential in terms of case-control Euclidean distance. 
BB showed that with respect to their Bayesian model it is possible to explain agreements and disagreements 
between the literature based SNPs and the important SNPs obtained from their model in terms of
gene-gene and SNP-SNP interactions. In our case, although it turned out that gene-gene interactions are insignificant,
there are still substantial SNP-SNP interactions with respect to case-control Euclidean distances.
Indeed, we illustrate that such SNP-SNP correlations %associated with case-control Euclidean distances 
play important roles in this regard. %although actual gene-gene correlations are insignificant.
Recall that %(see also panel (c) of Figure \ref{fig:DPL_genes_euclidean}) 
$APOC1$ consisting of the singleton locus $rs4420638$, is the most influential with respect
to the Euclidean metric in terms of case-control Euclidean distance.
The correlation between the case-control Euclidean distances associated with the literature-cited locus $rs1121980$ of $FTO$ and
$rs4420638$ of $APOC1$ is $-0.198163$, and this negative correlation with the most influential SNP is responsible
for low influence of $rs1121980$ in comparison with the DPL $rs1051336$, which 
the correlation $-0.1162004$ with $APOC1$.

As regards $PHACTR1$, the correlation between the literature based $rs12526453$ and $rs2820223$ of $ANKS1A$
is $-0.3860316$. Since $rs2820223$ is also the DPL of $ANKS1A$, it is not at all unlikely that $rs12526453$
would turn out to be less significant because of the negative correlation. However, the correlation of
the DPL of $ANKS1A$ with the DPL $rs10265116$ of $PHACTR1$ is $-0.4036174$, which is more negative than
than that with the literature based SNP. To comprehend this counter-intuitive phenomenon, note that
$APOC1$ exerts more positive influence on the DPL (correlation $0.3131022$) than on the literature based locus
(correlation $0.279255$), so that overall the DPL seems to have more influence. 

The same locus $rs2820223$ of $ANKS1A$ also exerts negative influence on $rs4846914$ of $GALNT2$, with correlation
$-0.2414285$, and on $rs17321515$ of $RP11-136O12.2$, with correlation $-0.03827756$, taking away much of the influences
of the aforementioned literature based loci. The correlations of the DPL of $ANKS1A$ with
the DPLs of $GALNT2$ and $RP11-136O12.2$ are $-0.2182731$ and $-0.01800756$, respectively, which are larger
than the correlations with the literature based SNPs. Furthermore, $APOC1$ has correlations $0.2921335$
and $0.100273$ with the DPLs of $GALNT2$ and $RP11-136O12.2$ and correlations $0.2887141$ and $0.07980527$
with the literature based SNPs, which are consistent with the order associated with the correlations with
the DPL of $ANKS1A$. 

On the other hand, the singleton $rs4420638$ of $APOC1$ has correlation $-0.09018503$ with the 
literature based $rs17609940$ of $ANKS1A$ making it somewhat less influential compared to
the DPL $rs2820223$, which has correlation $-0.05074254$ with $APOC1$.

For gene $ADCY5$, the DPL $rs6492261$ and the literature based locus $rs3742207$ are somewhat close
in terms of their case-control Euclidean distances. Indeed, in this case, the DPL of $ANKS1A$
has almost same positive correlations $0.02089441$ and $0.02126861$ with the DPL and the literature
based SNPs of $ADCY5$. Consistent with these observations, it is seen that the correlations
of $APOC1$ with these two SNPs of $ADCY5$ are $0.4463449$ and $0.453281$, respectively.
These seem to provide an explanation for $rs6492261$ and $rs3742207$ to be relatively consistent with each other.

A mathematical explanation of such influences based on the interactions has been provided in BB.
However, as in BB here also it is useful to remark that our above explanations, 
even though focussed on a very small number of genes and SNPs,
may still be inadequate; indeed it is not feasible to explain precisely the complex
influences the SNPs have on one another which might be responsible for the discrepancies between the DPLs that we obtained
and the so-called influential SNPs cited in the literature.

%\begin{figure}%[htp]
%\centering
%\subfigure[Presence/absence of gene-gene interaction.]{ \label{fig:ggi_indicator_plot}
%\includegraphics[width=16cm,height=16cm]{plots_realdata/ggi_indicator_plot-crop.pdf}}
%\caption{{\bf Presence/absence of gene-gene interactions}: Blue denotes presence and white
%represents absence of gene-gene interaction.}
%\label{fig:ggi_plots1}
%\end{figure}

\section{{\bf Posteriors of the number of distinct mixture components}}
\label{sec:no_of_components}
%The discrepancies between the sets of genes %$\{10,13,25,30\}$, $\{6,19,27\}$ 
%$\{HLA-DRA, RP11-306G20.1, PCSK9, ADCY5\}$,
%$\{AP 006216.10, APOC1, OR4A48P\}$ 
%and the remaining genes
%are also reflected in the
%posterior distributions of the number of mixture components associated with them.
%For the remaining genes, here we consider only %$\{1,2,4,8\}$, 
%$\{WDR12, FTO, SMARCA4, ZNF652\}$
%for the sake of brevity.

Unlike BB, under the current study the posteriors of the number of distinct components associated with all the genes turn out to be almost identical. Figure \ref{fig:ggi_comp_realdata1},
shows the posteriors of the number of distinct components associated with three of the relatively influential genes, $FTO$, $PHACTR1$ and $GALNT2$. Observe that the posteriors are almost identical for all the genes, with the mode at $5$ components, and $4$ receiving the next highest probability. Recall that in case of BB, the genes turned out to be highly significant with significant interactions among them and they were associated with different posteriors. After incorporating the environmental factor, the genes seem to play very little role in causing MI and also the posteriors of the number of distinct subpopulations associated with the genes are similar. Our results are also consistent with the four broad sub-populations composed of Caucasians, Han Chinese,
Japanese and Yoruban.

\addtocounter{figure}{7}
\begin{figure}%[htp]
\centering

\subfigure[Posterior of $\tau_{2,0}$.]{ \label{fig:gene_control_2}
\includegraphics[width=6cm,height=5cm]{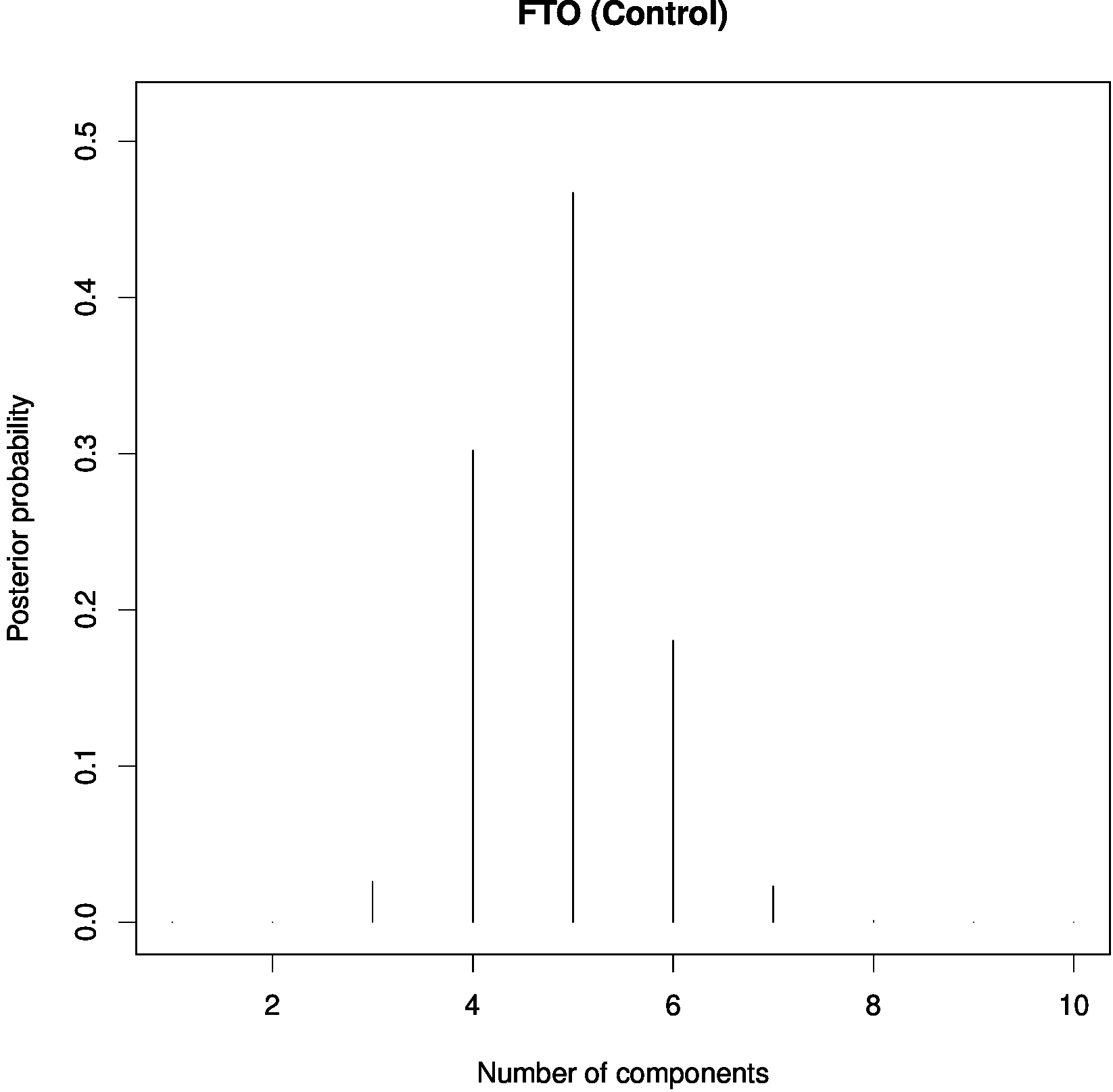}}
\hspace{2mm}
\subfigure[Posterior of $\tau_{2,1}$.]{ \label{fig:gene_case_2}
\includegraphics[width=6cm,height=5cm]{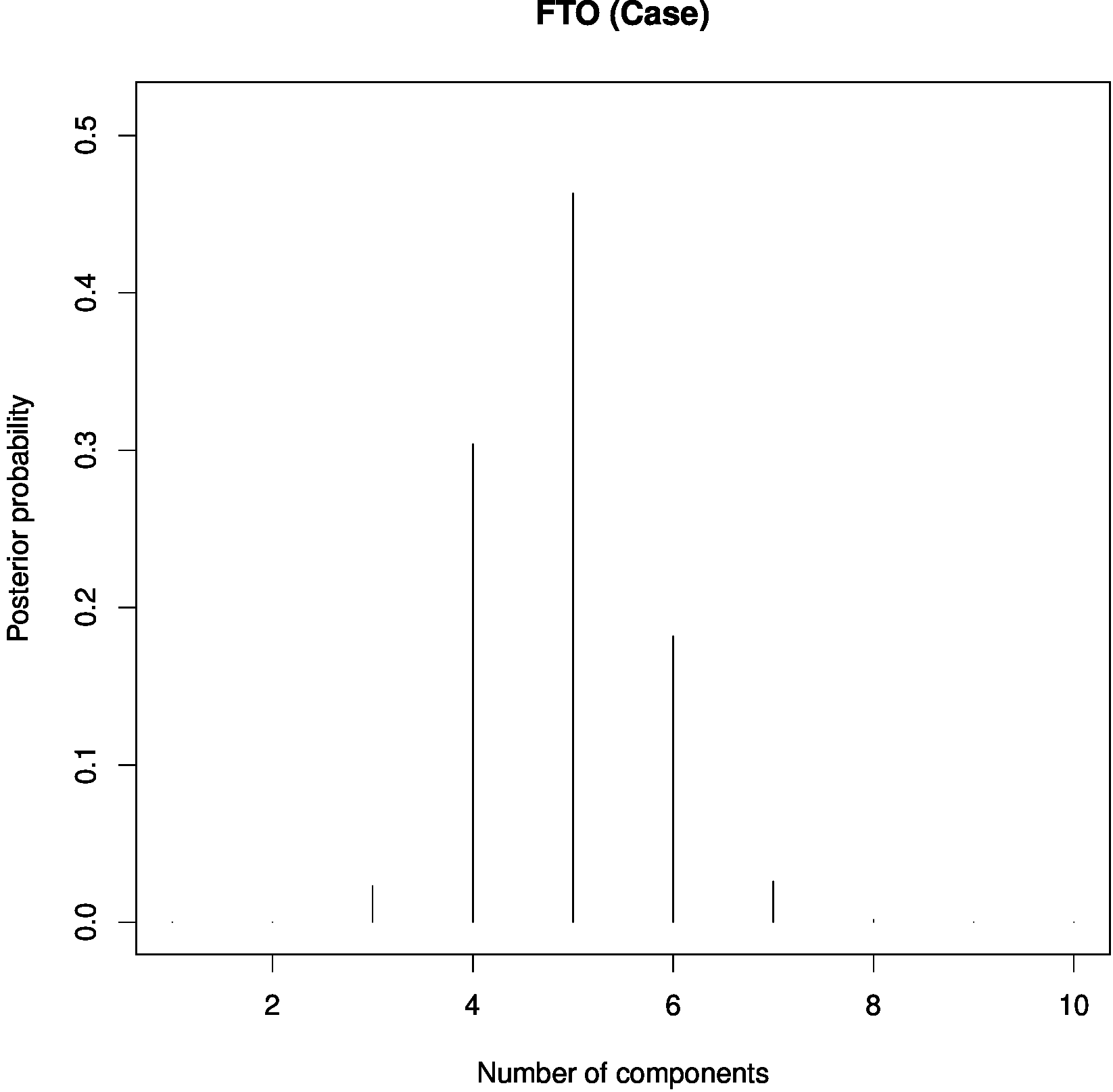}}\\
\vspace{2mm}
\subfigure[Posterior of $\tau_{5,0}$.]{ \label{fig:gene_control_5}
\includegraphics[width=6cm,height=5cm]{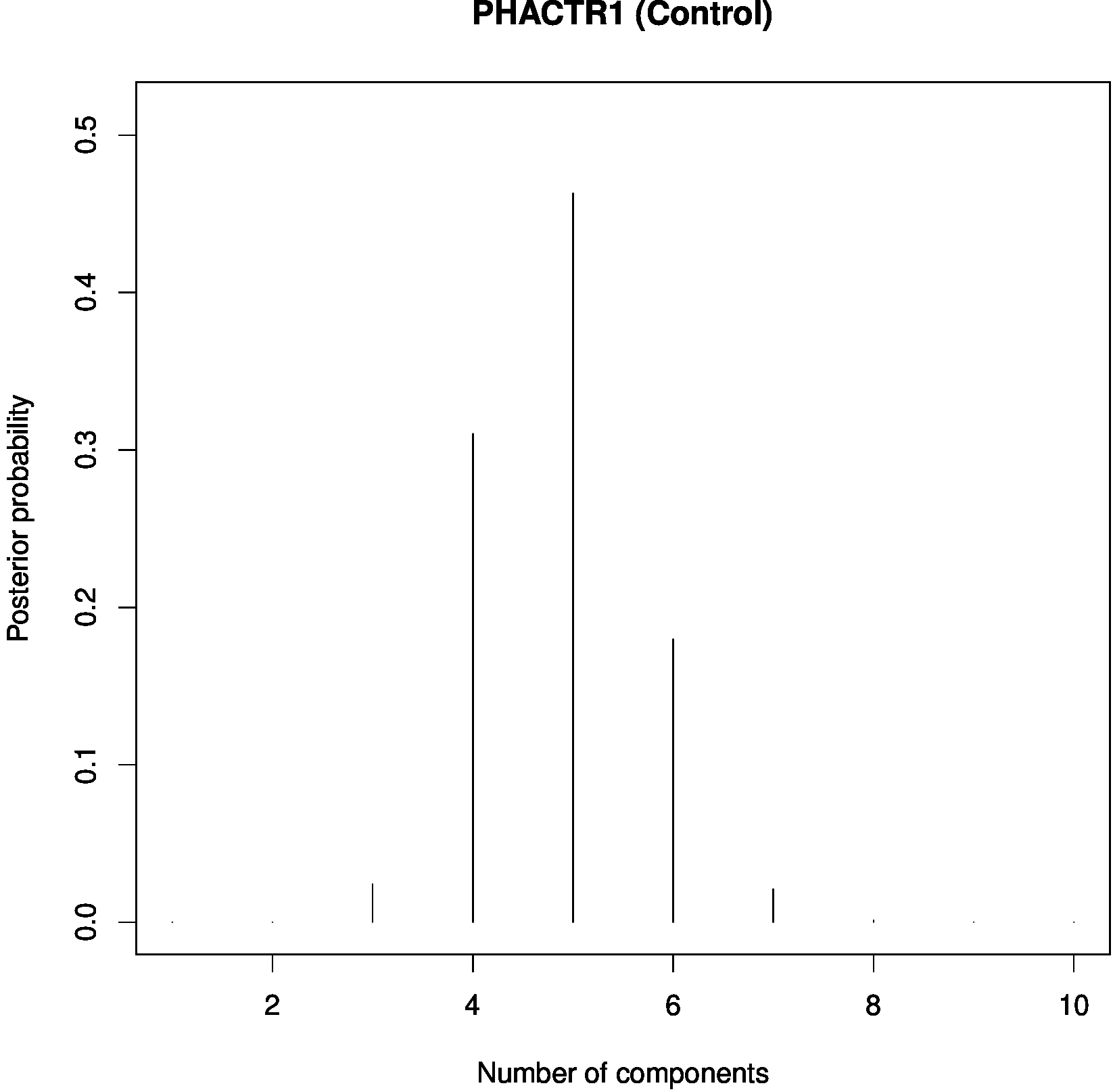}}
\hspace{2mm}
\subfigure[Posterior of $\tau_{5,1}$.]{ \label{fig:gene_case_5}
\includegraphics[width=6cm,height=5cm]{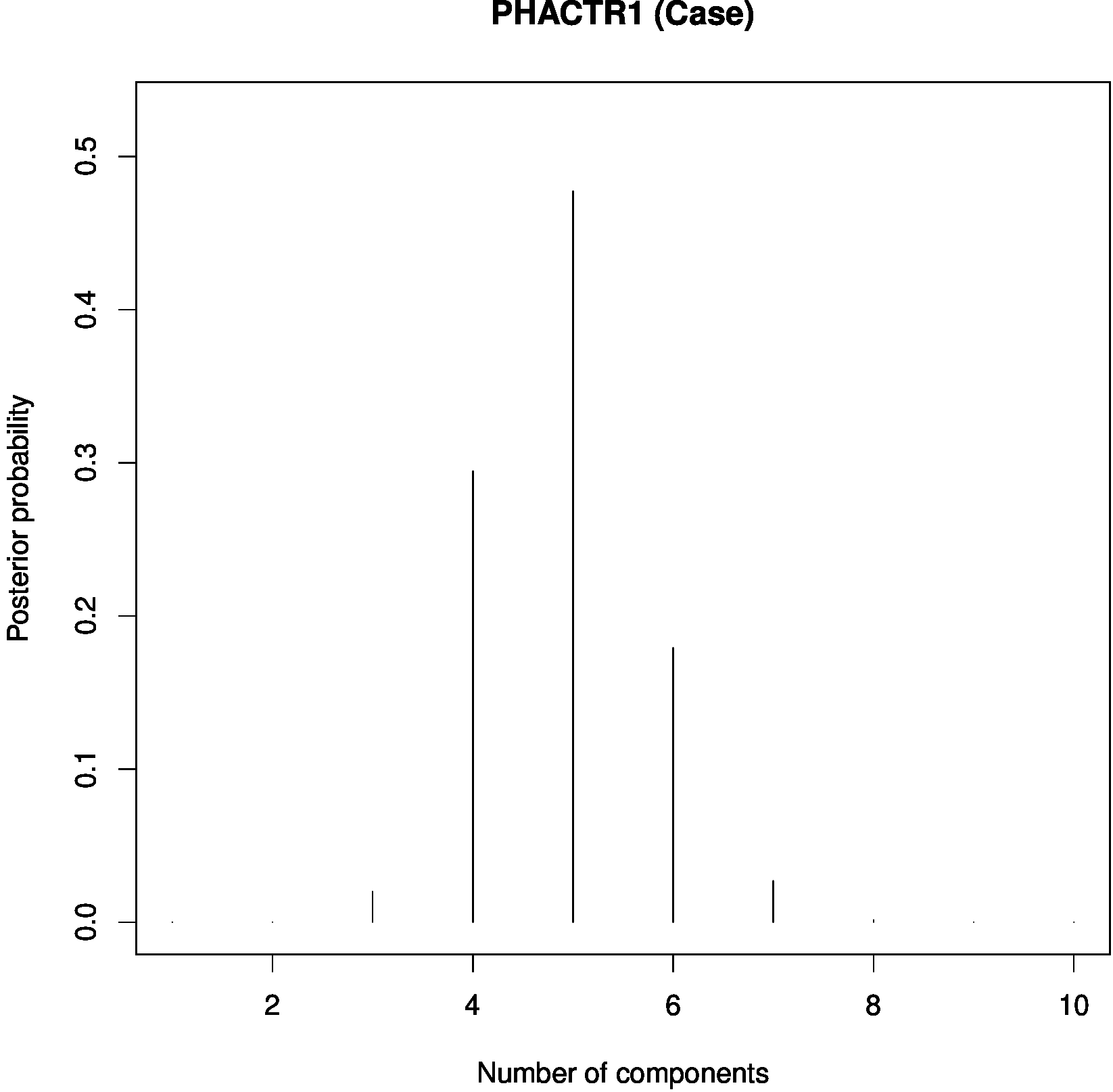}}
\vspace{2mm}
\subfigure[Posterior of $\tau_{14,0}$.]{ \label{fig:gene_control_14}
\includegraphics[width=6cm,height=5cm]{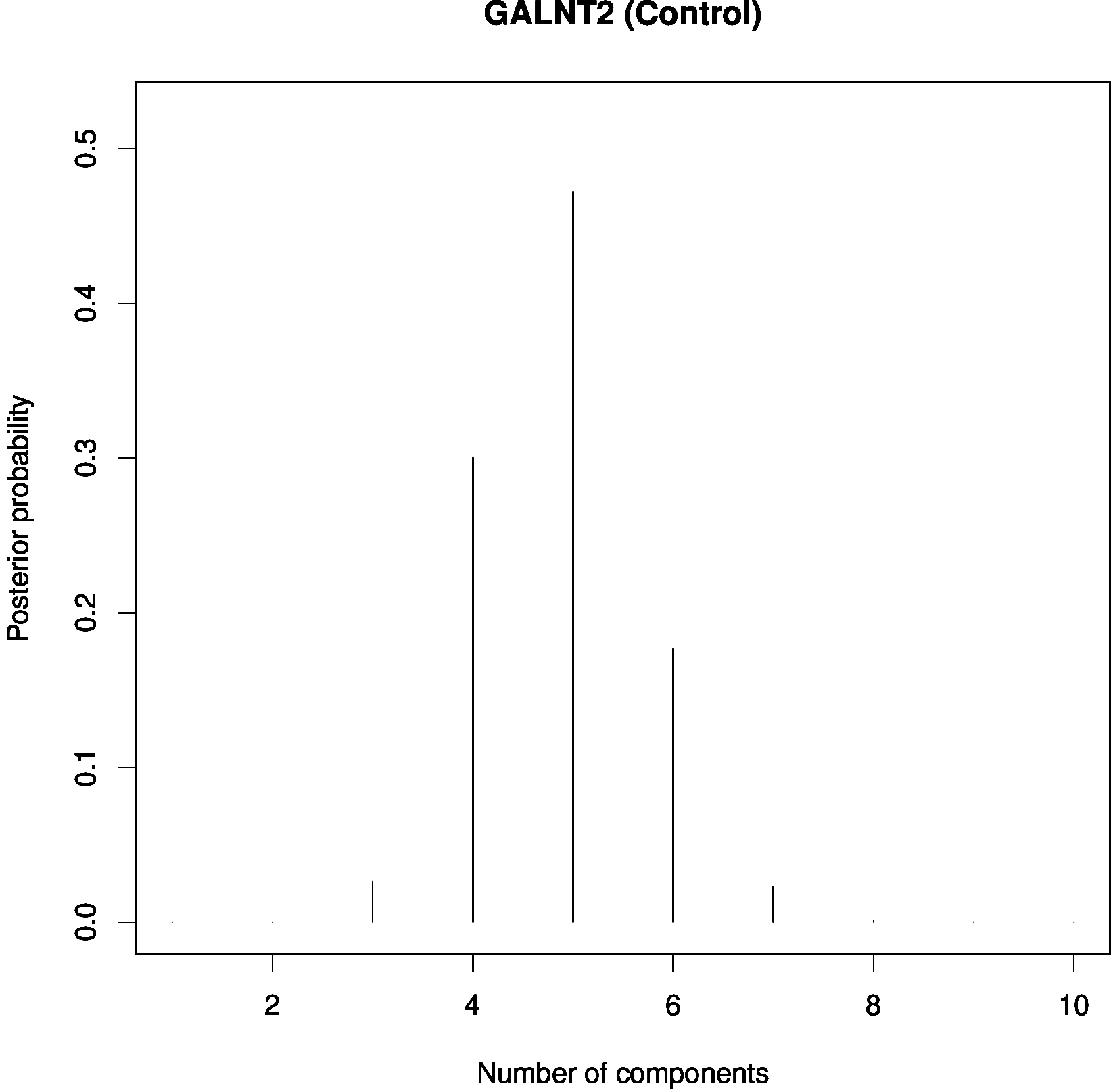}}
\hspace{2mm}
\subfigure[Posterior of $\tau_{14,1}$.]{ \label{fig:gene_case_14}
\includegraphics[width=6cm,height=5cm]{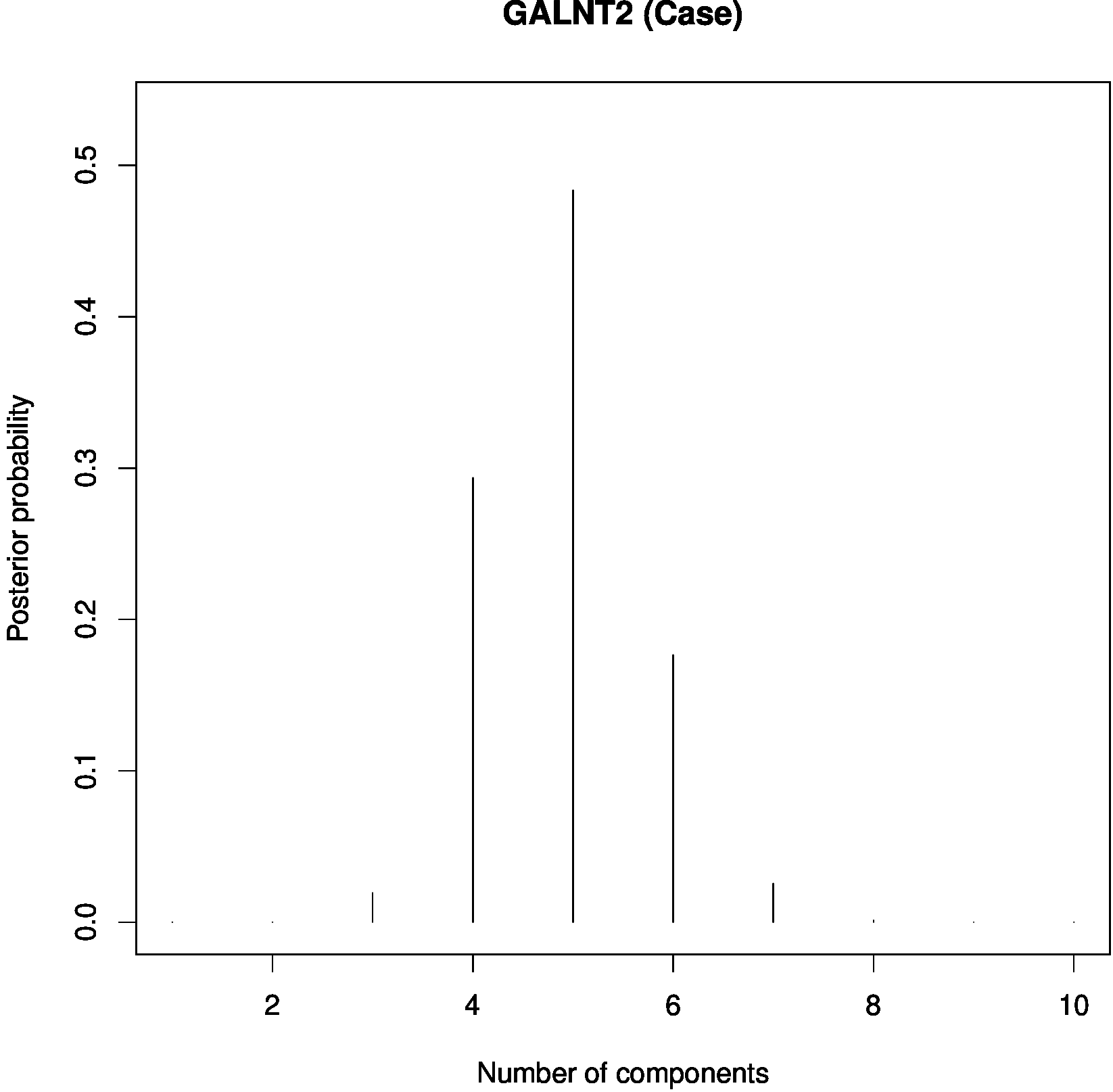}}
\caption{{\bf Posterior of number of components:} Posterior distributions of the number of distinct components $\tau_{j,k}$
for each pair ($j,k$); j=2,5,14; $k=0,1$. %$j=1,2,4,8$; $k=0,1$. 
The left and right panels show the posteriors associated with cases
and controls, respectively.}
\label{fig:ggi_comp_realdata1}
\end{figure}

%\newpage
\pagebreak
%\renewcommand\baselinestretch{1.3}
%\normalsize
\bibliographystyle{ECA_jasa}
\bibliography{irmcmc}

\end{document}